\documentclass[english,10pt]{article}
\usepackage[utf8]{inputenc}
\usepackage{graphicx}
\usepackage{color}
\usepackage{float}
\usepackage{slashed}
\usepackage{subfigure}
\usepackage{amsthm,amsmath,amssymb,mathrsfs,xcolor}
\usepackage{setspace}
\allowdisplaybreaks
\usepackage{jheppub}
\usepackage{xcolor} 
\usepackage{babel}
\usepackage{booktabs}
\usepackage{mathtools}
\usepackage{braket}
\usepackage{cancel}

\usepackage[normalem]{ulem}

\newcommand{\mchieff}{m_\text{$\chi$,eff}}

\newcommand{\xcfo}{x_\text{cfo}}
\newcommand{\xcfohat}{\hat{x}_\text{cfo}}

\newcommand{\omegaeff}{\omega_{\rm eff}}

\newcommand{\centeron}[2]{{\setbox0=\hbox{#1}\setbox1=\hbox{#2}\ifdim
		\wd1>\wd0\kern.5\wd1\kern-.5\wd0\fi \copy0
		\kern-.5\wd0\kern-.5\wd1\copy1\ifdim\wd0>\wd1
		\kern.5\wd0\kern-.5\wd1\fi}}
\newcommand{\ltap}{\>\centeron{\raise.35ex\hbox{$<$}}
	{\lower.65ex\hbox{$\sim$}}\>}
\newcommand{\gtap}{\>\centeron{\raise.35ex\hbox{$>$}}
	{\lower.65ex\hbox{$\sim$}}\>}
\newcommand{\gsim}{\mathrel{\gtap}}

\begin{document}
\title{Dark Matter Coherent Freeze-Out Mechanism: \\
A Comprehensive Analysis}

\author[a,b]{Steven Ferrante,}
\emailAdd{sef87@cornell.edu}
\author[a]{Lillian Luo,}
\emailAdd{yl823@cornell.edu}
\author[a]{Maxim Perelstein,}
\emailAdd{m.perelstein@cornell.edu}
\author[a]{and Bingrong Yu}
\emailAdd{bingrong.yu@cornell.edu}
\affiliation[a]{Department of Physics, LEPP, Cornell University, Ithaca, NY 14853, USA}
\affiliation[b]{Department of Physics, University of Toronto, Toronto, ON, M5S 1A7, Canada}

\abstract{Both Weakly-Interacting Massive Particle (WIMP) and Axion-Like Particle (ALP) are attractive particle dark matter candidates. In this paper we consider a scenario in which both WIMP and ALP are present. The two species interact via a non-derivative quadratic coupling modeled on the axion-nucleon interaction in QCD. We show that even a small (Planck-suppressed) coupling can drastically modify the cosmological history of both species. In the very early Universe, the ALP field acquires a vacuum expectation value, and the system undergoes an inverse phase transition at a temperature around or below the weak scale. WIMPs may remain in equilibrium far beyond the usual freeze-out time, eventually decoupling via a novel ``coherent freeze-out" mechanism. WIMP annihilation cross sections several orders of magnitude above the canonical thermal target value can be consistent with the observed relic density in the coherent freeze-out scenario.
At the same time, coherent oscillations of the ALP field do not begin until after the phase transition, and the initial conditions for these oscillations are set by the thermal physics of the phase transition. The ALP relic density is then predicted unambiguously in terms of the WIMP mass and coupling. We term this scenario ``ALP miracle", since the observed relic density is obtained when these parameters are around their natural values, the weak and the Planck scale. The ALP miracle scenario prefers the ALP mass above about $10^{-3}$~eV, far larger than other popular ALP dark matter models. The ALP miracle parameter space overlaps with a currently allowed region of the QCD axion band, a large fraction of which is expected to be probed by the proposed IAXO experiment.
}

\maketitle

\section{Introduction}
\label{sec:introduction}

The microscopic nature of dark matter (DM) is a central unsolved puzzle in fundamental physics. Among many candidates proposed over the years, two paradigms emerged as particularly promising from the theoretical point of view. Weakly-interacting massive particle (WIMP) hypothesis relates DM to the physics of electroweak interactions, with the characteristic energy scales of $10-1000$ GeV. Axion-like particle (ALP) proposal interprets DM as consisting of light, pseudo-scalar particles, such as the axion particle originally introduced to solve the strong CP problem of the Standard Model (SM). Cosmology and phenomenology of these two scenarios have been studied extensively in the literature, and experimental searches for non-gravitational signatures of WIMP and ALP dark matter are well advanced; see, e.g., Refs.~\cite{Cirelli:2024ssz,Marsh:2015xka} for recent reviews.   

Given that both WIMP and ALP are independently motivated by both the DM puzzle and by other shortcomings of the SM, it is reasonable to consider a scenario where {\it both} of these particles are present. Recently, three of the present authors (SF, MP and BY) pointed out that even a tiny (e.g. Planck-suppressed) interaction coupling the WIMP and the ALP can have a dramatic effect on their cosmological history, and hence on the phenomenological predictions of the model~\cite{Ferrante:2025avp}. Specifically, we considered an effective quadratic coupling between a fermion $\chi$ (the WIMP) and a real (pseudo-)scalar $\phi$ (the ALP):
\begin{align}
{\cal L}_{\rm int} = \frac{1}{\Lambda}\bar{\chi} \chi \frac{\phi^2}{2}\;. \label{eq:Leff}   
\end{align}  
This interaction is motivated by the QCD axion, whose leading non-derivative coupling to nucleons has exactly this form, including the sign.\footnote{A pion-nucleon interaction of this form is generated in the chiral Lagrangian due to the explicit breaking of chiral symmetry by quark masses, and it is then inherited by the axion thanks to the axion-pion mixing~\cite{Epelbaum:2008ga,Machleidt:2011zz,Hook:2017psm,Bigazzi:2019hav,DiLuzio:2020wdo,Fukuda:2021drn,Bauer:2023czj,Grossman:2025cov}. Unlike the case of a linear coupling $\phi\bar{\chi}\chi$, the sign of the quadratic coupling $\phi^2 \bar{\chi} \chi$ plays a crucial role in the dynamics of $\phi$.
Only a positive sign (in the Lagrangian) leads to a cosmological phase transition in the scalar potential once $\chi$ forms a thermal bath.} 
The scale $\Lambda$ is assumed to be high enough to prevent the two species from ever coming into thermal equilibrium. Nevertheless, they do affect each other's cosmological history in an essential way. At high temperatures, the ALP effective mass-squared term becomes negative due to its coherent scattering off of the thermal bath of WIMPs. The ALP then acquires a vacuum expectation value (vev), which in turn shifts the WIMP mass. Eventually, as the Universe expands and cools, the ALP vev returns back to zero in an ``inverse" (symmetry-restoring) phase transition,
while the WIMP freezes out. The coupled cosmological dynamics of the two species was studied in Ref.~\cite{Ferrante:2025avp}. It was found that, depending on the model parameters, the symmetry restoration may occur either via a first-order phase transition (FOPT) or via a more gradual crossover transition. Both have interesting phenomenological consequences: 

\begin{itemize}
    \item {\it First-Order Phase Transition:} In this case, the WIMP freeze-out may be significantly delayed compared to the standard WIMP evolution: this scenario was termed {\bf ``Coherent Freeze-Out"}. Correct WIMP relic density may be obtained with an annihilation cross section much larger than the standard value. To avoid overclosure, ALP has to be unstable in this scenario. 

    \item {\it Crossover:} In this case, both ALP and WIMP are viable as DM candidates. Unlike the standard misalignment scenario, the ALP field initial conditions are effectively set by the dynamics of the phase transition, leading to a parametrically different prediction for ALP DM relic density. Interestingly, the prediction is independent of the ALP mass within a broad range of parameters, and is close to the observed value for natural values of the other parameters (the WIMP mass close to the weak scale and the scale $\Lambda$ near the Planck mass). This coincidence was termed {\bf ``ALP miracle"}. 

\end{itemize}

Building on the analysis of Ref.~\cite{Ferrante:2025avp}, this paper provides a comprehensive analysis of the interacting WIMP-ALP scenario outlined above. First, in Sections~\ref{sec:frame} and~\ref{sec:crossover} we present an improved analysis of the phase transition dynamics which incorporates the full one-loop thermal potential for the ALP throughout the calculations. While the qualitative picture is unchanged from Ref.~\cite{Ferrante:2025avp}, our analysis here provides more precise predictions for various physical quantities of interest. Second, in Section~\ref{subsec:experiments-for-ALP-Miracle} we consider the possibility that our ALP is also coupled to the SM. Such coupling is not required by our scenario, but if it exists, it leads to interesting phenomenology. In this context, we establish the existence of a region in the parameter space where the ALP miracle is compatible with the axion that solves the SM strong-CP problem (the QCD axion). We then discuss the novel experimental target for ALP searches provided by the ALP miracle scenario, with ALP masses $m_\phi\gsim 10^{-3}$~eV, significantly heavier than in the usual axion-DM scenarios. Intriguingly, the overlap of this target region with the QCD axion band will be largely probed by the upcoming IAXO experiment~\cite{IAXO:2019mpb,IAXO:2025ltd}.

In Sections~\ref{sec:FOPT} and~\ref{sec:vacuumdomination}, we study several aspects of the physics and parameter regions that were not analyzed in Ref.~\cite{Ferrante:2025avp}. First, we calculate the tunneling rate in the FOPT regime, and analyze the dynamics of the transition. We find that the system becomes highly supercooled before the phase transition occurs. However, the very rapid nature of the transition unfortunately implies that there is no observable gravitational-wave signal. Second, we analyze the WIMP freeze-out if WIMP annihilation is mediated by a light mediator, as opposed to the point-like interaction assumed in Ref.~\cite{Ferrante:2025avp}. Finally, we consider the region in the parameter space where the Universe becomes dominated by the ALP field vacuum energy during the phase transition, leading to a burst of inflation. We show that this regime may in fact lead to viable cosmology, with WIMP dark matter yield parametrically different from both the standard freeze-out and the non-inflating coherent freeze-out of Ref.~\cite{Ferrante:2025avp}. In the inflating regime, very large WIMP annihilation cross sections, up to six orders of magnitude above the canonical thermal target, may produce the observed relic density. We conclude in Section~\ref{sec:conclusion}.

\section{Theoretical Framework of Coherent Freeze-Out}
\label{sec:frame}

We start with the effective quadratic coupling in Eq.~(\ref{eq:Leff}).
At the level of the effective theory, our setup involves three characteristic mass scales: the cutoff scale $\Lambda$, the fermion mass $m_\chi$, and the scalar mass $m_\phi$. We take $m_\chi$ to be around the weak scale, $\Lambda$ to lie well above it, and $m_\phi$ to be much lighter. This implies a clear hierarchy among the three scales:
\begin{align}
\Lambda \gg m_\chi \gg m_\phi\;.    
\end{align}
In addition, we require that scattering with $\chi$ never thermalizes $\phi$ throughout the cosmic evolution. This imposes the condition
\begin{align}
\Gamma(\chi\phi \to \chi \phi) \sim \frac{T^3}{\Lambda^2} \ll H \quad \Longrightarrow \quad \Lambda \gg \sqrt{T_{\rm RH} M_{\rm Pl}}\;,\label{eq:nonthermalizedcondition}
\end{align}
where $H$ is the Hubble expansion rate,
$T_{\rm RH}$ is the reheating temperature of the Universe after the end of inflation, and $M_{\rm Pl}\approx 1.2 \times 10^{19}~{\rm GeV}$ is the Planck mass. Under this condition, $\phi$ can be treated as a homogeneous classical scalar field throughout its cosmic evolution, while $\chi$ is assumed to remain in thermal equilibrium with the SM plasma at high temperatures.

In the following, we refer to $\chi$ as the WIMP and $\phi$ as the ALP for brevity, although the underlying dynamics apply more generally to any thermalized fermionic species coupled to a light scalar that can be described as a classical field. 

In Sec.~\ref{subsec:potential}, we first compute the full ALP effective potential induced by its coupling to the WIMP thermal bath, and derive the corresponding equation of motion (EOM) for the ALP field. The temperature dependence of the potential gives rise to an inverse phase transition, which underlies the coherent freeze-out mechanism; its essential physics is discussed in Sec.~\ref{subsec:physicalorigin}. Depending on the WIMP and ALP masses, as well as the strength of their coupling, the phase transition can proceed through different patterns, leading to qualitatively distinct cosmological scenarios. A systematic classification of the resulting phase diagram is presented in Sec.~\ref{subsec:phasestructure}.

\subsection{Effective Potential and Field Dynamics}
\label{subsec:potential}

For convenience, we introduce the following dimensionless variables:
\begin{align}
x\equiv \frac{m_\chi}{T}\;,\qquad
\varphi\equiv \frac{\phi}{\sqrt{m_\chi\Lambda}}\;,\qquad
\gamma\equiv \left|1-\frac{\varphi^2}{2}\right| \;.\label{eq:boostfactordef}
\end{align}
We also define two dimensionless parameters:
\begin{align}
\kappa &\equiv \frac{2\pi^2}{g_\chi} \frac{m_\phi^2 \Lambda}{m_\chi^3}\;,\label{eq:kappadef}\\
\eta &\equiv \frac{g_\chi}{2\pi^2} \frac{1}{1.66^2 g_*} \frac{M_{\rm Pl}^2}{m_\chi \Lambda}\;,\label{eq:etadef}
\end{align}
where $g_\chi = 2$ (or $g_\chi =4$) if $\chi$ is a Majorana (or Dirac) fermion, and $g_*$ denotes the number of relativistic degrees of freedom at the relevant temperature.

At high temperatures, $\chi$ is in thermal equilibrium with the SM plasma. Although the coupling in Eq.~(\ref{eq:Leff}) is too weak to thermalize $\phi$, it nevertheless modifies the dispersion relation of $\phi$ through coherent forward scattering as $\phi$ propagates in the $\chi$ bath.
The resulting effective ALP potential can be computed using finite-temperature field theory. At the one-loop level, it is given by~\cite{Ferrante:2025avp}
\begin{align}
V(\phi,T) = \frac{g_\chi}{2\pi^2}m_\chi^4 {\cal V}(\varphi,x)\;,\label{eq:Vfulldimensional}
\end{align}
with
\begin{align}
{\cal V}(\varphi,x) = -\frac{1}{x^4} \int_0^\infty {\rm d}y\,y^2 \log\left[1+e^{-\sqrt{y^2+\gamma^2 x^2}}\right]+\kappa\frac{\varphi^2}{2}\;,\label{eq:Vfull}   
\end{align}
where the first (second) term in Eq.~(\ref{eq:Vfull}) represents the thermal (vacuum) contribution.\footnote{For simplicity, we assume that the zero-temperature potential is well described by a quadratic term $m_\phi^2 \phi^2/2$; see Appendix~\ref{app:naturalness} for a discussion of naturalness.} 
At high temperatures, the thermal potential generates a negative quadratic term, proportional to $-T^2 \phi^2$, which leads to spontaneous breaking of the $Z_2$ symmetry of the theory ($\phi\to-\phi$). As the temperature decreases, the bare mass term eventually dominates, restoring the symmetry. This ``inverse phase transition'' constitutes a key ingredient of the coherent freeze-out mechanism (see Sec.~\ref{subsec:physicalorigin}).

\begin{figure}[t!]
\centering
\includegraphics[width=3.0in]
{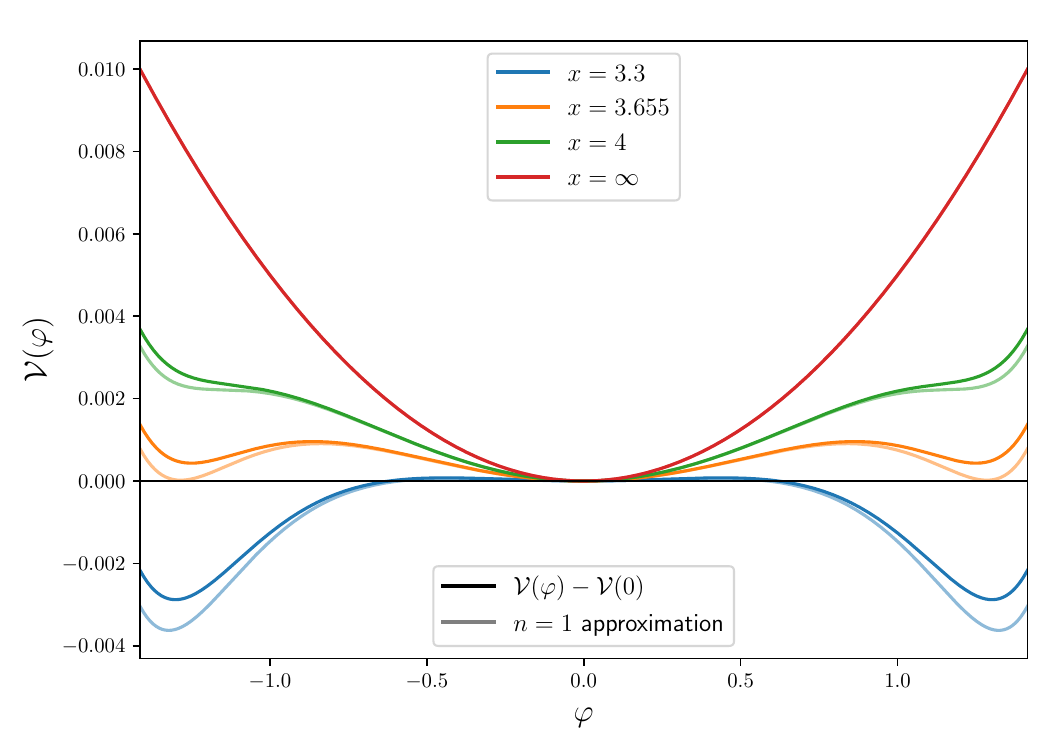}
\includegraphics[width=3.0in]
{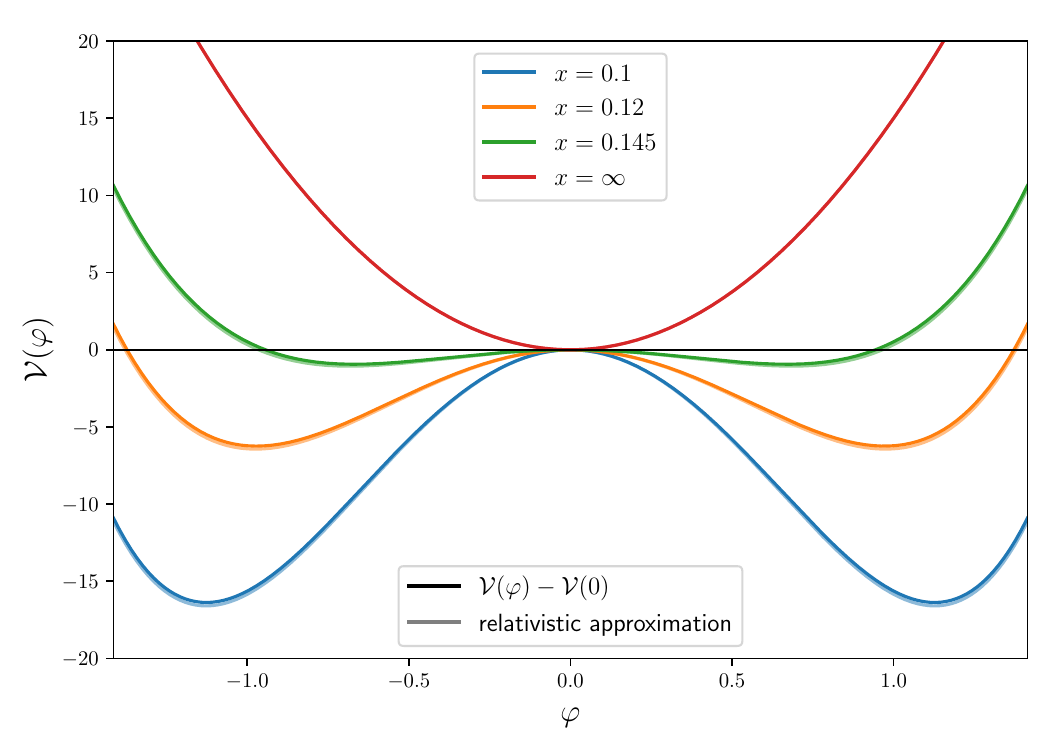}
\caption{\label{fig:potential}
The normalized ALP potential in a thermal bath of WIMPs at several temperatures (Eq.~(\ref{eq:Vfull})).  
For $\kappa < \kappa_c \approx 0.2$, the phase transition is first order (left panel), while for $\kappa > \kappa_c$ it becomes a crossover (right panel). In the lighter curves, we show two different approximations: the $n=1$ approximation (Eq.~(\ref{eq:potentialn1})) for the FOPT regime and the relativistic approximation (Eq.~(\ref{eq:VhighT})) for the crossover regime.}
\end{figure}

The potential in Eq.~(\ref{eq:Vfull}) is shown in Fig.~\ref{fig:potential}. Depending on the coupling strength between the WIMP and the ALP, characterized by the dimensionless parameter $\kappa$ defined in Eq.~(\ref{eq:kappadef}), the symmetry restoration proceeds either through a FOPT or through a crossover (see Sec.~\ref{subsec:phasestructure} for a detailed classification of the phase structure).

\vspace{0.2cm}
The integral in Eq.~(\ref{eq:Vfull}) can be worked out by changing the variable $y \to u \equiv \sqrt{y^2+\gamma^2 x^2}$, 
\begin{align}
\int_0^\infty {\rm d}y\,y^2 \log\left[1+e^{-\sqrt{y^2+\gamma^2 x^2}}\right]
&=  \int_{\gamma x}^\infty {\rm d}u\,u\sqrt{u^2-\gamma^2 x^2}\,\log\left[1+e^{-u}\right]\nonumber\\
& = \sum_{n=1}^\infty \frac{(-1)^{n-1}}{n} \int_{\gamma x}^\infty {\rm d}u\,u\sqrt{u^2-\gamma^2 x^2}\, e^{-nu}\nonumber\\
& = \gamma^2x^2 \sum_{n=1}^\infty \frac{(-1)^{n-1}}{n^2} K_2(n\gamma x)\;, 
\end{align}
where $K_n$ is the modified Bessel function of the second kind. The potential in Eq.~(\ref{eq:Vfull}) can then be recast into a more explicit form
\begin{align}
{\cal V}(\varphi,x)  &= -\frac{\gamma^2}{x^2} \sum_{n=1}^\infty \frac{(-1)^{n-1}}{n^2} K_2(n\gamma x) +\kappa\frac{\varphi^2}{2}\;.\label{eq:Vfullseries}  
\end{align}
Two limits of the potential are useful:
\begin{itemize}
    \item Relativistic limit $\gamma x \ll 1$: this corresponds to the standard high-temperature expansion in Eq.~(\ref{eq:Vfull}), giving
\begin{align}
{\cal V}(\varphi,x) = - \frac{1}{x^4}\left(\frac{7\pi^4}{360}  - \frac{\pi^2}{24} \gamma^2 x^2  + {\cal O}(\gamma^4 x^4) \right)+ \kappa \frac{\varphi^2}{2}\,.     
\end{align}    
Ignoring terms independent of $\varphi$ and terms with higher powers of $\gamma^2 x^2$, one obtains
\begin{align}
{\cal V}(\varphi,x) \approx \frac{\pi^2 }{96 x^2} \varphi^4 - \left(\frac{\pi^2}{24 x^2}-\frac{\kappa}{2}\right)\varphi^2\;.\label{eq:VhighT}    
\end{align}
In this region, the symmetry is broken at $x < x_c = \pi /(2\sqrt{3 \kappa})\sim 1/\sqrt{\kappa}$ and is restored for $x>x_c$.

\item  Non-relativistic limit $\gamma x \gtrsim 1$: in this case, it is more convenient to use Eq.~(\ref{eq:Vfullseries}). Since $K_2(z) \sim e^{-z}/\sqrt{z}$ for $z \gtrsim 1$, one can ignore all modes except the $n=1$ mode:
\begin{align}
{\cal V}(\varphi,x) \approx - \frac{\gamma^2}{x^2}K_2(\gamma x) + \kappa \frac{\varphi^2}{2}\;.
\label{eq:potentialn1}
\end{align}
\end{itemize}
The comparison between the full potential (\ref{eq:Vfull}) and the approximate expressions (\ref{eq:VhighT})-(\ref{eq:potentialn1}) at different temperatures is shown in Fig.~\ref{fig:potential}.

\subsubsection{Dynamics of the ALP}
Given the ALP potential in Eqs.~(\ref{eq:Vfulldimensional})--(\ref{eq:Vfull}), the EOM of the ALP in a flat  Friedmann-Robertson-Walker (FRW) Universe is given by
\begin{align}
\Ddot{\phi} + 3 H \dot{\phi} + \frac{\partial V (\phi,T)}{\partial \phi}=0\;,
\end{align}
where the dot denotes the derivative with respect to time. Assuming the radiation-dominated era and changing the variable from $t$ to $x$ and using the normalized field defined in Eq.~(\ref{eq:boostfactordef}), one obtains
\begin{align}
\varphi'' + \frac{2}{x} \varphi' + \eta \,x^2\,\frac{\partial {\cal V} (\varphi,x)}{\partial \varphi} = 0\;, \label{eq:EOMgeneral}   
\end{align}
where the prime denotes the derivative with respect to $x$, and we neglect the temperature dependence of $\eta$.

Substituting the potential (\ref{eq:Vfullseries}) into (\ref{eq:EOMgeneral}), we arrive at
\begin{align}
\varphi'' + \frac{2}{x}\varphi' - \eta\,x\,\varphi\,{\rm sgn}\left(1-\frac{\varphi^2}{2}\right) \sum_{n=1}^{\infty} \left[\frac{(-1)^{n-1}}{n} \gamma^2 K_1(n \gamma x)\right] +\eta\,\kappa\,x^2 \varphi =0\;, \label{eq:EOMgeneral2}   
\end{align}
where the sign function is defined as ${\rm sgn}(x)\equiv x/|x|$.

\begin{figure}[t!]
\centering
\includegraphics[width=3.0in]
{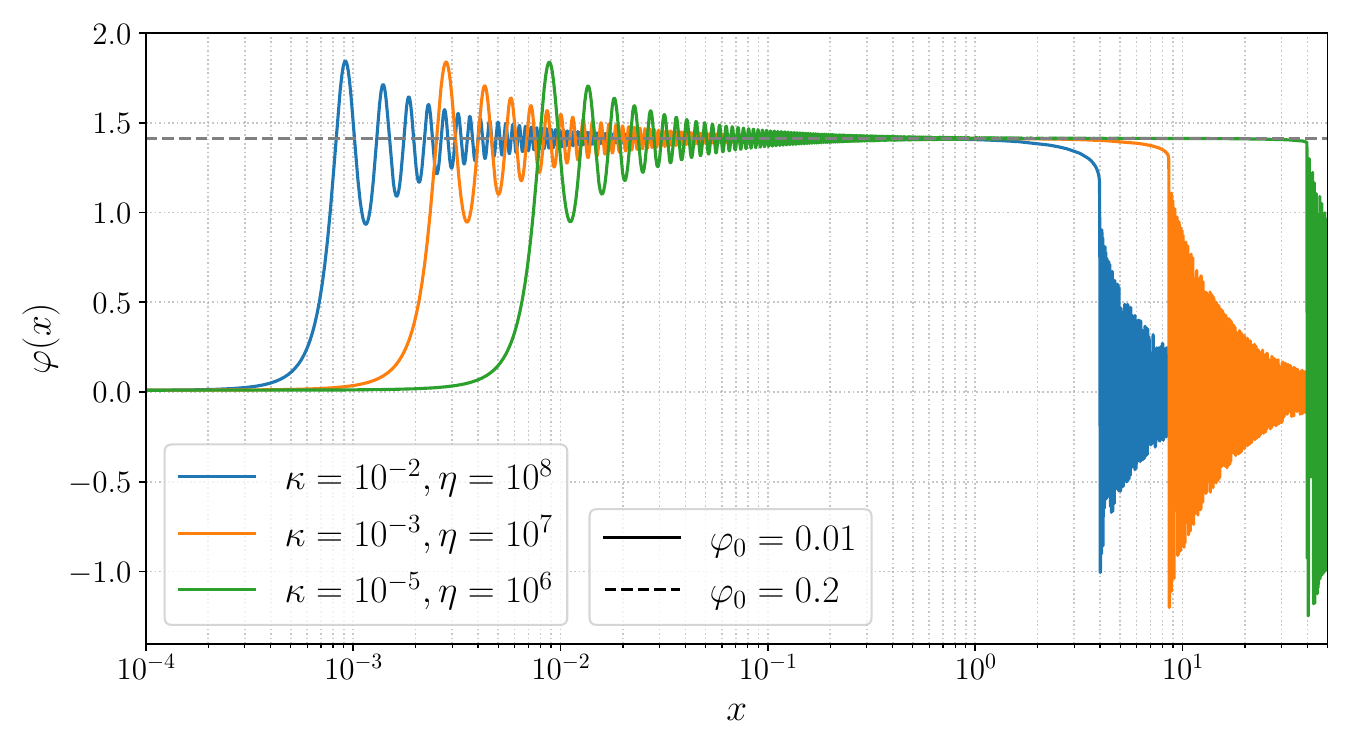}
\includegraphics[width=3.0in]
{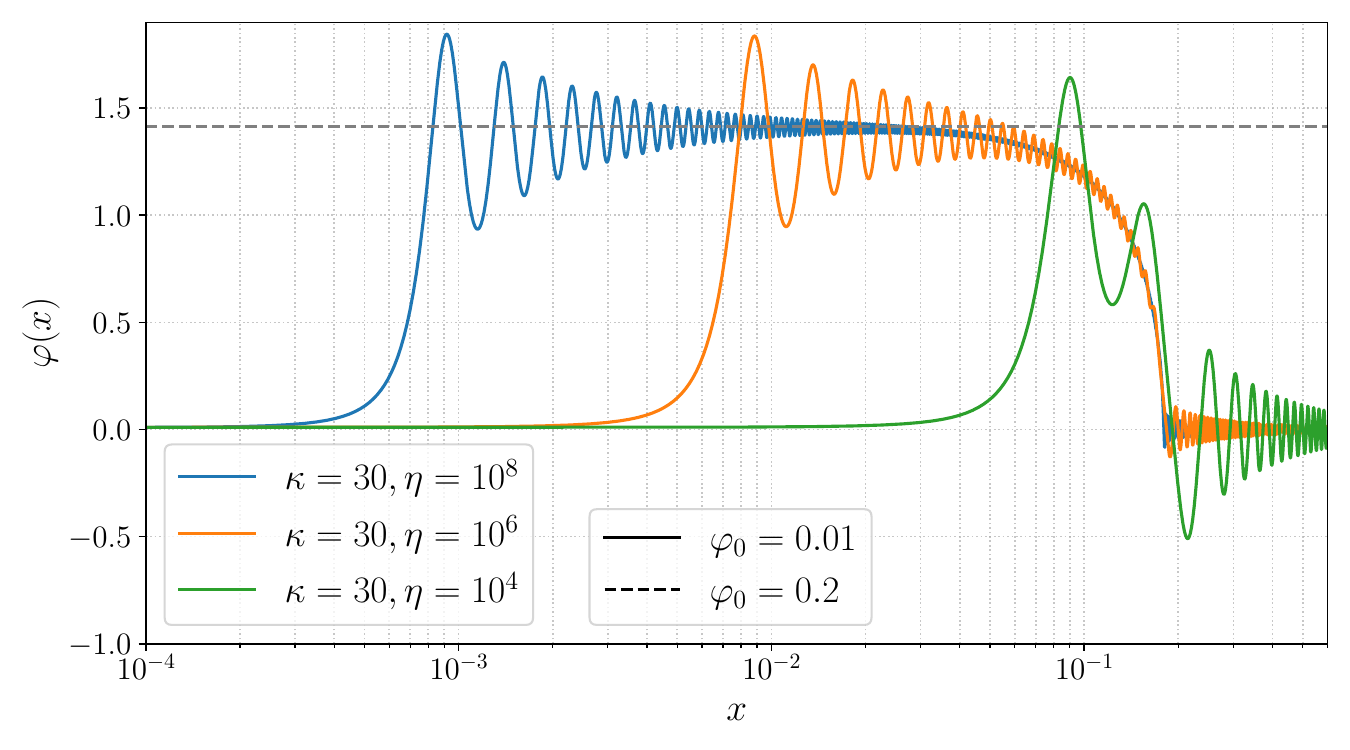}
\caption{\label{fig:solutions1st2nd}
Evolution of the ALP field in the first-order phase transition regime (left panel) and the crossover regime (right panel) (see Sec.~\ref{subsec:phasestructure} for classification of the phase diagram). 
Different colors correspond to different choices of $\kappa$ and $\eta$, while different line styles denote different initial field displacements.}
\end{figure}

The general EOM of the ALP in Eq.~(\ref{eq:EOMgeneral2}) depends on only two dimensionless parameters, $\kappa$ and $\eta$, defined in Eqs.~(\ref{eq:kappadef})--(\ref{eq:etadef}).  Physically, they characterize the competition between the relevant scales at a given inverse temperature $x\equiv m_\chi/T$: 
$\kappa$ compares the bare ALP mass with the thermal mass induced by the WIMP bath, while $\eta$ compares the thermal mass with the Hubble friction. To make this more explicit, we recall that in the high-temperature limit the ALP thermal mass squared scales as $m_T^2 \sim -m_\chi T^2/\Lambda$. Combining this relation with Eqs.~(\ref{eq:kappadef})--(\ref{eq:etadef}) and the Hubble rate during radiation domination, we obtain
\begin{align}
\kappa \sim \frac{m_\phi^2}{\left|m_T^2\right| x^2}\;,\qquad
\eta \sim \frac{\left|m_T\right|^2}{H^2 x^2}\;.\label{eq:kappaetaphysical}
\end{align}

The numerical solution of Eq.~(\ref{eq:EOMgeneral2}) is shown in Fig.~\ref{fig:solutions1st2nd} for representative values of $\kappa$ and $\eta$, and the initial field displacement. 
The qualitative features of the field evolution can be readily understood from Eq.~(\ref{eq:kappaetaphysical}):
at sufficiently high temperatures, the field is frozen at its initial value because the Hubble friction dominates over the effective mass. As the temperature decreases, the negative thermal mass squared eventually overcomes the Hubble friction, driving the field toward the symmetry-breaking vacuum at $\varphi = \sqrt{2}$,\footnote{Note that although the curvature at the origin is negative due to the thermal contribution, the potential remains bounded from below thanks to the back-reaction effect, which is encoded in the nonlinear terms of Eq.~(\ref{eq:EOMgeneral2}).} thereby erasing the dependence on the initial field displacement. 
As the temperature decreases further, the bare ALP mass eventually becomes dominant, restoring the symmetry, after which the field oscillates around the symmetric vacuum at $\varphi=0$.
We defer a detailed quantitative analysis of the field evolution in the different regimes to Secs.~\ref{sec:crossover}--\ref{sec:vacuumdomination}.

\subsection{The Physical Origin of Coherent Freeze-Out}
\label{subsec:physicalorigin}

The coherent freeze-out (CFO) mechanism is driven by an inverse phase transition, in which the symmetry is broken at high temperatures and restored as the temperature decreases.
In general, the thermodynamically preferred phase minimizes the free energy
\begin{align}
    F = E - TS \;,
\end{align}
where $E$ and $S$ denote the energy and entropy of the system, respectively. In an ordinary phase transition, the symmetric (disordered) phase carries a larger entropy than the broken (ordered) phase and is therefore favored at sufficiently high temperatures. By contrast, an inverse phase transition occurs when the broken phase instead possesses a larger entropy, for example, because it supports more low-energy excitations. Examples of inverse phase transitions in particle-physics contexts can be found in Refs.~\cite{Mohapatra:1979vr,Dvali:1995cc,Pietroni:1996zj,Riotto:1997tf,Espinosa:2004pn,Shakya:2025mdh}, while an example in a condensed-matter system is discussed in Ref.~\cite{PhysRevE.72.046107}.

In our setup, this happens because a nonzero ALP vev \emph{reduces} the effective WIMP mass.
As a result, WIMPs remain relativistic for longer in the broken phase, delaying the Boltzmann suppression of their number density and thereby increasing the entropy of the thermal bath relative to the symmetric phase.
This entropy gain outweighs the energy cost of the symmetry-breaking vacuum, stabilizing the broken phase at high temperatures. As the temperature decreases, the thermal contribution gradually weakens, the bare ALP mass eventually dominates, and the symmetry is restored, thereby triggering the decoupling of the two sectors. In the following, we make this picture quantitative.

\begin{figure}[t!]
\centering
\includegraphics[width=3in]
{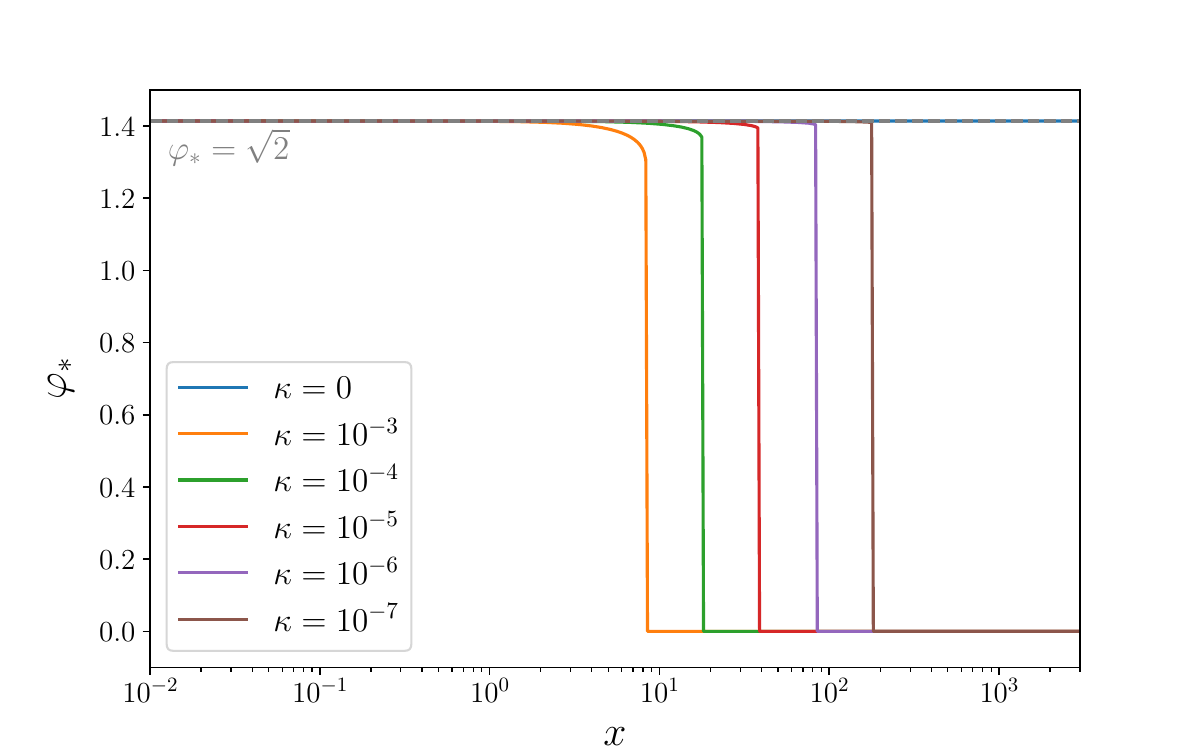}
\includegraphics[width=3in]
{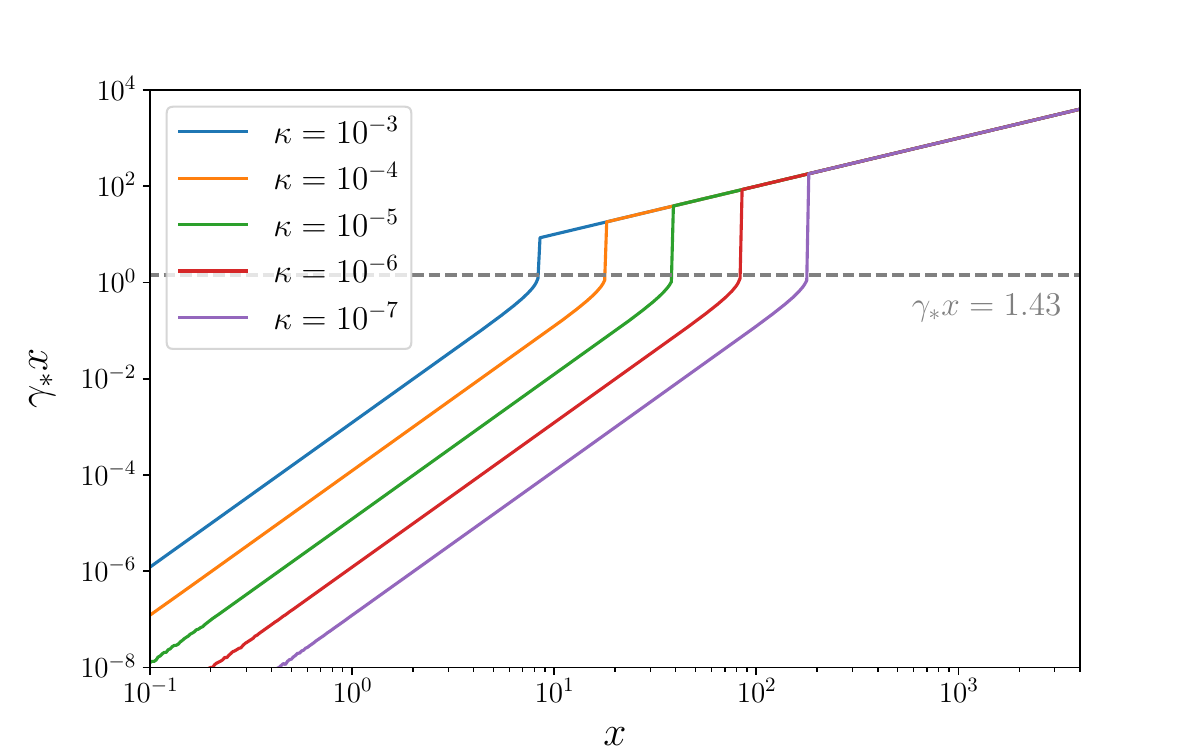}
\caption{\label{fig:vacuumevolution} The evolution of the symmetry-breaking vacuum $\varphi_*$ (left) and the boost factor $\gamma_* \equiv |1-\varphi_*^2 / 2| $ (right) as a function of the inverse temperature $x \equiv m_\chi / T$ according to the full thermal potential (\ref{eq:Vfull}). Before coherent freeze-out, the ratio $m_{\chi, \text{eff}} / T = \gamma_* x$ remains below $\mathcal{O}(1)$ and scales approximately as $x^{3}$. Once $x$ reaches $\xcfo$, $\varphi_*$ drops sharply to zero and $\gamma_*$ rises sharply to unity, signaling the onset of coherent freeze-out.}
\end{figure}

As the solution of the EOM (\ref{eq:EOMgeneral2}) indicates, at high temperatures, the ALP is trapped in the symmetry-breaking vacuum $\varphi_* \neq 0$. The evolution of $\varphi_*$ as a function of the inverse temperature can be described by a simple analytical expression (see Appendix~\ref{app:falsevacuum} for details):
\begin{align}
\varphi_*(x) = \sqrt{2\left(1-f(x)\, \kappa x^2\right)}\;,\label{eq:phistarevolution} 
\end{align}
where $f(x) \sim {\cal O}(1)$ is a monotonically increasing function with the asymptotic behavior $f(x\ll \xcfo) = 12/\pi^2$  and $f(\xcfo) = c_1/c_2$. Here $c_1 \approx 1.429$ and $c_2 \approx 0.585$ arise from numerical summation over Bessel functions. The quantity $\xcfo$ is defined as the temperature at which the false minimum at $\varphi_*$ disappears, i.e., 
\begin{align}
\frac{\partial{\cal V}(\varphi,\xcfo)}{\partial \varphi} \Bigg|_{\varphi=\varphi_*}=\frac{\partial^2{\cal V}(\varphi,\xcfo)}{\partial \varphi^2} \Bigg|_{\varphi=\varphi_*} = 0\;.\label{eq:xcfoexpression}  
\end{align}
Using the expression of ${\cal V}$ in Eq.~(\ref{eq:Vfullseries}), one finds (see Appendix~\ref{app:falsevacuum})
\begin{align}
\xcfo = \left(c_2/\kappa\right)^{1/3} \approx 0.836\,\kappa^{-1/3}\;. \label{eq:xcfotilde}    
\end{align}
By definition, the local minimum disappears at $\xcfo$, beyond which $\varphi_* = 0$. The evolution of $\varphi_*(x)$ is shown in Fig.~\ref{fig:vacuumevolution}.

For $x>\xcfo$, the ALP field  relaxes to the symmetric vacuum at $\varphi=0$ and redshifts as ordinary matter,\footnote{For the range of ALP masses considered in this paper, we have $m_\phi \gg H(\xcfo)$. Therefore, once the local minimum at $\varphi_*$ disappears, the field is no longer frozen by Hubble friction and rapidly evolves toward the origin.} $\varphi(x) \propto x^{-3/2}$. Therefore, the ALP field value can be approximated as a piecewise function:
\begin{align}
\varphi(x) = 
\begin{cases}
\varphi_*(x)  & \text{for}\;x < \xcfo\;,\\[10pt]
\sqrt{2\left(1-\dfrac{c_1}{\xcfo}\right)}\left(\dfrac{\xcfo}{x}\right)^{3/2}  & \text{for}\;x > \xcfo\;.
\end{cases}\label{eq:ALPevolution}
\end{align}
The back-reaction of the ALP field on the WIMP bath reduces the WIMP mass from $m_\chi$ to $\mchieff = \gamma m_\chi$. Using Eq.~(\ref{eq:ALPevolution}), one obtains
\begin{align}
\mchieff/T = \gamma x = 
\begin{cases}
c_2f(x)\,\dfrac{x^3}{\xcfo^3} & \text{for}\;x < \xcfo\;,\\[15pt]  
x- \dfrac{\xcfo^3}{x^2} + c_1 \dfrac{\xcfo^2}{x^2} & \text{for}\;x > \xcfo\;. \label{eq:mchieffoverT}
\end{cases}
\end{align}

The crucial point is that $\mchieff/T$ remains below $\mathcal{O}(1)$ for $x < \xcfo$. Consequently, $\chi$ cannot undergo conventional non-relativistic freeze-out prior to $\xcfo$, regardless of how large $x \equiv m_\chi/T$ becomes.\footnote{In principle, the scalar field can tunnel through the potential barrier before the false minimum disappears classically. However, as we show in Sec.~\ref{subsec:bounceaction}, the tunneling rate is highly suppressed until very close to $\xcfo$. As a result, the nucleation temperature is approximately equal to $\xcfo$.} More precisely, for $x<\xcfo$, $\mchieff/T$ increases monotonically with $x$, saturating at $x=\xcfo$ at a value $c_1 \approx 1.429$; for $x> \xcfo$, the relaxation of the ALP field toward the origin rapidly restores $\mchieff$ to its bare mass $m_\chi$, thereby triggering freeze-out. 

This is the central idea of the coherent freeze-out mechanism: unlike conventional freeze-out, which is driven by the expansion of the Universe, coherent freeze-out is dynamically controlled by a phase transition in the ALP potential and is  triggered by the onset of coherent oscillations of the ALP field. This mechanism leads to two notable effects in both the WIMP and ALP sectors:
\begin{itemize}
    \item The WIMP freeze-out can be significantly delayed, occurring at $\xcfo\gg \mathcal{O}(25)$ for  sufficiently small $\kappa$,  thereby substantially modifying the predicted relic abundance compared to the standard freeze-out scenario.

    \item The initial dependence on the ALP displacement is erased by thermal dynamics. The ALP begins its coherent oscillation with a predictable initial field value, resulting in a relic abundance that can differ markedly from and is more predictive than that obtained via the standard misalignment mechanism.
\end{itemize}

\subsection{Phase Structure and Regime Classification}
\label{subsec:phasestructure}
\begin{figure}[t!]
\centering
\includegraphics[width=5in]
{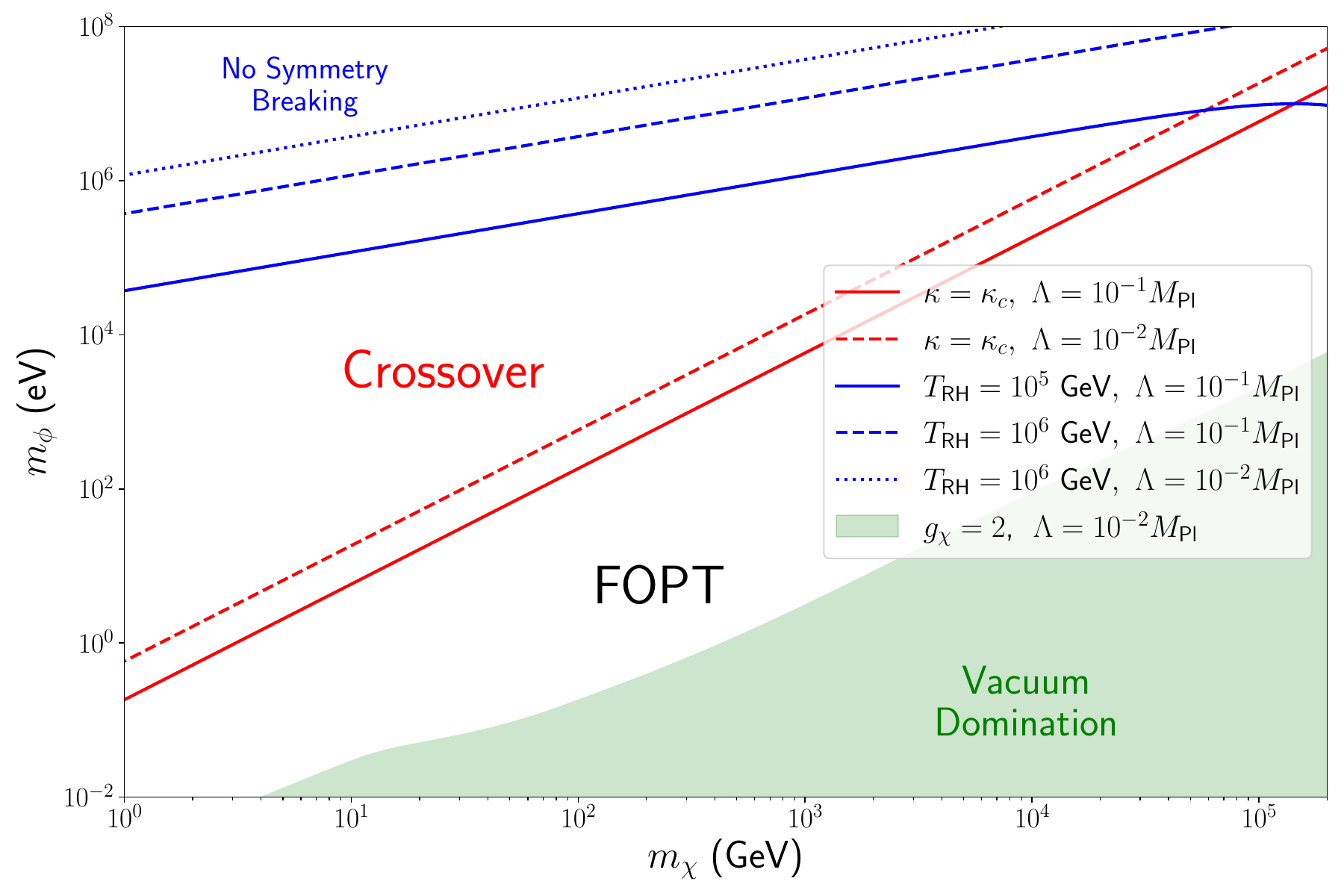}
\caption{\label{fig:phasediagram}
Phase diagram in the $(m_\chi, m_\phi)$ plane. 
Below the $\kappa = \kappa_c \approx 0.2$ contour (red lines), symmetry restoration proceeds through a first-order phase transition (FOPT); above it, the transition is a crossover. For small enough $\kappa$, vacuum energy can dominate and lead to an inflationary phase (green shaded region). For sufficiently large $\kappa$ (above the blue lines), the symmetry is never broken during the cosmic evolution. The crossover, FOPT (without vacuum domination), and vacuum domination regimes are discussed in Secs.~\ref{sec:crossover}, \ref{sec:FOPT}, and \ref{sec:vacuumdomination}, respectively.
}
\end{figure}

The potential in Eq.~(\ref{eq:Vfull}) is controlled by a single parameter $\kappa$. The order of the phase transition depends on this parameter, and we can find the critical values of $\kappa$ and $x$, $\kappa_c$ and $x_\text{tri}$, that split the FOPT and crossover regimes (also known as the tricritical point). Expanding Eq.~(\ref{eq:Vfull}) to the quartic order in $\varphi$ around $\varphi = 0$,  
\begin{align}
    \mathcal{V}(\varphi, x) \approx 
    \frac{1}{2!}a(x, \kappa)\varphi^2 + 
    \frac{1}{4!}b(x)\varphi^4\;,
\end{align}
where
\begin{align}
    a(x, \kappa) &= \kappa - 
    \sum_{n=1}^{\infty}
    \frac{(-1)^{n-1}}{nx}
    K_1(n x)\;,
\\
    b(x) &= 
    3
     \sum_{n=1}^{\infty}
    (-1)^{n-1}
    \bigg( \frac{K_1(nx)}{nx}-
    K_0(nx)  
    \bigg)\;.
\end{align} 
The tricritical point is determined by the simultaneous conditions
\begin{align}
    a(x_{\rm tri},\kappa_{c}) = 0\;,
    \qquad
    b(x_{\rm tri}) = 0\;.
\end{align}
Starting with the quartic coefficient, which is independent of $\kappa$, the tricritical temperature is found numerically to be
\begin{align}
    x_{\rm tri} \approx 1.43\;.
\end{align}
The corresponding critical value $\kappa_{c}$ is then fixed by requiring 
that the quadratic coefficient also vanishes at this temperature, numerically giving:
\begin{align}
    \kappa_c \approx 0.2 \;.
\end{align}
In Fig.~\ref{fig:phasediagram}, this value $\kappa_c$ is denoted by the red lines, which split the parameter space into regions of FOPT and crossover. 

For sufficiently large $\kappa$, the symmetry may never have been broken in the first place. This occurs if the quadratic coefficient $a(x, \kappa)$ remains positive even at the reheating temperature $T_{\rm RH}\equiv m_\chi/x_\text{RH}$. The corresponding condition, $a(x_\text{RH}, \kappa) > 0$, implies a maximum value of $\kappa$ we denote as $\kappa_\text{max}$: 
\begin{align}
    \kappa_\text{max} = \sum_{n=1}^{\infty}
    \frac{(-1)^{n-1}}{nx_\text{RH}}
    K_1(nx_\text{RH}) .
\end{align}
The condition for symmetry breaking is then $\kappa < \kappa_\text{max}$ and is shown in Fig.~\ref{fig:phasediagram} as the region below the blue lines. 

For sufficiently small $\kappa$, the ALP can remain trapped in the false vacuum long enough to have its vacuum energy dominate over the total energy density of the Universe, which would otherwise be radiation-dominated during this epoch. 
This occurs when the ratio of the vacuum energy density to the radiation energy density (assuming $x \gg 1$)
\begin{align}
    \alpha(x) \equiv \frac{\rho_\phi}{\rho_\text{rad}} 
    \approx \frac
    {\frac{g_\chi}{2\pi^2} m_\chi^4\kappa}
    {\frac{\pi^2}{30} g_* T^4}
    = 
    \frac{15 g_\chi}{\pi^4 g_*}\kappa x^4\;,
\label{eq:alphadef}
\end{align}
exceeds unity at $x=\xcfo$. In terms of $\kappa$ or $\xcfo$, vacuum domination occurs when
\begin{align}
    \kappa & < \kappa_{\rm vac} \equiv \left(\frac{0.836^4\cdot 15 g_\chi}{\pi^4 g_*}\right)^3
    \approx
    5 \times 10^{-9} 
    \left(\frac{g_\chi/2}{g_*/90}\right)^3, \label{eq:kappavac}
\end{align}
or equivalently, 
\begin{align}
\xcfo &>x_{\rm vac}\equiv \frac{\pi^4 g_*}{0.836^3\cdot 15g_\chi}\approx 
    500
    \left(\frac{g_*/90}{g_\chi/2}\right),\label{eq:xvac}
\end{align}
where $g_*$ is evaluated at $\xcfo$.
This region is shaded in green in Fig.~\ref{fig:phasediagram}, where the Universe undergoes a period of vacuum domination. We defer a detailed study of this inflationary phase to Sec.~\ref{sec:vacuumdomination}.

To summarize, depending on the value of $\kappa$, the cosmological history of the model falls into one of the four distinct scenarios (see Fig.~\ref{fig:phasediagram}):
\begin{enumerate}
    \item $\kappa > \kappa_{\rm max}(x_{\rm RH})$: The symmetry remains unbroken throughout the cosmic evolution, and the WIMP and ALP sectors evolve independently. Since this regime is phenomenologically trivial, we do not discuss it further.

    \item $\kappa_c < \kappa < \kappa_{\rm max}(x_{\rm RH})$: The symmetry is broken at high temperatures and restored at low temperatures through a crossover. This regime is studied in Sec.~\ref{sec:crossover}.

    \item $\kappa_{\rm vac} < \kappa < \kappa_c$: The symmetry is broken at high temperatures and restored at low temperatures through a FOPT. The two sectors decouple before vacuum domination sets in. This regime is studied in Sec.~\ref{sec:FOPT}.

    \item $\kappa < \kappa_{\rm vac}$: As in regime 3, the symmetry is restored through a FOPT, but the decoupling occurs only after the vacuum energy has come to dominate the Universe. Consequently, the Universe undergoes a period of vacuum-dominated inflation. This regime is discussed in Sec.~\ref{sec:vacuumdomination}.
\end{enumerate}

\section{Crossover Regime}
\label{sec:crossover}

In the crossover regime ($\kappa > \kappa_c \approx 0.2$), the symmetry restoration of the ALP potential typically occurs at $x \lesssim {\cal O}(1)$. As a result, the ALP decouples from the WIMP bath well before the standard freeze-out at $x_{\rm fo}\approx 25$, and therefore has a negligible impact on the WIMP freeze-out. By contrast, the WIMP thermal bath significantly modifies the dynamics of the ALP, leading to a relic abundance that is parametrically different from the standard misalignment prediction, as we compute in this section.

In this regime, it is a good approximation to take the relativistic limit $\gamma x \ll 1$ in Eq.~(\ref{eq:Vfull}), yielding
\begin{align}
{\cal V}(\varphi,x) \approx \frac{\pi^2 }{96 x^2} \varphi^4 - \left(\frac{\pi^2}{24 x^2}-\frac{\kappa}{2}\right)\varphi^2\;.\label{eq:Vfullrelativistic}  
\end{align}
Symmetry restoration occurs at the critical temperature
\begin{align}
x_c = \frac{\pi}{2\sqrt{3}} \frac{1}{\sqrt{\kappa}} = \frac{\pi}{2\sqrt{3}} \left(\frac{g_\chi}{2\pi^2}\right)^{1/2} \frac{m_\chi^{3/2}}{m_\phi\,\Lambda^{1/2}}
\;,\label{eq:xccrossover}    
\end{align}
where in the second equality we have used Eq.~(\ref{eq:kappadef}).

For $x < x_c$, the symmetry-breaking vacuum evolves as 
\begin{align}
\varphi_*(x) = \sqrt{2\left(1-\frac{x^2}{x_c^2}\right)}\;.\label{eq:varphistar}   
\end{align}
This immediately leads to a simple expression for the effective WIMP mass in the broken phase:
\begin{align}
\mchieff(x) = \left(1-\frac{\varphi_*^2}{2}\right) m_\chi = m_\chi \frac{x^2}{x_c^2}\;, \qquad \text{for $x < x_c$}\;.    
\end{align}
Therefore, the effective WIMP mass scales as $\mchieff\propto T^{-2}$ in the broken phase and becomes increasingly suppressed at higher temperatures.
For $x> x_c$, the symmetry is restored and the vacuum relaxes to $\varphi=0$, so that $\mchieff$ rapidly returns to its bare value $m_\chi$.

Substituting Eq.~(\ref{eq:Vfullrelativistic}) into Eq.~(\ref{eq:EOMgeneral}), the EOM of the ALP reduces to
\begin{align}
\varphi'' + \frac{2}{x} \varphi' - \frac{\pi^2}{12} \left(1-\frac{x^2}{x_c^2}\right)\eta\,\varphi + \frac{\pi^2}{24}\,\eta\,\varphi^3 = 0\;. \label{eq:EOMcrossover}    
\end{align}
The numerical solution of the EOM in the crossover regime is shown in the right panel of Fig.~\ref{fig:solutions1st2nd}. The ALP evolution can be divided into four distinct periods. In the following, we analyze the dynamics in each period in detail.

\subsection{Nearly Adiabatic Evolution of the ALP}
A characteristic feature of the crossover regime is that the ALP evolution is nearly adiabatic. To see this, we compute the effective frequency of the ALP oscillation around the vacuum
\begin{align}
\omegaeff^2 \equiv \frac{\partial^2 V(\phi,T)}{\partial\phi^2}\Bigg|_{\phi=\phi_*} = \frac{g_\chi}{2\pi^2} \frac{m_\chi^3}{\Lambda} \frac{\partial^2 {\cal V}(\varphi,x)}{\partial\varphi^2}\Bigg|_{\varphi=\varphi_*}  =    \frac{g_\chi}{12}\frac{m_\chi^3}{\Lambda} \frac{1}{x^2}\left(1-\frac{x^2}{x_c^2}\right) = 2m_\phi^2 \left(\frac{x_c^2}{x^2}-1\right)
\end{align}
for $x < x_c$; for $x > x_c$, the vacuum is at $\varphi =0$ and
we instead have $\omegaeff=m_\phi \sqrt{1-x_c^2/x^2}$. 

The adiabatic approximation holds when $\omegaeff$ varies slowly compared to itself, i.e.,
\begin{align}
\left|\frac{\dot{\omega}_{\rm eff}}{\omegaeff}\right| \ll \omegaeff \;.  
\end{align}
Evaluating this condition explicitly, we obtain
\begin{align}
\left|\frac{\omegaeff^2}{\dot{\omega}_{\rm eff}} \right| = \frac{\omegaeff}{H}\left| \left(1-\frac{x^2}{x_c^2}\right) \right|= 
\begin{cases}
\dfrac{\pi \sqrt{\eta}}{\sqrt{6}}\,x\left(1-\dfrac{x^2}{x_c^2}\right)^{3/2} & \text{for $x < x_c$}\;,\\[1.5em]
\dfrac{\pi \sqrt{\eta}}{2\sqrt{3}}\,x\left(\dfrac{x^2}{x_c^2}-1\right)^{3/2} & \text{for $x > x_c$}\;.
\end{cases}
\label{eq:adiabaticrato}    
\end{align}
The adiabatic approximation is valid when the ratio in Eq.~(\ref{eq:adiabaticrato}) is much greater than unity. The ALP evolution can then be divided into four characteristic periods. 

\vspace{0.2cm}
\noindent\textbf{Period 1:} $x \ll 1/\sqrt{\eta} \ll x_c$.\footnote{Note that $\eta$ is parametrically large. Using Eq.~(\ref{eq:etadef}), one finds $\eta \sim 10^{15} \left(\frac{100~{\rm GeV}}{m_\chi}\right) \left(\frac{10^{-2} M_{\rm Pl}}{\Lambda}\right)$. In the parameter region of interest, the hierarchy $\eta \gg \kappa$ holds, and consequently $1/\sqrt{\eta} \ll x_c$ is always well satisfied.} The ALP is frozen at its initial displacement due to Hubble friction at sufficiently high temperatures. This occurs because the Hubble rate $H \propto T^2$ grows with temperature faster than the thermal mass term $m_T \propto T$.

\vspace{0.2cm}
\noindent\textbf{Period 2:} $1/\sqrt{\eta} \ll x \ll x_c$. The ALP evolves toward the symmetry-breaking vacuum $\varphi_* \approx \sqrt{2}$ and oscillates around it. The evolution is adiabatic, since the ratio in Eq.~(\ref{eq:adiabaticrato}) is large. Expanding around the minimum, $\delta \varphi \equiv \varphi-\varphi_*$, the EOM becomes
    \begin{align}
    \delta \varphi'' + \frac{2}{x} \delta\varphi' + \eta \left(\frac{\pi^2}{6} \delta \varphi+ \frac{\pi^2}{8}\sqrt{2}\,\delta \varphi^2  + \frac{\pi^2}{24} \delta \varphi^3\right) = \frac{3\sqrt{2}}{x_c^2}\;.    
    \end{align}
Neglecting the nonlinear terms, the solution is $\delta\varphi \sim e^{i \omega x}/x + 3\sqrt{2}/(x_c^2 \omega^2)$, with $\omega \equiv \sqrt{\eta \pi^2/6}$. Thus, the field oscillates around $\varphi_* \approx \sqrt{2}$ with an amplitude that decreases as $1/x$.

\vspace{0.2cm}    
\noindent\textbf{Period 3:}  $ x_c  - \delta x < x < x_c +\delta x$ with $\delta x \ll x_c$. This is the most subtle period. As $x\to x_c$ from below, the vacuum approaches the origin, $\varphi_* \to 0$, with a divergent velocity $\varphi_*' = -2x/(x_c^2 \varphi_*)$ and acceleration $\varphi_*'' = -4x^2/(x_c^4 \varphi_*^3) $. As a result, the actual field value can no longer track the evolution of the vacuum. For $x = x_c(1-\epsilon)$ with $\epsilon \to 0^+$, the EOM for $\delta\varphi = \varphi -\varphi_*$ becomes
\begin{align}
    \delta \varphi'' + \frac{2}{x} \delta\varphi' + \eta \frac{\pi^2}{24} \delta \varphi^3 = \frac{1}{2 x_c^2} 
    \frac{1}{\epsilon^{3/2}}\;.
\end{align}
As $x \to x_c$ from below,  a strong repulsive force appears in the EOM for $\delta\varphi$,\footnote{More precisely, this is a pseudo-force in the frame following $\varphi_*$. The large acceleration of $\varphi_*$ toward the origin produces an oppositely directed inertial force on $\delta\varphi$, analogous to the forward push experienced when a car brakes suddenly.} preventing the field from continuing to track the rapidly decreasing minimum. Consequently, $\varphi$ lags behind $\varphi_*$ and remains displaced from the origin as $\varphi_*$ approaches zero.

This loss of tracking reflects the breakdown of adiabaticity near $x_c$, where the thermal and bare mass contributions cancel and the effective frequency vanishes. However, because $\eta \gg 1/x_c^2$, the non-adiabatic interval is narrow. We define its approximate boundaries, $x_c^{\pm} \equiv x_c \pm \delta x^{\pm}$, by
$\left|\dot{\omega}_{\rm eff}(x_c^\pm)\right|\equiv\omegaeff^2(x_c^\pm)$, such that adiabaticity is lost during the interval $x_c^{-} < x < x_c^{+}$.
Using Eq.~(\ref{eq:adiabaticrato}), we obtain, to leading order,
     \begin{align}
        \frac{\delta x^{+}}{x_c} = 2^{1/3}\,
       \frac{\delta x^-}{x_c} \approx \left(\frac{18}{\pi^4}\right)^{1/3} \left(\frac{\kappa}{\eta}\right)^{1/3} =\frac{1}{2}\left(\frac{H(x_c)}{m_\phi}\right)^{2/3}
       \ll 1\;, \label{eq:loss-of-adiabaticity-window}
    \end{align}
where the last equality follows from Eqs.~(\ref{eq:kappadef}) and (\ref{eq:etadef}). The fractional width of the non-adiabatic interval is therefore of order $(\kappa/\eta)^{1/3}\ll 1$, so the breakdown of adiabaticity during symmetry restoration is brief compared to the Hubble timescale.

Before entering this interval, the field approximately tracks the evolving minimum. Using Eq.~(\ref{eq:varphistar}), we estimate its value at $x_c^-$ as
\begin{align}
\varphi(x_c^{-}) \approx \varphi_*(x_c^{-}) \approx 2 \left(\frac{\delta x_-}{x_c}\right)^{1/2}.   
\end{align}
Thus, relative to its value $\varphi\approx \sqrt{2}$ at $x\ll x_c$, the field amplitude has already decreased by a factor
\begin{align}
\xi_1 \equiv \frac{\varphi(x_c^{-})}{\sqrt{2}} \approx \frac{1}{2^{1/6}} \left(\frac{H(x_c)}{m_\phi}\right)^{1/3} = \left(\frac{72}{\pi^4}\right)^{1/6}\left(\frac{\kappa}{\eta}\right)^{1/6} \ll 1\;.\label{eq:dropfactor1}
\end{align}

After crossing $x_c$, the effective frequency $\omegaeff =m_\phi\sqrt{1-x_c^2/x^2}$ rapidly increases from zero to a value of order $m_\phi$. Once adiabaticity is restored at $x_c^+$, the oscillation amplitude obeys the adiabatic invariant $\omegaeff\,\varphi^2a^3 = \mathrm{const}$, corresponding to conservation of the comoving ALP number density. Here $\varphi$ denotes the oscillation envelope and $a$ the cosmic scale factor. For $x>x_c^+$ we have
\begin{align}
\omegaeff(x)\,\varphi^2(x)\,a^3(x) = \omegaeff(x_c^+)\,\varphi^2(x_c^+)\,a^3(x_c^+)\;.
\end{align}
Using Eq.~(\ref{eq:loss-of-adiabaticity-window}), we find $\omegaeff(x_c^+)/m_\phi \approx \left[H(x_c)/m_\phi\right]^{1/3}\ll 1$. The subsequent increase in frequency at $x>x_c^+$ therefore further suppresses the amplitude. For $m_\phi\gg H(x_c)$, a large suppression develops near $x_c$ before cosmological redshift becomes important, giving an effective sudden-drop factor
\begin{align}
    \xi_2\equiv 
    \left[\frac{\omegaeff(x_c^+)}{m_\phi}\right]^{1/2} \approx \left[\frac{H(x_c)}{m_\phi}
    \right]^{1/6} = \left(\frac{12}{\pi^2}\right)^{1/6} \left(\frac{\kappa}{\eta}\right)^{1/12} \ll 1\;.
\label{eq:dropfactor2} 
\end{align}

Taking $\varphi(x_c^+)\approx \varphi(x_c^-)$ and combining Eqs.~(\ref{eq:dropfactor1}) and (\ref{eq:dropfactor2}), we obtain the total effective drop factor relative to the field amplitude at $x\ll x_c$:
\begin{align}
\xi \equiv \xi_1\xi_2 \approx \frac{1}{2^{1/6}}\left[\frac{H(x_c)}{m_\phi}\right]^{1/2} = \frac{2^{5/6}\sqrt{3}}{\pi}\left(\frac{\kappa}{\eta}\right)^{1/4}\ll 1\;.
\label{eq:dropfactor}
\end{align}
Note that since we have not analytically tracked the field evolution through the non-adiabatic interval $x_c^-<x<x_c^+$, this estimate is subject to an ${\cal O}(1)$ uncertainty in its overall coefficient.

\begin{figure}[t!]
\centering
\includegraphics[width=3.5in]
{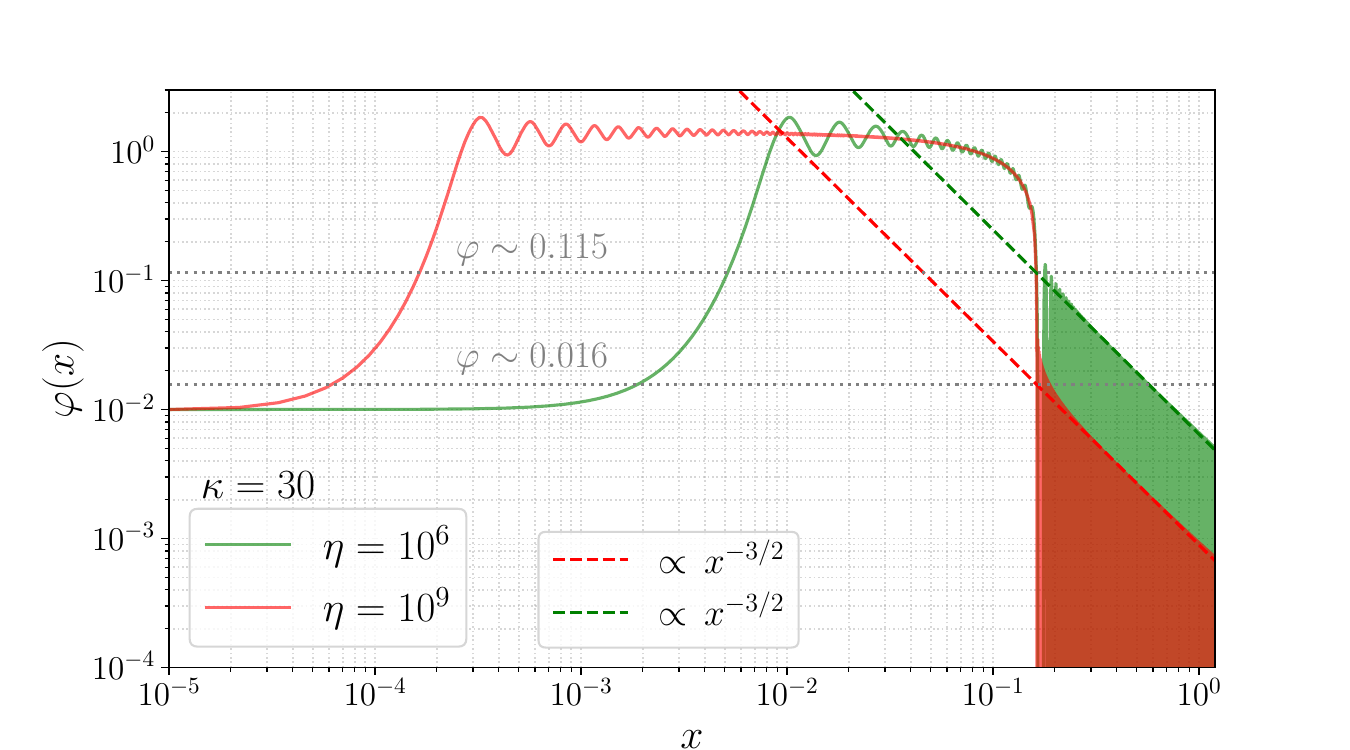}
\includegraphics[width=2.5in]
{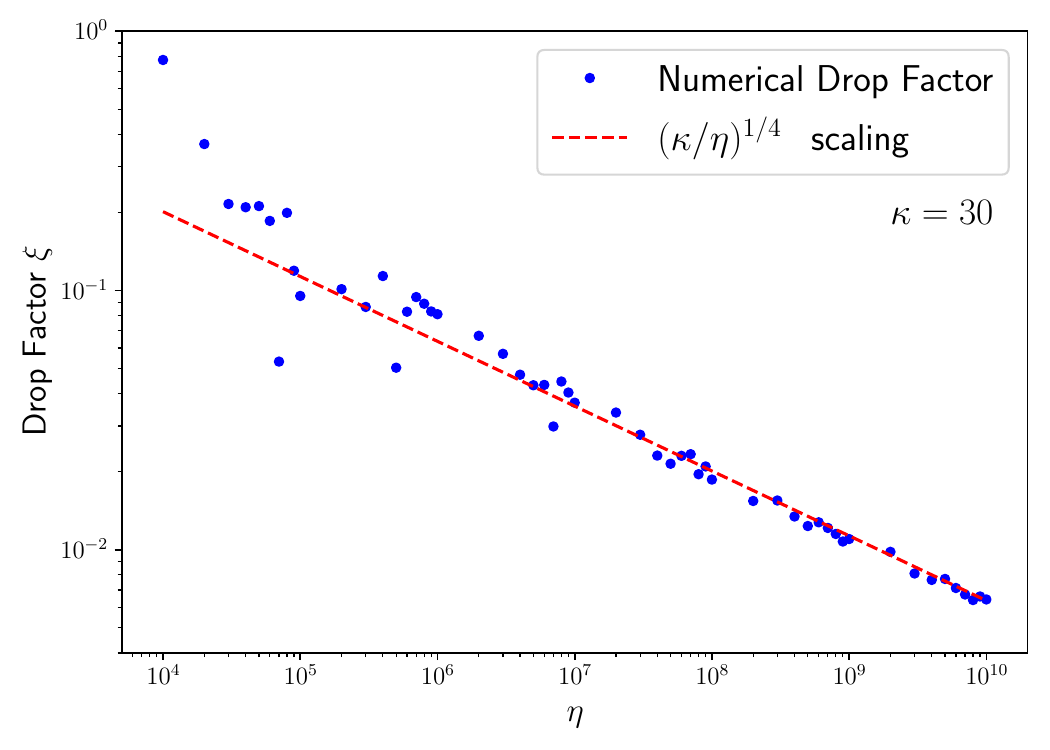}
\caption{\label{fig:zoom}
\textbf{Left}: The ALP evolution as a function of $x$ for $\eta = 10^6$ (green) and $\eta = 10^9$ (red) in the crossover regime with $\kappa$ fixed to be 30. Dashed lines indicate the matter-like scaling behavior. \textbf{Right}: Numerical drop factor $\xi$ as a function of $\eta$, with $\kappa$ fixed to be 30. Blue dots show the results from numerically solving the ALP equation of motion, and the red dashed line shows the scaling $(\kappa/\eta)^{1/4}$ up to an overall prefactor. 
}
\end{figure}

We have verified that Eq.~(\ref{eq:dropfactor}) is indeed consistent with the numerical solutions to the ALP equation of motion. 
In the left panel of Fig.~\ref{fig:zoom}, we plot a zoomed-in version of the numerical solutions for two values of $\eta$. For $\eta = 10^6\, (10^9)$ and $\kappa=30$, the expected drop factor is given by Eq.~(\ref{eq:dropfactor}) (multiplied by an $\mathcal{O}(1)$ prefactor $\sim0.88$ so that the right panel of Fig.~\ref{fig:zoom} shows agreement for large $\eta$) to be 0.064 (0.011), so the expected field values after the drop are 0.090 (0.016). In the plot we find numerically that the field after the drop is 0.115 (0.016). As expected, the agreement is better for larger $\eta$. 

In the right panel of Fig.~\ref{fig:zoom}, we further verify this scaling for more values of $\eta$. Here we plot Eq.~(\ref{eq:dropfactor}) up to an overall factor (red dashed line) and compare it to numerical results (blue dots). For each value of $\eta$, the blue dots are computed as $\varphi_{\rm isect} / \sqrt{2}$, where $\varphi_{\rm isect}$ is found using a numerical procedure that involves finding the oscillation peaks of each solution, fitting their envelope with a power law close to $x^{-3/2}$, and finding the intersection between this fit and the first drop of $\varphi$.  The right panel of Fig.~\ref{fig:zoom} demonstrates that the agreement improves for larger $\eta$. Therefore, for values of $\eta$ of phenomenological relevance to us ($\eta \sim 10^{13} - 10^{17})$, the scaling predicted in Eq.~(\ref{eq:dropfactor}) is an excellent description of the drop factor.  

\vspace{0.2cm}
\noindent\textbf{Period 4:} $x \gg x_c$. The symmetry is restored, and the field oscillates adiabatically around $\varphi =0$:
    \begin{align}
     \varphi'' + \frac{2}{x} \varphi' + \frac{\pi^2}{12} \frac{x^2}{x_c^2}\,\eta\,\varphi + \frac{\pi^2}{24}\,\eta\, \varphi^3 = 0\;.   
    \end{align}
Neglecting the cubic term, this reduces to the standard EOM for a harmonic oscillator in an expanding FRW background. Averaging over oscillations, the field exhibits matter-like redshift behavior, $\varphi \sim 1/x^{3/2}$.

\subsection{Calculation of the ALP Relic Abundance}
\label{subsec:miracle}

Using the above analysis in \textbf{Periods 3} and \textbf{4}, we can compute the ALP relic abundance analytically. The present-day energy density at $x_0 \equiv m_\chi/T_0$ (where $T_0$ is the present-day temperature) is obtained from the energy density before the drop, multiplied by the square of the drop factor in Eq.~(\ref{eq:dropfactor}) and the subsequent redshift, namely,
\begin{align}
\rho_\phi (x_0) &= \frac{1}{2}m_\phi^2 \phi^2(x_0) \approx \frac{1}{2}m_\phi^2 \phi_{i}^2\,\xi^2 \left(\frac{x_c}{x_0}\right)^3 \frac{g_{*S}(x_0)}{g_{*S}(x_c)}\;,
\end{align}
where $\phi_i \approx \sqrt{2 m_\chi \Lambda}$ is the field amplitude around $x_c$ before the drop, $g_{*S}$ is the entropy-weighted number of relativistic degrees
of freedom, and we have used $x_c^+ \approx x_c$ for $\eta \gg \kappa$. 
Using Eqs.~(\ref{eq:xccrossover}) and (\ref{eq:dropfactor}), we arrive at 
\begin{align}
\rho_\phi(x_0) &= \frac{1}{2}m_\phi^2 \times 2m_\chi \Lambda \times \frac{H(x_c)}{2^{1/3}m_\phi} \times \frac{x_c^3}{x_0^3} \times\frac{g_{*S}(x_0)}{g_{*S}(x_c)}\nonumber\\
& = m_\phi^2 \times m_\chi \Lambda \times \frac{1.66 \sqrt{g_*(x_c)}\,m_\chi^2}{2^{1/3}M_{\rm Pl}\,m_\phi} \times \frac{x_c}{x_0^3} \times\frac{g_{*S}(x_0)}{g_{*S}(x_c)}\nonumber\\
& = \cancel{m_\phi^2} \times m_\chi \Lambda \times \frac{1.66 \sqrt{g_*(x_c)}\,m_\chi^2}{2^{1/3}M_{\rm Pl}\,\cancel{m_\phi}} \times \frac{\pi}{2\sqrt{3}}\left(\frac{g_\chi}{2\pi^2}\right)^{1/2}\frac{m_\chi^{3/2}}{\Lambda^{1/2} \cancel{m_\phi} x_0^3} \times\frac{g_{*S}(x_0)}{g_{*S}(x_c)}\nonumber\\
& = \frac{1.66}{2^{4/3}} \sqrt{\frac{g_\chi g_*(x_c)}{6}}\frac{g_{*S}(x_0)}{g_{*S}(x_c)} \frac{m_\chi^{3/2} \Lambda^{1/2} T_0^3}{M_{\rm Pl}}\;.
\end{align}

Surprisingly, the dependence on $m_\phi$ cancels exactly in the present-day energy density. This cancellation arises from the combined effect of the initial ALP energy density at the time of the phase transition, its sudden drop at $x_c$, and the subsequent redshift. A simple physical interpretation can be obtained in terms of the ALP yield. Symmetry restoration occurs when the bare ALP mass becomes comparable to the thermal mass, at $T_c\propto m_\phi$ [see Eq.~(\ref{eq:xccrossover})]. Near symmetry restoration, the curvature of the effective potential becomes small and the ALP can no longer adiabatically follow the rapidly evolving minimum. The residual oscillation is therefore set by the Hubble timescale at the point where adiabaticity is restored after the transition. Indeed, the drop factor $\xi^2\sim H(x_c)/m_\phi$ implies
$n_\phi(T_c)\sim m_\phi \phi^2_i\xi^2\sim \phi^2_iH(x_c)$,
so that the ALP number density generated by the transition is set by $H(x_c)$ rather than directly by the zero-temperature ALP mass. Since $\phi_i$ is independent of $m_\phi$, the conserved ALP yield scales as
$Y_\phi\propto H(T_c)/s(T_c)\propto 1/T_c\propto 1/m_\phi$
as long as the transition occurs during radiation domination, thereby yielding an $m_\phi$-independent present-day energy density; see Table~\ref{tab:misalignment-comparison}. Note that in order to obtain this result, we have assumed that the bare ALP mass $m_\phi$ is generated above the scale of symmetry restoration, $T_c =m_\chi/ x_c \sim m_\chi \sqrt{\kappa}$, and is temperature-independent at and below $T_c$.

\begin{table}[t]
\centering
\renewcommand{\arraystretch}{1.5}
\begin{tabular}{c|c|c}
\hline\hline
 & Standard misalignment & ALP miracle scenario \\
\hline
Onset of relic oscillations
& $H(T_{\rm osc})\sim m_\phi$
& $\lvert m_T(T_c)\rvert\sim m_\phi$
\\
Oscillation temperature
& $T_{\rm osc}\propto m_\phi^{1/2}$
& $T_c\propto m_\phi$
\\
Field value before oscillation
& $\phi_i$~[UV-sensitive]
& $\phi_i\approx\sqrt{2m_\chi\Lambda}$
\\
Initial relic-oscillation amplitude
& $\phi_{\rm osc}\sim\phi_i$
& $\phi_{\rm osc}\sim\xi\phi_i\propto m_\phi^{1/2}$
\\
Number density at production
& $n_\phi\sim m_\phi\phi_i^2\propto m_\phi$
& $n_\phi\sim H(T_c)\phi^2_i\propto m_\phi^2$
\\
Yield, $Y_\phi\equiv n_\phi/s$
& $Y_\phi\propto m_\phi^{-1/2}$
& $Y_\phi\propto m_\phi^{-1}$
\\
Present-day relic abundance
& $\Omega_\phi\propto m_\phi^{1/2}$
& $\Omega_\phi$~independent of~$m_\phi$
\\
Initial-condition dependence
& Yes (set by UV-dependent $\phi_i$)
& No (erased by thermal evolution)
\\
\hline\hline
\end{tabular}
\caption{Comparison between the standard misalignment mechanism and the ALP miracle scenario considered here. For standard misalignment, we assume a temperature-independent ALP mass and radiation domination at the onset of coherent oscillations.}
\label{tab:misalignment-comparison}
\end{table}

Normalized by the critical density $\rho_c = 3H_0^2 M_{\rm Pl}^2/(8\pi)$, we obtain the ALP relic abundance:
\begin{align}
\Omega_\phi &= \frac{1.66 \times 8\pi}{3 \times 2^{4/3}}\sqrt{\frac{g_\chi g_*(x_c)}{6}}\frac{g_{*S}(x_0)}{g_{*S}(x_c)} \frac{m_\chi^{3/2} \Lambda^{1/2} T_0^{3}}{H_0^2 M_{\rm Pl}^3}\nonumber\\
& \approx 0.16 \left(\frac{m_\chi}{10~{\rm GeV}}\right)^{3/2}\left(\frac{\Lambda}{0.1~M_{\rm Pl}}\right)^{1/2},\label{eq:ALPrelicabundancecrossover}
\end{align}
where, in obtaining the numerical value in the second line, we have taken $g_\chi=2$ and used $g_*(x_c) =g_{*S}(x_c) = 106.75$, $g_{*S}(x_0)=3.93$, $H_0 = 1.4 \times 10^{-42}~{\rm GeV}$, and $T_0 = 2.35 \times 10^{-13}~{\rm GeV}$. 

We have obtained an exceptionally simple expression for the ALP relic abundance, determined by only two free parameters: the mass of the thermalized fermion to which the ALP couples, and the strength of the quadratic coupling. Remarkably, for a weak-scale fermion and a Planck-suppressed coupling -- both well-motivated energy scales -- the resulting relic abundance is of the same order as the observed DM density. We refer to this coincidence as the \textbf{ALP miracle}.\footnote{The ALP miracle is robust under variations of $\Lambda$, requiring only modest changes in the fermion mass around the weak scale. For example, decreasing $\Lambda$ from $M_{\rm Pl}$ to $10^{-3}M_{\rm Pl}$ increases the $m_\chi$ needed to reproduce the observed DM abundance only from approximately $6$ to $60~{\rm GeV}$.}

In contrast to the standard misalignment mechanism, the ALP abundance in Eq.~(\ref{eq:ALPrelicabundancecrossover}) is insensitive to both the initial misalignment angle and the ALP mass (provided the latter lies within the crossover regime); see Table~\ref{tab:misalignment-comparison} for a comparison. As a result, the prediction is largely insensitive to the UV initial conditions inherited from inflation, making our scenario significantly more predictive.

On the other hand, the WIMP freeze-out is essentially unaffected in the crossover regime, since the ALP decouples from the thermal bath at $x\ll {\cal O}(25)$. The WIMP relic abundance therefore remains determined by the standard freeze-out mechanism,
\begin{align}
\Omega_\chi \sim \Omega_{\rm DM} \frac{\langle \sigma v \rangle_{\rm th}}{\langle\sigma v\rangle}\;,    
\end{align}
where $\Omega_{\rm DM} \approx 0.26$ is the observed DM fractional density,
$\langle \sigma v\rangle_{\rm th} \approx 3\times 10^{-26}~{\rm cm^3/s}$ is the canonical thermal annihilation cross section, and $\langle \sigma v\rangle$ is the actual WIMP annihilation cross section into SM particles, which is generally independent of the WIMP-ALP coupling.

Consequently, the crossover regime naturally realizes a two-component DM scenario. Without introducing any intermediate mass scale between the weak and the Planck scales, the mechanism predicts that \emph{both} the ALP and the WIMP contribute an ${\cal O}(1)$ fraction of the observed DM density,
\begin{align}
\Omega_\phi \sim \Omega_\chi \sim \Omega_{\rm DM}\;.    
\end{align}
Defining $\epsilon_\phi\equiv \Omega_\phi/\Omega_{\rm DM}$ and $\epsilon_\chi\equiv \Omega_\chi/\Omega_{\rm DM}$ as the corresponding fractional relic abundances, we have $\epsilon_\phi+\epsilon_\chi=1$, with $\epsilon_\phi\sim\epsilon_\chi\sim{\cal O}(1)$ for natural choices of the model parameters.

\subsection{Experimental Targets for the ALP Miracle Window}
\label{subsec:experiments-for-ALP-Miracle}
\begin{figure}[t!]
\centering
\includegraphics[width=3.0in]
{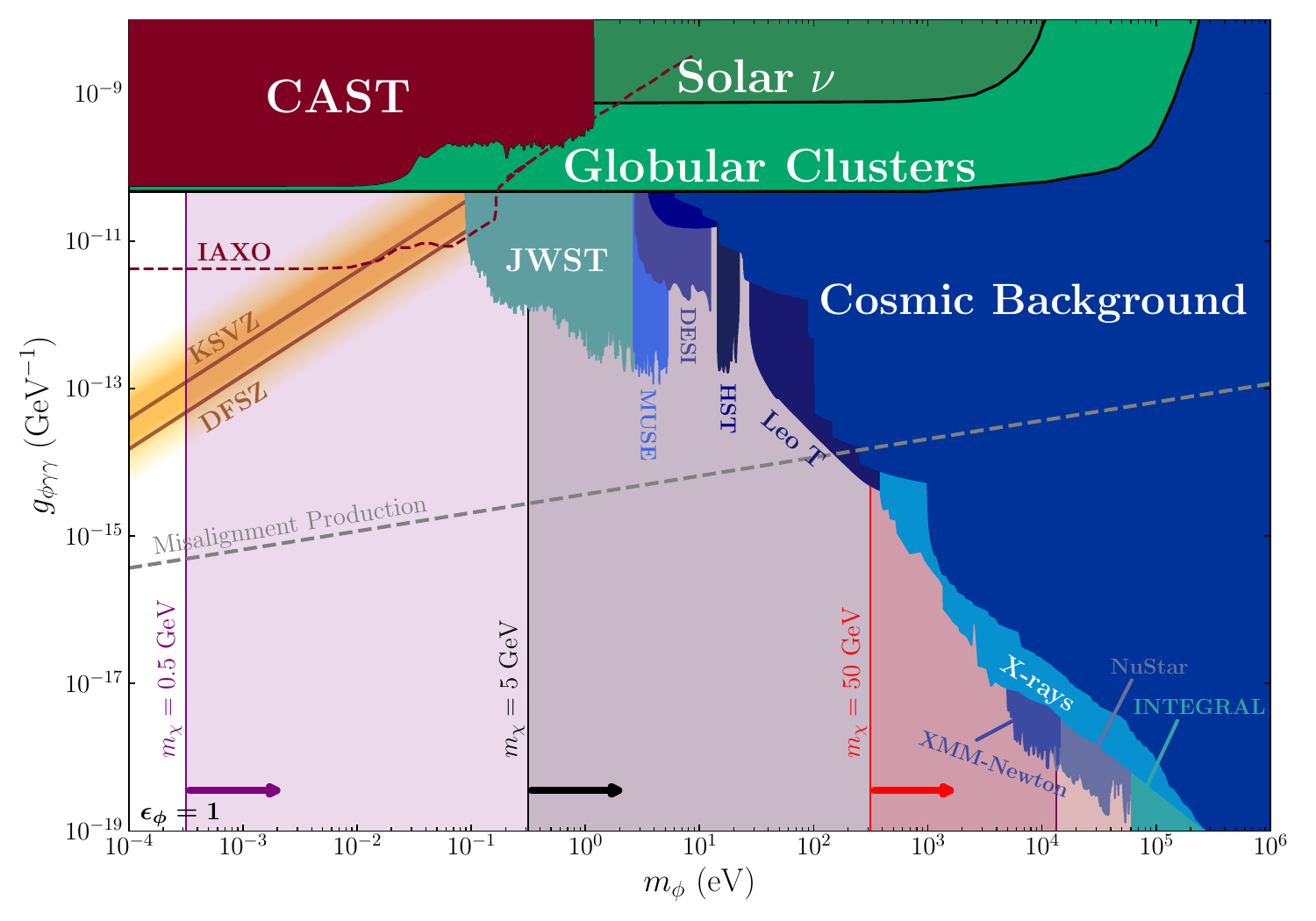}
\includegraphics[width=3.0in]
{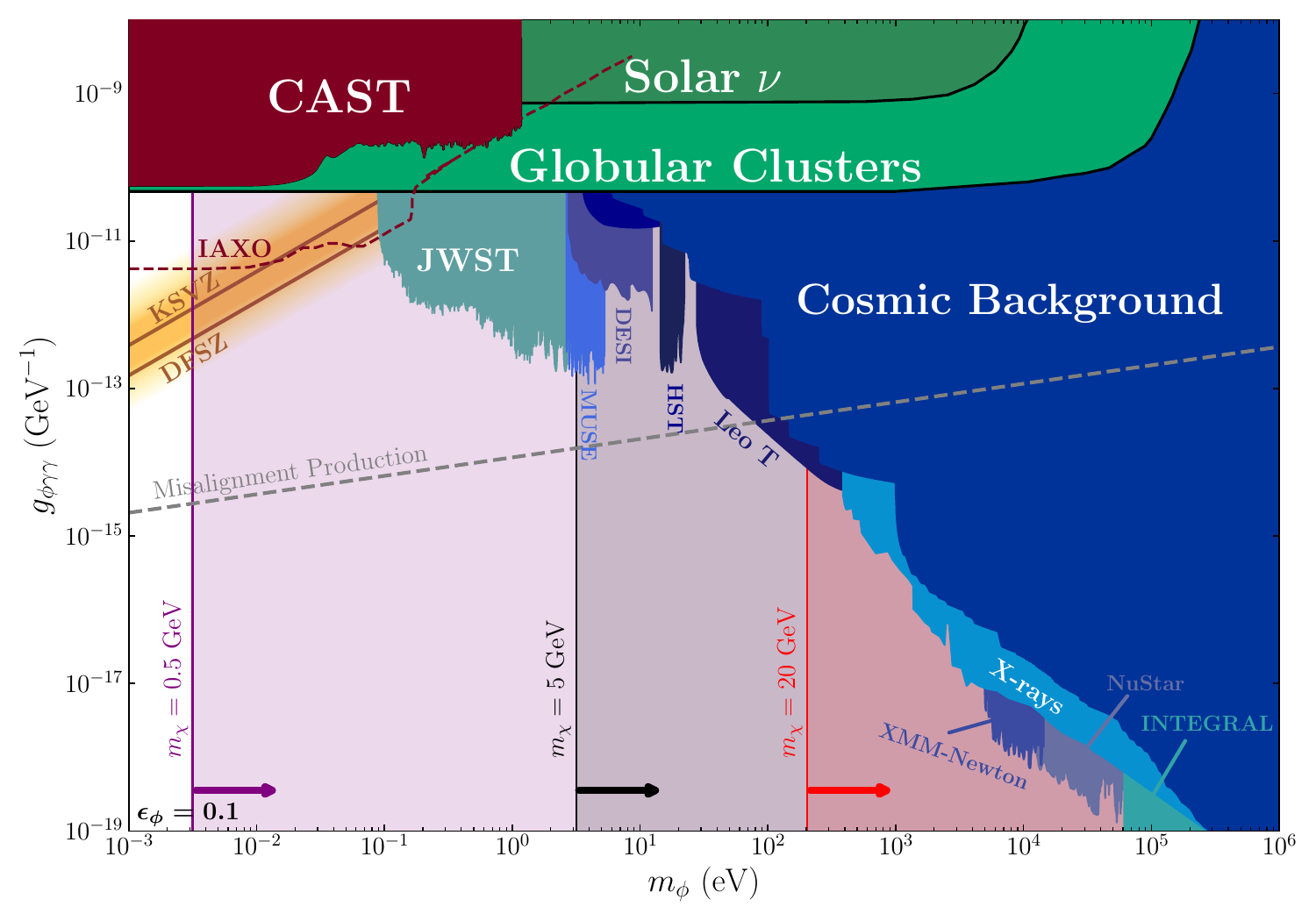}
\caption{\label{fig:ALPMiracleWindow}
The allowed ALP miracle windows for fixed ALP fractional abundances $\epsilon_\phi=1$ (left) and $\epsilon_\phi=0.1$ (right). The purple, black, and red vertical bands indicate the allowed windows for $m_\chi=0.5$, $5$, and $50~{\rm GeV}$ in the left panel, and $m_\chi=0.5$, $5$, and $20~{\rm GeV}$ in the right panel, respectively. 
For each $m_\chi$, the corresponding value of $\Lambda$ is fixed by the ALP relic abundance (see text for details). 
The left boundary of each band is set by the minimum ALP mass in the crossover regime, corresponding to $\kappa=\kappa_c =0.2$. 
The yellow band denotes the QCD axion band.
For comparison, the gray dashed line shows the parameter space yielding the indicated ALP fraction through the misalignment mechanism, assuming an ${\cal O}(1)$ initial misalignment angle.
The remaining shaded regions show the experimental constraints on ALP-photon interactions, taken from Ref.~\cite{AxionLimits}. The red dashed line shows the projected sensitivity of the IAXO experiment \cite{IAXO:2025ltd}.}
\end{figure}

In this subsection, we discuss experimental probes of the ALP dark matter predicted by our mechanism. We emphasize that ALP production in our framework relies only on the WIMP-ALP coupling and is independent of any coupling to SM particles. Detecting the ALP beyond its gravitational effects, however, requires a coupling between the ALP and the SM, provided that it does not interfere with the production mechanism discussed above. The simplest possibility is to couple the ALP to photons via ${\cal L}\supset-g_{\phi\gamma\gamma}\phi F\tilde F/4$, where $F$ and $\tilde F$ denote the electromagnetic field strength and its dual, respectively. Since the thermal expectation value of the topological term $F\tilde F$ vanishes in the cosmological plasma of the early Universe, this interaction does not induce an additional ALP potential and therefore does not affect the ALP dynamics or production mechanism discussed above.\footnote{Similarly, a linear derivative coupling of the ALP to an SM fermion $\psi$, of the form $\partial_\mu\phi\,\bar{\psi}\gamma^\mu\gamma^5\psi$, does not induce an ALP potential in a thermal fermion bath, as it preserves the ALP shift symmetry.}

In Fig.~\ref{fig:ALPMiracleWindow}, we show the ALP miracle window together with the current experimental exclusion landscape for ALP-photon couplings, for two representative choices: $\epsilon_\phi = 1$ (left) and $\epsilon_\phi = 0.1$ (right). The purple, black, and red shaded regions indicate the allowed parameter space in the ALP miracle scenario. The left boundary of each region is set by the critical value $\kappa_c = 0.2$, below which the system enters the FOPT regime.  For each choice of $m_\chi$, the corresponding value of $\Lambda$ is fixed by the ALP relic abundance in Eq.~(\ref{eq:ALPrelicabundancecrossover}). In the left panel, $m_\chi/{\rm GeV}=\left\{0.5,5,50\right\}$ corresponds to $\Lambda/M_{\rm Pl}\approx\left\{2100,\,2.1,\,2.1\times10^{-3}\right\}$, while in the right panel, $m_\chi/{\rm GeV}=\left\{0.5,\,5,\,20\right\}$ corresponds to $\Lambda/M_{\rm Pl}\approx\left\{21, \,2.1\times10^{-2},\, 3.3\times10^{-4}\right\}$. Indeed, using Eq.~(\ref{eq:ALPrelicabundancecrossover}), one finds $\Lambda/M_{\rm Pl}\approx \epsilon_\phi^2\left(6.4~{\rm GeV}/m_\chi\right)^3$. Combining it with $\kappa = \kappa_c$ gives the lower bound of the ALP mass for the ALP miracle window: 
\begin{align}
m_\phi \gtrsim m_\phi^{\rm min} \approx \frac{0.3~{\rm eV}}{\epsilon_\phi} \left(\frac{m_\chi}{5~{\rm GeV}}\right)^{3}.\label{eq:minimalALPmass}    
\end{align}
Thus, lowering the ALP fraction $\epsilon_\phi$ shifts the ALP miracle window toward larger ALP masses for a fixed $m_\chi$, while simultaneously decreasing $\Lambda$.

The other shaded regions in Fig.~\ref{fig:ALPMiracleWindow} denote the existing constraints on the ALP-photon coupling, taken from Ref.~\cite{AxionLimits}. We do not separately show the stability bound from requiring the ALP lifetime against decay into photons to exceed the age of the Universe, as it is weaker than the constraints displayed in the figure.
Most of these bounds are derived assuming $\epsilon_\phi=1$ and would be relaxed for $\epsilon_\phi<1$. Therefore, the constraints shown in the right panel of Fig.~\ref{fig:ALPMiracleWindow} should be understood as conservative. (Note that the limits from CAST and the projected sensitivity of IAXO do not depend on $\epsilon_\phi$, since both experiments look for axions produced in the Sun.) For comparison, we also show the misalignment prediction as the gray dashed line, obtained using the mapping $g_{\phi\gamma\gamma}=\alpha/(2\pi f_\phi)$, where $\alpha$ is the fine-structure constant and $f_\phi$ is the ALP decay constant. For illustration, we take the initial field displacement for misalignment production to be $\phi_i=f_\phi$, corresponding to an ${\cal O}(1)$ initial misalignment angle.

\subsubsection{QCD Axion}
\label{subsub:QCDaxion}

Interestingly, the ALP miracle parameter space shown in Fig.~\ref{fig:ALPMiracleWindow} overlaps with the QCD axion band, motivating an exploration of whether $\phi$ can be identified with the QCD axion.

The QCD axion obtains its mass from non-perturbative QCD effects. Its mass approaches the zero-temperature value below the QCD phase transition, at $T_{\rm QCD}\sim200~{\rm MeV}$, and decreases rapidly as $(T_{\rm QCD}/T)^4$ at higher temperatures~\cite{DiLuzio:2020wdo}. Our treatment, however, assumes that the $\phi$ mass remains constant below the symmetry restoration temperature; see Sec.~\ref{subsec:miracle}. Therefore, within this approximation, our constant-mass treatment applies to a QCD axion if and only if  the symmetry restoration occurs \emph{after} the QCD phase transition:\footnote{Symmetry restoration may also occur for $T_c>T_{\rm QCD}$, since the QCD axion mass remains nonzero above $T_{\rm QCD}$. In this case, however, the relic abundance calculation in Sec.~\ref{subsec:miracle} cannot be applied directly, as one needs to fully take into account the variation of the $\phi$ mass between the production period around $T_c$ and the present day. We leave the exploration of this interesting possibility to future work.} $T_{\rm QCD}\gtrsim T_c=m_\chi/x_c\sim m_\chi\sqrt{\kappa}$, where we have used Eq.~(\ref{eq:xccrossover}). Since the crossover regime requires $\kappa>\kappa_c\approx0.2$, this gives the approximate upper bound on the $\chi$ mass:
\begin{align}
m_\chi \lesssim T_{\rm QCD}/\sqrt{\kappa_c} \sim 0.5~{\rm GeV}\;.\label{eq:QCDaxionbound}   \end{align}
Provided that this bound is satisfied, the ALP miracle may be realized with a QCD axion. (Note also that in some non-minimal models, the QCD axion mass may be generated at a temperature well above $T_{\rm QCD}$, without modifying the zero-temperature mass~\cite{Heurtier:2021rko}. In such models, the upper bound on $m_\chi$ may be weaker.) 

Taking $m_\chi=0.5~{\rm GeV}$ to saturate the  upper bound in Eq.~(\ref{eq:QCDaxionbound}), we obtain the lower edge of the ALP miracle window at this WIMP mass:
\begin{align}
m_\phi^{\rm min} \approx 0.6 \times 10^{-3}~{\rm eV}\left(\frac{0.5}{\epsilon_\phi}\right).    
\end{align}
This suggests a possible QCD axion target at masses of order millielectronvolts, significantly heavier than in the familiar cases of production through traditional misalignment~\cite{Dine:1982ah,Preskill:1982cy,Arias:2012az} or axion string fragmentation (see e.g.~\cite{Buschmann:2021sdq}). Thus, the ALP miracle scenario provides a novel target for the QCD axion and ALP searches, which would be clearly distinguishable from other production mechanisms if axion DM is observed experimentally. Intriguingly, a large fraction of the overlap with the QCD axion band is expected to be probed by the IAXO experiment~\cite{IAXO:2019mpb,IAXO:2025ltd}; see Fig.~\ref{fig:ALPMiracleWindow}.

In non-minimal models, the mass and the coupling of the pseudoscalar responsible for solving the SM strong-CP problem may lie far outside of the canonical QCD axion band, and its mass can be generated at temperatures substantially higher than $T_{\rm QCD}$. We leave an exploration of compatibility of such models with the ALP miracle scenario to future work.



\subsection{Comparison with Literature}
The scenario in which a light coherent scalar field is coupled to a thermal bath has been explored in a variety of contexts~\cite{Piazza:2010ye,Moroi:2013tea,Lillard:2018zts,Ramazanov:2021eya,Batell:2021ofv,Cyncynates:2024bxw, Cyncynates:2024ufu,Banerjee:2025nvs,Graham:2025gtd,Banerjee:2025qyg}. In particular, the possibility of erasing the dependence of the relic abundance of a light scalar on its initial field value through early-Universe thermal effects has been explored in other models, such as the ``thermal misalignment'' mechanism proposed in Ref.~\cite{Batell:2021ofv}, where $\phi$ is linearly coupled to the thermal bath. A key difference lies in the structure of the thermal potential. A linear coupling primarily shifts the minimum of the scalar potential and, within the regime considered in Ref.~\cite{Batell:2021ofv}, does not induce a phase transition, since the temperature-dependent correction does not produce a negative $\phi^2$ term. By contrast, the quadratic coupling considered in our case generates a negative $T^2 \phi^2$ contribution, which destabilizes the origin at high temperatures and restores the symmetry at low temperatures, leading to a genuine (inverse) phase transition.

This distinction results in different field dynamics and, consequently, different predictions for the relic abundance. Although both mechanisms render the final abundance insensitive to the initial field value, the $\phi$ relic density in Ref.~\cite{Batell:2021ofv} retains a dependence on the scalar mass, whereas in our case this dependence also drops out, as shown in Eq.~(\ref{eq:ALPrelicabundancecrossover}).

Finally, while Ref.~\cite{Batell:2021ofv} incorporates the $\phi$-dependence of the thermal free energy, it assumes that the thermal bath remains in equilibrium and does not account for the dynamical back-reaction of $\phi$ on the bath. In contrast, in our case the back-reaction of $\phi$ plays a central role in modifying the freeze-out process of the thermal species, leading to a novel coherent freeze-out scenario, as we discuss in the next section.

\section{First-Order Phase Transition Regime}
\label{sec:FOPT}
In this section, we investigate the regime in which the symmetry is restored via a first-order phase transition (FOPT), corresponding to $\kappa < \kappa_c \approx 0.2$. Unlike the crossover regime discussed in the previous section -- where the decoupling occurs at $x_c \sim 1/\sqrt{\kappa} \lesssim {\cal O}(1)$ -- the decoupling of the ALP from the thermal bath in the FOPT regime can take place much later, at $x\gg x_{\rm fo} \approx 25$, and can therefore significantly affect the freeze-out of $\chi$ particles.

As we will show in Sec.~\ref{subsec:bounceaction}, the critical temperature $x_c$ in the FOPT regime, at which the potential develops degenerate vacua, scales as $x_c \sim \kappa^{-1/4}$. On the other hand, the false minimum $\varphi_*$ disappears at $\xcfo \sim \kappa^{-1/3}$ (see Eq.~(\ref{eq:xcfotilde})). We will also show in  Sec.~\ref{subsec:bounceaction} that, due to the hierarchy between the large field value in the false vacuum and the height of the potential barrier, i.e., $ \phi_* \sim \sqrt{m_\chi \Lambda} \gg V^{1/4}_{\rm eff} \sim m_\chi$, quantum tunneling of the scalar through the barrier is highly suppressed until very close to the disappearance of the false vacuum. This results in a nucleation temperature $x_n \approx \xcfo \sim \kappa ^{-1/3}$. Therefore, for $\kappa \ll 1$, the potential inevitably leads to strong supercooling, with nucleation occurring much later than the critical temperature, $x_n \gg x_c$.
As discussed in Sec.~\ref{subsec:physicalorigin}, $\chi$ cannot undergo freeze-out prior to $\xcfo$ due to the suppression of its effective mass. Consequently, freeze-out is triggered by the tunneling of $\phi$, rather than by the Hubble expansion of the Universe as in the standard scenario, and can therefore occur at $x \gg {\cal O}(25)$. As a result, the relic abundance of $\chi$ can be substantially enhanced compared to the conventional freeze-out case.

The goal of this section is to quantitatively establish the physics outlined above. In Sec.~\ref{subsec:bounceaction}, we analytically compute the bounce action and the nucleation temperature in the strong supercooling regime ($\kappa \ll 1$). In Sec.~\ref{subsec:WIMPyield}, we determine the resulting WIMP relic abundance inherited from coherent freeze-out.

\subsection{Bounce Action and Nucleation Temperature}
\label{subsec:bounceaction}
At high temperatures, the $Z_2$ symmetry of the ALP potential $V(\phi)$ is spontaneously broken. As the Universe cools, the symmetry-restored phase eventually becomes energetically favorable, and the ALP sector undergoes a FOPT. The critical temperature $T_c \equiv m_\chi/x_c$ of the phase transition is defined by 
\begin{equation}
    \mathcal{V} (\varphi_*, x_c)=\mathcal{V} (0,x_c)\;.
\end{equation}
We examine the full potential in Eq.~(\ref{eq:Vfullseries}). Near the critical temperature, $\gamma_* x \ll 1$, so for small $n$ such that $n\gamma_* x \ll 1$, we can use the approximation of Bessel functions 
\begin{equation}
    K_\nu(z) \approx \frac{1}{2}\Gamma(\nu)\left(\frac{z}{2}\right)^{-\nu} \implies \frac{(-1)^{n-1}}{n^2} K_2(n\gamma_* x) \approx \frac{(-1)^{n-1}}{n^4} \frac{2}{(\gamma_* x)^2}\;.
\end{equation}
The contribution to the potential is then suppressed by $n^{-4}$ for small $n$, while larger $n$ terms are even further suppressed as $K_\nu(z) \ll 1/z$ for large $z$. Thus, truncating the infinite sum of the full potential at $n=1$ is a good approximation:
\begin{equation}
    {\cal V}(\varphi_*,x) \approx \kappa(1-\gamma_*) -\frac{\gamma_*^2}{x^2} K_2(\gamma_* x) \approx \kappa(1-\gamma_*) - \frac{2}{x^4}\;,
\end{equation}
and similarly for Eq.~(\ref{eq:phistarEOM}),
\begin{equation}
    \kappa \approx \frac{\gamma_*^2}{x} K_1(\gamma_* x) \approx \frac{\gamma_*}{x^2}\;.
\end{equation}
Combining the above equations, we obtain the critical temperature in the FOPT regime for $\kappa\ll 1$
 \begin{equation}
     T_c \approx \frac{m_\chi}{\sqrt[4]{2}}\kappa^{1/4}\;.
 \end{equation}
Retaining the full series instead of truncating at $n=1$ preserves the leading $\kappa^{1/4}$ scaling and changes the numerical prefactor only slightly, from $2^{-1/4}\approx0.841$ to $(7\pi^4/360)^{-1/4}\approx0.852$.

Below the critical temperature, the phase transition completes when the bubble nucleation rate, $\Gamma$, is larger than $H^4$, where $H \sim T^2/M_{\rm Pl}$ is the Hubble parameter. The bubble nucleation rate $\Gamma \sim T^4 e^{-S_B}$ is computed from the Euclidean bounce action $S_B$ for a finite nucleation temperature $T < T_c$. This imposes an upper bound on the bounce action, \begin{equation}
    S_B \lesssim 4 \log \left(\frac{M_{\rm Pl}}{T}\right).
\end{equation}

\subsubsection{Thick-Wall Limit}
\begin{figure}[t!]
    \centering
    \includegraphics[width=0.8\linewidth]{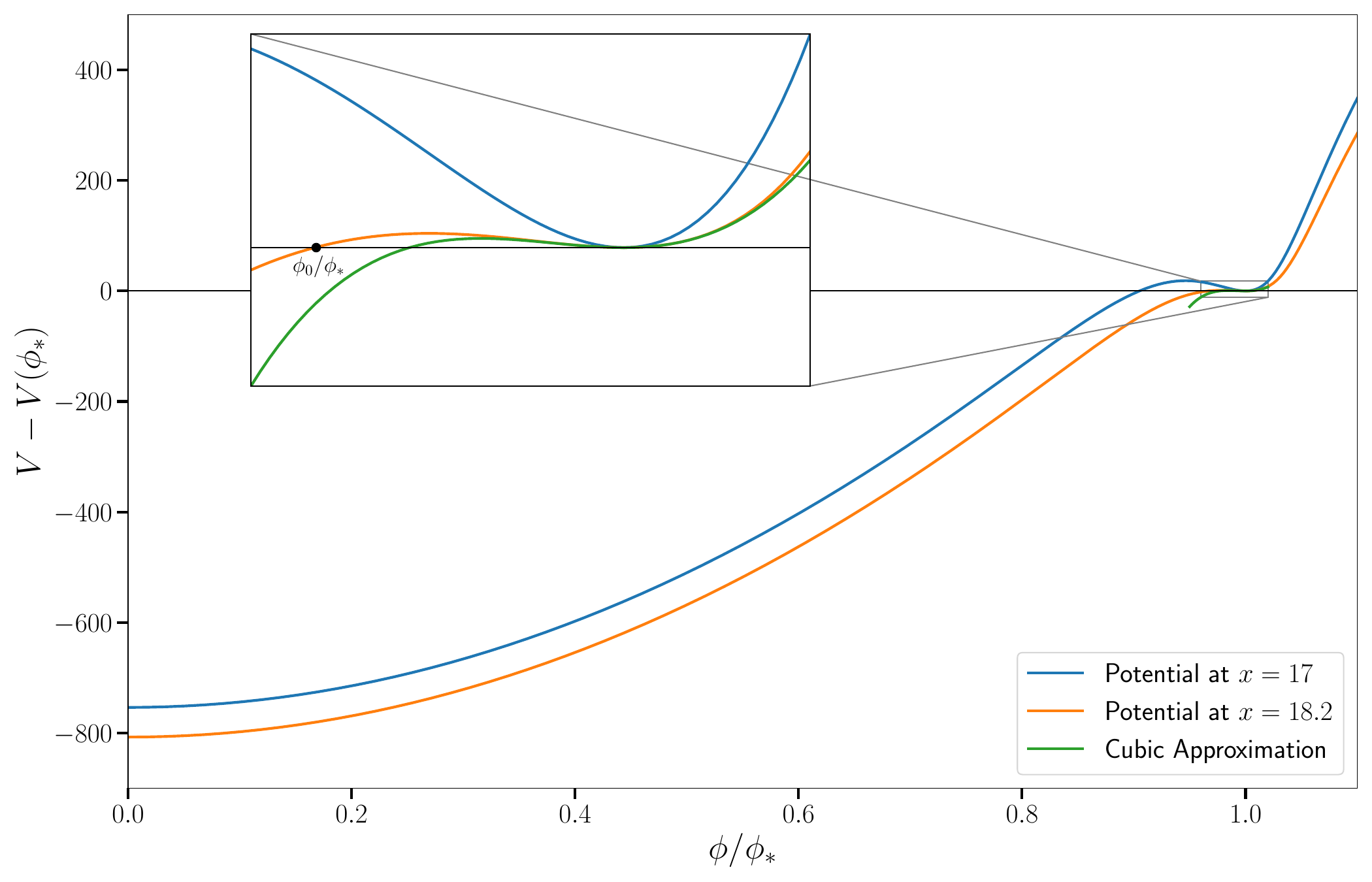}
    \caption{The full potential $V(\phi, x)$ as the inverse temperature $x$ approaches $\xcfo$. The potential barrier separating the false vacuum at $\phi=\phi_*$ from the true vacuum at $\phi =0$ becomes small near $\xcfo$, such that the thick-wall approximation becomes applicable. In this regime, the potential near the false vacuum is well approximated by a cubic form, which is sufficient to solve the bounce equation without requiring detailed knowledge of the potential much beyond the release point near $\phi_0$.}
    \label{fig:thickwallV}
\end{figure}

The large overall scaling factor $\sqrt{m_\chi \Lambda}$ of the physical field $\phi$ generally produces a very large bounce action, such that the phase transition cannot complete. We thus only consider nucleation temperatures close to $\xcfo$, where the potential barrier is sufficiently small compared to the depth of the true vacuum at $\phi=0$, shown in Fig. \ref{fig:thickwallV}. This is the thick-wall limit studied in \cite{Linde:1981zj}. 

To find the bounce solution for this form of potential, we interpret the problem in semi-classical language, where a particle from the false vacuum tunnels through the potential barrier and is released at the escape point. In the thick-wall limit, the escape point is approximately $\phi_0$ with $V(\phi_0) = V(\phi_*) = 0$, and thus it is sufficient to solve the bounce equation using the potential in the immediate region around the potential barrier, $V(\phi \approx \phi_*)$. We follow the convention for phase-transition studies and adjust the constant in the potential $V(\phi) \rightarrow V(\phi) - V(\phi_*)$ for simplicity. Using a series expansion of $V$ near $\phi_*$,

\begin{equation}
    V(\phi,T) \approx \frac{1}{2} \frac{\partial^2 V}{\partial \phi^2}\Bigg|_{\phi_*} (\phi-\phi_*)^2 + \frac{1}{6} \frac{\partial^3 V}{\partial \phi^3}\Bigg|_{\phi_*} (\phi-\phi_*)^3\,,
\end{equation}
where the 0th and 1st order terms both vanish by definition. The bounce action for this form of cubic potential in the thick-wall limit is given by
\begin{equation}
    S_B = \frac{S_3}{T} = \tilde{S}_3\frac{9M^3}{4\delta^2 T}\,, \label{eq:thickbounce}
\end{equation}
where $\tilde{S}_3 \approx 19.4$, $M^2 \equiv \frac{\partial^2 V}{\partial \phi^2}\big|_{\phi_*}$, and $\delta \equiv \frac{1}{2}\frac{\partial^3 V}{\partial \phi^3}\big|_{\phi_*}$.\\

We define the small dimensionless parameter $\Delta \equiv \xcfo - x$ and examine the leading-order behavior of $S_B$ in $\Delta$ as the supercooled symmetry-breaking phase approaches the $\xcfo$ limit. In terms of $\Delta$ and the parameters $m_\chi, \Lambda, \kappa$, we find 
\begin{equation}
    S_B \approx A \left(\frac{\Lambda}{m_\chi}\right)^{3/2}\kappa^{1/4} \Delta^{3/4}\,, \label{eq:thickwallbouncescaling}
\end{equation}
where $A \approx 1300$ (for $g_\chi=4$)  or 1900 (for $g_\chi=2$), independent of $m_\chi$, $\Lambda$, and $\kappa$. A detailed derivation is presented in Appendix~\ref{app:thickwall}. It follows that, for the phase transition to complete, the nucleation temperature must satisfy
\begin{equation}
    \Delta \lesssim \left[\frac{4}{A}\log \left(\frac{M_{\mathrm{Pl}}}{T_\mathrm{cfo}}\right)\right]^{4/3} \left(\frac{m_\chi}{\Lambda}\right)^{2}\kappa^{-1/3},
\end{equation}
where $T_{\rm cfo}\equiv m_\chi/\xcfo$.

\begin{figure}[t!]
    \centering
    \includegraphics[width=0.8\linewidth]{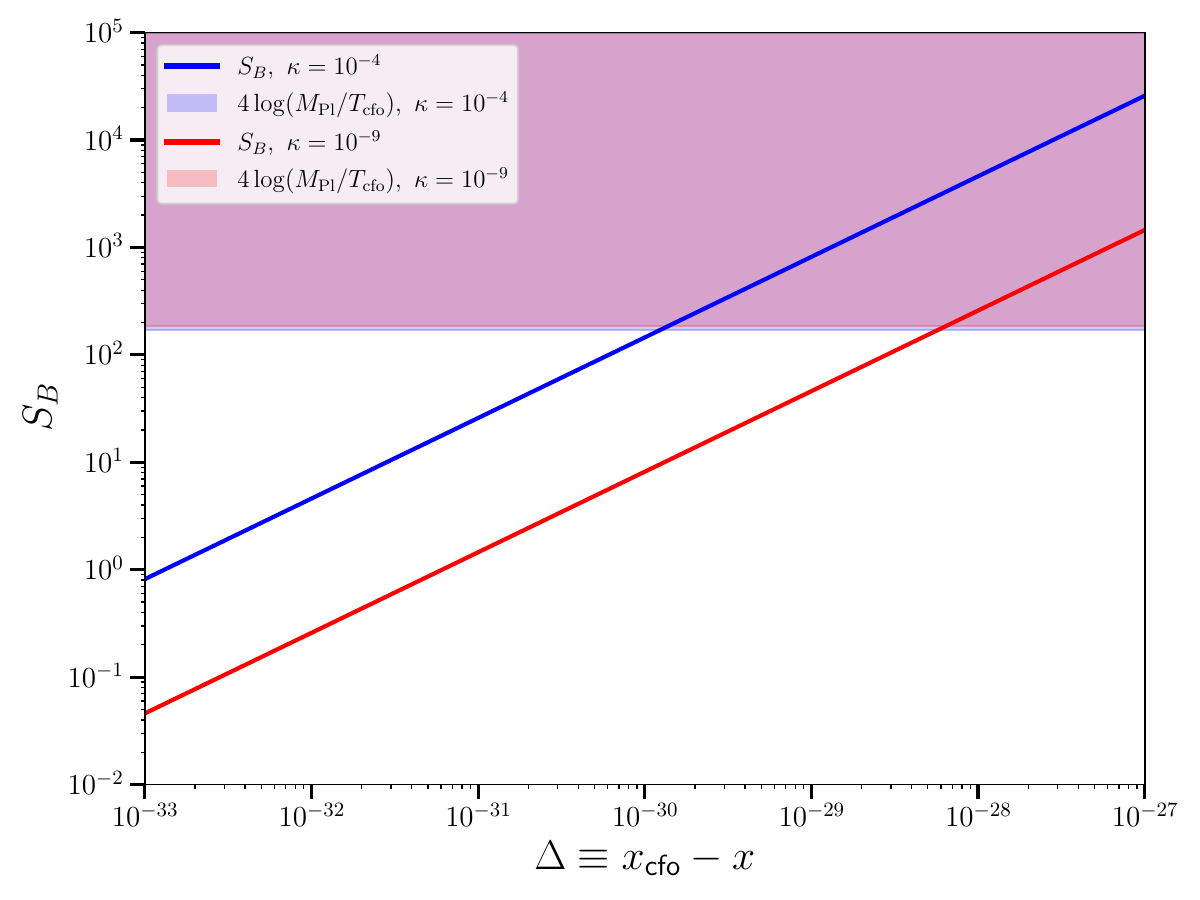}
    \caption{The bounce action $S_B$ near $\xcfo$ for the benchmark parameters $\Lambda = 10^{-2}M_\mathrm{Pl}$, $m_\chi = 100$ GeV, $g_\chi= 2$, with $\kappa \sim 10^{-4} - 10^{-9}$ in the FOPT regime. In the shaded region, the phase transition cannot complete. Consequently, the phase transition occurs only when $x$ is very close to $\xcfo$.}\label{fig:ptbound}
\end{figure}

In Fig.~\ref{fig:ptbound}, we show the dependence of the bounce action on the nucleation temperature over the parameter range in the FOPT regime relevant for DM phenomenology, $\kappa \sim 10^{-4} - 10^{-9}$. The upper limit, $\kappa\sim 10^{-4}$, corresponds to the onset of delayed WIMP freeze-out ($\xcfo \gtrsim 25$), while the lower limit, $\kappa\sim 10^{-9}$, is set by the requirement of avoiding vacuum domination. We see that the phase transition can proceed only after maximal supercooling, when the potential barrier nearly vanishes, with bubble nucleation occurring at a temperature extremely close to $\xcfo$.

\subsubsection{Gravitational-Wave Signal}

It is well known that cosmological first-order phase transitions may produce a stochastic background of gravitational waves, which can survive until the present time. FOPT-produced gravitational waves have been put forward as a potential explanation of the signal reported by pulsar timing arrays in 2023~\cite{NANOGrav:2023gor,Xu:2023wog,Reardon:2023gzh,NANOGrav:2023hvm}, and form an important target for the proposed future observatories such as LISA~\cite{Caprini:2015zlo,Caprini:2019egz}. In this subsection, we investigate whether the FOPT in the coherent freeze-out DM scenario can be associated with an observable gravitational wave signal.

The two relevant parameters for gravitational wave production are 
\begin{align}
\alpha = \frac{\Delta \rho}{\rho_{\rm rad}}\Bigg|_{T_*}, \qquad
\frac{\beta}{H_*}  = T \frac{\rm d}{{\rm d}T} \left(\frac{S_3}{T}\right)\Bigg|_{T_*}, 
\end{align}
where $H_* \equiv H(T_*)$ and $T_*$ is the transition temperature. As explained above, we can take $T_* \approx T_n$, the nucleation temperature, to an excellent approximation. The released energy at a fixed temperature is given by
\begin{align}
\Delta \rho = \Delta V - T \frac{\partial \Delta V}{\partial T}\;, \qquad
\text{with}\;\Delta V \equiv V(\phi_{\rm false}, T) - V(\phi_{\rm true},T)\;,
\end{align}
where the first term represents the vacuum energy difference between the false and true vacua, while the second term corresponds to the thermal (entropy) contribution, which reduces to the latent heat at $T_c$.
At the critical temperature $T_c$, $\Delta V=0$, so $\Delta \rho$ is entirely given by the latent heat. In contrast, in the strong supercooling regime ($T_n \ll T_c$), one typically has $\Delta \rho \approx \Delta V$, since the thermal contribution is suppressed by the low nucleation temperature.\\

As we have established, the phase transition only proceeds with near maximal supercooling. Taking the thick-wall bounce action of Eq.~(\ref{eq:thickwallbouncescaling}), we find
\begin{align}
    \frac{\beta}{H_*} &\approx A\left(\frac{\Lambda}{m_\chi}\right)^{3/2}\kappa^{1/4} T\frac{\mathrm{d}}{\mathrm{d}T}\Delta^{3/4} \nonumber\\
    &= \frac{3}{4} c_2^{1/3}A \left[\frac{4}{A}\log \left(\frac{M_{\mathrm{Pl}}}{T_\mathrm{cfo}}\right)\right]^{-1/3}\left(\frac{\Lambda}{m_\chi}\right)^{2},
\end{align}
with $c_2^{1/3}\approx 0.836$. Notably, the inverse duration of the phase transition only has an implicit, logarithmically suppressed dependence on $\kappa$ from $\log(T_\mathrm{cfo})$, and is almost entirely determined by the ratio of the scales $\Lambda$ and $m_\chi$. The scale $\Lambda$ is bounded from below by the requirement that the scalar particle does not come to thermal equilibrium with the WIMP, see Eq.~(\ref{eq:nonthermalizedcondition}). With this constraint, we obtain $\beta/H_* \gtrsim 10^{20}$. This means that the phase transition in our scenario is {\it extremely fast}, and completes essentially instantaneously compared to the Hubble timescale. This result makes sense intuitively: significant bubble nucleation only begins when the false vacuum and the potential barrier are both on the verge of disappearing.

The peak fractional abundance and the peak frequency of the gravitational waves are roughly estimated to be 
\begin{align}
\Omega_{\rm GW}h^2 &\sim 10^{-6} \left(\frac{H_*}{\beta}\right) \left(\frac{\alpha}{1+\alpha}\right)^2 \left(\frac{100}{g_*}\right)^{1/3},\\
f_{\rm peak} & \sim 10^{-5}~{\rm Hz} \left(\frac{\beta}{H_*}\right)\left(\frac{T_*}{100~{\rm GeV}}\right)\left(\frac{g_*}{100}\right)^{1/6}.
\end{align}
The LISA experiment can probe the fractional abundance as low as $ \Omega_{\rm GW}h^2 \sim 10^{-12}-10^{-11}$ within its peak sensitivity range of $10^{-3}~{\rm Hz} \lesssim f_{\rm peak} \lesssim 10^{-2}~{\rm Hz}$~\cite{Caprini:2019egz}. The factor $\frac{\alpha}{1+\alpha}$ is at most $\mathcal{O}(1)$ in the strong supercooling region that we consider here. The smallness of $H_*/\beta$ implies that the abundance of gravitational waves from the phase transition in our scenario is unfortunately well below the sensitivity reach of LISA. This conclusion is entirely due to the extremely short duration of the phase transition.   

\subsection{Calculation of the WIMP Relic Abundance}
\label{subsec:WIMPyield}
We now proceed to compute the WIMP relic abundance in the coherent freeze-out scenario. For definiteness, we assume that the freeze-out of $\chi$ is dominated by $2 \to 2$ annihilations into SM species. The Boltzmann equation then takes the form:
\begin{align}
\dot{n}_\chi + 3H n_\chi = - \langle \sigma v\rangle_{\rm eff} \left(n_\chi^2-n_{\chi,{\rm eq}}^2\right),\label{eq:BoltzmannEquation}    
\end{align}
where $n_\chi$ and $n_{\chi,{\rm eq}}$ denote the number density and its equilibrium value, respectively, and $\langle\sigma v\rangle_{\rm eff}$ is the effective annihilation cross section.

The back-reaction of ALP on the WIMP bath affects Eq.~(\ref{eq:BoltzmannEquation}) in two ways. First, it modifies the equilibrium distribution $n_{\chi,{\rm eq}}$ through a shift in the WIMP mass, $m_\chi \to \mchieff = \gamma m_\chi$, where the field-dependent boost factor $\gamma$ is defined in Eq.~(\ref{eq:boostfactordef}). Assuming a Maxwell-Boltzmann distribution, the equilibrium number density is given by
\begin{align}
n_{\chi,{\rm eq}} = \frac{g_\chi}{2\pi^2} \mchieff^2 T K_2\left(\frac{\mchieff}{T}\right).\label{eq:nchieq}   
\end{align}
As discussed in Sec.~\ref{subsec:physicalorigin} and shown in Eq.~(\ref{eq:mchieffoverT}), the ratio $\mchieff/T$ remains below ${\cal O}(1)$ for $x< x_n \approx \xcfo \sim \kappa^{-1/3}$, because the back-reaction of the  $\phi$ field reduces the effective $\chi$ mass while $\phi$ is trapped in the false vacuum. Consequently, the equilibrium distribution $n_{\chi,{\rm eq}}$ is not Boltzmann suppressed prior to $\xcfo$, regardless of how large $x$ becomes.

Second, the back-reaction of $\phi$ also modifies the effective annihilation cross section of $\chi$ through the shift in its mass. Let $m_{\rm M}\sim 100~{\rm GeV}$ denote the characteristic mass scale of the mediator responsible for interactions between $\chi$ and SM particles. The effective annihilation cross section relevant for freeze-out takes the parametric form
\begin{align}
\langle \sigma v \rangle_{\rm eff} \sim 
\begin{cases}
T^2/m_{\rm M}^4 & \text{for $\mchieff \lesssim T \lesssim m_{\rm M}$}\;,\\[5pt]
\mchieff^2/(m_{\rm M}^4 x^\ell) & \text{for $T \lesssim \mchieff \lesssim m_{\rm M}$}\;,\\[5pt]
1/(\mchieff^2 x^\ell) & \text{for $m_{\rm M}, T \lesssim \mchieff $}\;,
\end{cases} \label{eq:sigmaeff}
\end{align}
where $\ell = 0$ ($\ell =1$) corresponds to $s$-wave ($p$-wave) annihilation. For $x \lesssim \xcfo$, we have $\mchieff\lesssim T$,
and the $\chi$ particles remain relativistic,  so $\langle \sigma v \rangle_{\rm eff}$ takes the form given in the first line of Eq.~(\ref{eq:sigmaeff}). 
For $x > \xcfo$, the ratio $\mchieff/T$ rapidly exceeds ${\cal O}(1)$ as $\phi$ relaxes toward the origin (see Eq.~(\ref{eq:mchieffoverT}) and the right panel of Fig.~\ref{fig:vacuumevolution}), signaling the onset of coherent freeze-out. 

Throughout this section, we focus on the regime in which freeze-out is significantly delayed while tunneling still occurs early enough to avoid vacuum-energy domination, namely, $ {\cal O}(25) \lesssim \xcfo \lesssim {\cal O}(500)$ (corresponding to $10^{-9} \lesssim \kappa \lesssim 10^{-4}$).\footnote{For $ \xcfo > x_{\rm vac} \sim {\cal O}(500)$, where $x_{\rm vac}$ is given in Eq.~(\ref{eq:xvac}), the Universe enters a period of vacuum-energy domination before tunneling occurs, leading to a burst of inflation. See Sec.~\ref{sec:vacuumdomination} for further discussion of this regime.} The subsequent freeze-out dynamics depend on the form of $\langle \sigma v \rangle_{\rm eff}$ around $\xcfo$. Depending on the bare mass of $\chi$, the effective annihilation cross section is described by the second or third line of Eq.~(\ref{eq:sigmaeff}), respectively:
\begin{align}
\langle \sigma v \rangle_{\rm eff} (\xcfo) \sim 
\begin{cases}
\gamma^2\times m_\chi^2/(m_{\rm M}^4 x^\ell) & \text{for $m_\chi \lesssim \xcfo\,m_{\rm M}$}\;,\\[5pt]
\gamma^{-2} \times 1/(m_\chi^2 x^\ell) & \text{for $m_\chi \gtrsim \xcfo \,m_{\rm M}$}\;,
\end{cases} \label{eq:sigmaeff2}
\end{align}
where we have used the relation $\mchieff \sim T$, or equivalently $\gamma \sim 1/x$, for $x$ near $\xcfo$. Note that in Eq.~(\ref{eq:sigmaeff2}) we have factored out the field-independent part, corresponding to the bare annihilation cross section.
For example, for $m_{\rm M}\sim 100~{\rm GeV}$ and $\xcfo \sim {\cal O}(100)$, the boundary between the two regions is given by $\xcfo\, m_{\rm M} \sim {\cal O}(10)~{\rm TeV}$. For $m_\chi$ below (above) this boundary scale, the effective annihilation cross section around $\xcfo$ is suppressed (enhanced) relative to its bare value.
In the following, we discuss these two cases separately.

\subsubsection{Lighter mass region: $m_\chi \lesssim \xcfo \,m_{\rm M}$}
\label{subsubsec:lightermass}

The WIMP relic abundance in this region was previously computed in Ref.~\cite{Ferrante:2025avp}; here we briefly review the main steps relevant for the present discussion.

The effective annihilation cross section 
during freeze-out can be written as
\begin{align}
\langle \sigma v\rangle_{\rm eff} = \gamma^2 \langle \sigma v \rangle\;,  \end{align}
where $\langle \sigma v\rangle \sim m_\chi^2/(m_{\rm M}^4 x^\ell)$ denotes the bare annihilation cross section.
Defining the yield $Y_\chi \equiv n_\chi/s$, where $s$ is the entropy density, the Boltzmann equation (\ref{eq:BoltzmannEquation}) becomes
\begin{align}
Y_\chi'(x) = -\frac{\lambda \gamma^2}{x^2}\left[Y_\chi^2(x)-Y_{\chi,\rm eq}^2(x)\right],\label{eq:BoltzmannEquationlightregion}    
\end{align}
where 
$Y_{\chi,\rm eq} \equiv n_{\chi,{\rm eq}}/s = 45 g_\chi\gamma^2 x^2 K_2(\gamma x)/(4\pi^4 g_{*S})$, and 
\begin{align}
\lambda = \sqrt{\frac{\pi}{45}}\frac{g_{*S}}{\sqrt{g_*}} m_\chi M_{\rm Pl} \langle \sigma v\rangle \approx  \sqrt{\frac{\pi}{45}}\frac{g_{*S}}{\sqrt{g_*}} m_\chi M_{\rm Pl} \left(\sigma_0 + \frac{6}{x} \sigma_1\right),\label{eq:lambdadef} 
\end{align}
where in the last step we have used the partial-wave expansion of the thermally averaged annihilation cross section, with $\sigma_0$ ($\sigma_1$) corresponding to the $s$-wave ($p$-wave) annihilation, while neglecting terms beyond $p$-wave order.

The coherent freeze-out begins at $x>\xcfo$, where the boost factor $\gamma$ takes the form (see Eq.~(\ref{eq:mchieffoverT}))
\begin{align}
\gamma =  1 - \frac{\varphi_*^2(\xcfo)}{2} \frac{\xcfo^3}{x^3} = 1 - \frac{\xcfo^3}{x^3} + c_1 \frac{\xcfo^2}{x^3}\;,\label{eq:boostfactor}   
\end{align}
with $c_1 \approx 1.429$. As can be seen, $\gamma \to 1$ (or equivalently, $\mchieff \to m_\chi $) rapidly for $x>\xcfo$ as the ALP rolls back toward the origin, thereby driving the WIMP out of equilibrium. The coherent freeze-out temperature $\xcfohat$, defined as the time when the actual yield ceases to track its equilibrium value, $Y_\chi(\hat{x}_{\rm cfo}) - Y_{\chi,{\rm eq}}(\hat{x}_{\rm cfo}) \equiv Y_{\chi,{\rm eq}}(\hat{x}_{\rm cfo})$, can be computed analytically by combining Eqs.~(\ref{eq:BoltzmannEquationlightregion}) and (\ref{eq:boostfactor}), yielding~\cite{Ferrante:2025avp}
\begin{align}
\xcfohat \approx \xcfo + \frac{1}{3} \log C + \frac{7}{6} \log\left(\frac{1}{3} \log C +\frac{c_1}{3}\right) - \frac{c_1}{3}\;, \label{eq:xcfohat}
\end{align} 
where
\begin{align}
C \equiv \frac{27 g_\chi}{2\pi^3}\sqrt{\frac{15}{2g_*}} m_\chi M_{\rm Pl}\left(\sigma_0 + 6\, \sigma_1/\xcfo\right)\,\xcfo^{-4}\;.\label{eq:Cfactor}    
\end{align}
Note that the above expression holds in the limit of $e^{-(3\Delta x + c_1)} \ll 1$ with
$ \Delta x \equiv \xcfohat - \xcfo \ll \xcfo$.\footnote{In the limit $\xcfo \to \infty$ (corresponding to $\kappa \to 0$), the condition $3\Delta x + c_1 \gg 1$ no longer holds. In this case, one instead finds $\xcfohat \approx \xcfo$, i.e., coherent freeze-out completes almost instantaneously at $\xcfo$.} The final yield of $\chi$ is obtained by integrating Eq.~(\ref{eq:BoltzmannEquationlightregion}) from $\xcfohat$ to infinity:
\begin{align}
\int_{Y_{\chi}(\xcfohat)}^{Y_\chi^{\rm cfo}} \frac{{\rm d}Y_\chi}{Y_\chi^2} = - \int_{\xcfohat}^{\infty} {\rm d}x\, \frac{\lambda \gamma^2}{x^2}\;, \label{eq:yieldintegral}    
\end{align}
with $Y_\chi^{\rm cfo} \equiv Y_\chi(\infty)$. Substituting Eqs.~(\ref{eq:lambdadef}) and (\ref{eq:boostfactor}) into Eq.~(\ref{eq:yieldintegral}), performing the integral explicitly, and neglecting terms suppressed by higher powers of $\Delta x/\xcfo$, one obtains the WIMP yield in the coherent freeze-out scenario:
\begin{align}
Y_\chi^{\rm cfo} \approx \frac{14}{9} \sqrt{\frac{45 g_*(\xcfohat)}{\pi g_{*S}^2(\xcfohat)}}\frac{1}{m_\chi M_{\rm Pl}} \frac{\xcfohat}{\sigma_0 + 21\,\sigma_1/(10\,\xcfohat)}\;.\label{eq:yieldinlightregion}    
\end{align}
The approximate analytical expressions derived in Eqs.~(\ref{eq:xcfohat}) and (\ref{eq:yieldinlightregion}) show excellent agreement with numerical results in the small-$\kappa$ region. Indeed, their deviation from the full numerical solution of the Boltzmann equation is already below 10\% for $\kappa \lesssim 10^{-5}$ (corresponding to $\xcfohat \gtrsim 50$)~\cite{Ferrante:2025avp}. 

The delayed freeze-out leads to a parametric enhancement of the WIMP yield relative to the standard freeze-out scenario when $\xcfohat \gg 25$. Comparing Eq.~(\ref{eq:yieldinlightregion}) with the conventional freeze-out yield~\cite{Kolb:1990vq}, one finds
\begin{align}
\frac{Y_\chi^{\rm cfo}}{Y_\chi^{\rm fo}} =  \frac{14\,\Xi\left(\xcfohat,x_{\rm fo}\right)}{9}\frac{\xcfohat}{x_{\rm fo}} \frac{\sigma_0^{\rm fo} + 3\,\sigma_1^{\rm fo}/x_{\rm fo}}{\sigma_0^{\rm cfo} + 21\,\sigma_1^{\rm cfo}/(10\,\xcfohat)}\;,\label{eq:yieldratio}
\end{align}
where 
\begin{align}
\Xi\left(x_1,x_2\right) \equiv \sqrt{\frac{g_*(x_1)}{g_*(x_2)}} \frac{g_{*S}(x_2)}{g_{*S}(x_1)}\label{eq:Xifunction}  
\end{align}
counts the change in the relativistic
degrees of freedom between two temperatures,
$x_{\rm fo}\approx 25$,
and $\sigma_{i}^{\rm (c)fo}$ denotes the annihilation cross section in the (coherent) freeze-out scenario. 
Taking $\sigma_i^{\rm cfo} = \sigma_i^{\rm fo}$, the yield is parametrically enhanced by
\begin{align}
\frac{Y_\chi^{\rm cfo}}{Y_\chi^{\rm fo}} = \Xi\left(\xcfohat,x_{\rm fo}\right)\times\begin{cases}
\dfrac{14}{9} \dfrac{\xcfohat}{x_{\rm fo}}&\text{for $s$-wave}\;, \\[10pt]
\dfrac{20}{9} \dfrac{\xcfohat^2}{x_{\rm fo}^2}&\text{for $p$-wave}\;. 
\end{cases}    
\end{align}

Numerically, without entering a vacuum-dominated era (i.e., for $\xcfohat \approx \xcfo \lesssim {\cal O}(500)$) or violating the unitarity bound, the enhancement factor can reach $\sim 30$ for $s$-wave annihilation and $\sim 10^3$ for $p$-wave annihilation (cf.~Fig.~3 of Ref.~\cite{Ferrante:2025avp}).
Therefore, in this region, the coherent freeze-out scenario allows for a substantially larger annihilation cross section than in the standard freeze-out case while still yielding the correct relic abundance.

\subsubsection{Heavier mass region: $m_\chi \gtrsim \xcfo \,m_{\rm M}$}
\label{subsubsec:heaviermass}

Next, we consider the heavier WIMP region, $m_\chi \gtrsim {\cal O}(10)~{\rm TeV}$, which was not explored in Ref.~\cite{Ferrante:2025avp}. 
The main difference from the previous case is that $\mchieff$ remains larger than $m_{\rm M}$ at $\xcfo$. Consequently, the effective annihilation cross section is enhanced relative to its bare value, rather than suppressed. More explicitly,
\begin{align}
\langle \sigma v\rangle_{\rm eff} = \gamma^{-2} \langle \sigma v \rangle\;,    
\end{align}
where $\langle \sigma v \rangle \sim 1/(m_\chi^2 x^\ell)$ denotes the bare annihilation cross section. The Boltzmann equation in Eq.~(\ref{eq:BoltzmannEquationlightregion}) is then modified to
\begin{align}
Y_\chi'(x) = -\frac{\lambda}{\gamma^2x^2}\left[Y_\chi^2(x)-Y_{\chi,\rm eq}^2(x)\right],\label{eq:BoltzmannEquationheavyregion}   
\end{align}
where $\lambda$ is still given by Eq.~(\ref{eq:lambdadef}), with the appropriate bare cross section substituted in. 

The coherent freeze-out temperature $\xcfohat$ can still be computed analytically using a method similar to that of Ref.~\cite{Ferrante:2025avp}. We obtain an expression that differs slightly from the result in the lighter WIMP mass region, Eq.~(\ref{eq:xcfohat}):
\begin{align}
\xcfohat \approx \xcfo + \frac{1}{3} \log C' - \frac{1}{6} \log\left(\frac{1}{3} \log C' +\frac{c_1}{3}\right) - \frac{c_1}{3}\;,\label{eq:xcfohatheavy}  
\end{align}
where
\begin{align}
C' \equiv \frac{g_\chi}{6\pi^3}\sqrt{\frac{15}{2g_*}} m_\chi M_{\rm Pl}\left(\sigma_0 + 6\, \sigma_1/\xcfo\right).\label{eq:Cfactorheavy}    
\end{align}
Contrary to the $C$ factor in Eq.~(\ref{eq:Cfactor}), the quantity $C'$ in Eq.~(\ref{eq:Cfactorheavy}) remains finite in the $\xcfo\to\infty$ limit. Consequently, $\Delta x\equiv \xcfohat-\xcfo$ approaches an ${\cal O}(1)$ constant. For a typical thermal annihilation cross section, $\sigma_0\sim 3\times10^{-26}~{\rm cm^3/s}$, and $m_\chi\sim{\cal O}(100)~{\rm GeV}$, we find $\Delta x\approx 7$ in the large-$\xcfo$ limit.

Integrating Eq.~(\ref{eq:BoltzmannEquationheavyregion}), we obtain
\begin{align}
\int_{Y_{\chi}(\xcfohat)}^{Y_\chi^{\rm cfo}} \frac{{\rm d}Y_\chi}{Y_\chi^2} = - \int_{\xcfohat}^{\infty} {\rm d}x\, \frac{\lambda}{\gamma^2x^2}\;. \label{eq:yieldintegralheavy}     
\end{align}
Plugging Eqs.~(\ref{eq:lambdadef}) and (\ref{eq:boostfactor}) into Eq.~(\ref{eq:yieldintegralheavy}) and neglecting terms suppressed by higher powers of $\Delta x/\xcfo$, we arrive at
\begin{align}
Y_\chi^{\rm cfo} \approx  3 \sqrt{\frac{45 g_*(\xcfohat)}{\pi g_{*S}^2(\xcfohat)}}\frac{1}{m_\chi M_{\rm Pl}} \frac{c_1 + 3\Delta x}{\sigma_0 + 6\,\sigma_1/(\xcfohat)}\;.\label{eq:yieldinheavyregion}       
\end{align}
Comparing this result with the standard freeze-out yield, we find
\begin{align}
\frac{Y_\chi^{\rm cfo}}{Y_\chi^{\rm fo}} =  3 \,\Xi\left(\xcfohat,x_{\rm fo}\right)\frac{c_1+3\Delta x}{x_{\rm fo}} \frac{\sigma_0^{\rm fo} + 3\,\sigma_1^{\rm fo}/x_{\rm fo}}{\sigma_0^{\rm cfo} + 6\,\sigma_1^{\rm cfo}/\xcfohat}\;,\label{eq:yieldratioheavy}
\end{align}
where $\Xi$ is defined in Eq.~(\ref{eq:Xifunction}).
Taking $\sigma_i^{\rm cfo} = \sigma_i^{\rm fo}$, it leads to
\begin{align}
\frac{Y_\chi^{\rm cfo}}{Y_\chi^{\rm fo}} = \Xi\left(\xcfohat,x_{\rm fo}\right)\times\begin{cases}
\dfrac{3\left(c_1+3\Delta x\right)}{x_{\rm fo}}&\text{for $s$-wave}\;, \\[10pt]
\dfrac{3\left(c_1+3\Delta x\right)\xcfohat}{2x_{\rm fo}^2}&\text{for $p$-wave}\;. 
\end{cases}\label{eq:yieldratiocases}
\end{align}

\begin{figure}[t!]
	\centering
	\includegraphics[width=6.0in]
	{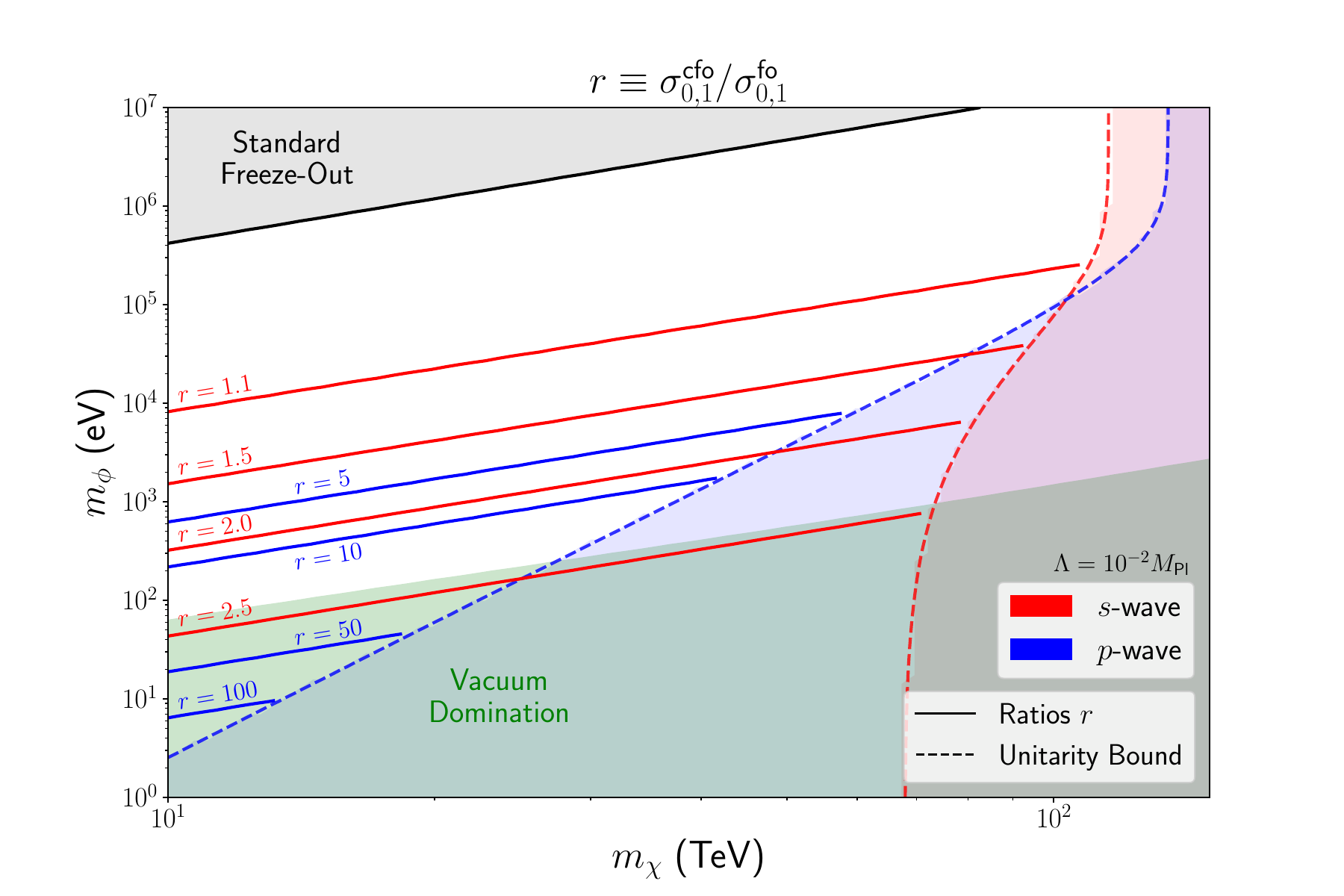}
    \caption{Parameter space of the coherent freeze-out mechanism in the heavier WIMP mass region, $m_\chi \gtrsim 10~{\rm TeV}$. The cutoff scale is fixed at $\Lambda = 10^{-2}M_\text{Pl}$. The solid red (blue) contour lines show the ratio of the WIMP annihilation cross sections in the coherent and standard freeze-out scenarios for $s$-wave ($p$-wave) annihilation, assuming that both scenarios reproduce the observed relic abundance. The red (blue) dashed curves denote the $s$-wave ($p$-wave) unitarity bounds, below which partial-wave unitarity is violated. The green shaded region denotes the parameter space in which the Universe undergoes a period of vacuum domination; see Sec.~\ref{sec:vacuumdomination} for details. In the gray shaded region, bounded by the critical value $\kappa=\kappa_c \approx 0.2$, the phase transition becomes a crossover and there is no enhancement of the annihilation cross section.}
    \label{fig:moneyplotFOPT}
\end{figure}

Interestingly, although the instantaneous annihilation cross section is strongly enhanced around $\xcfo$ by $\gamma^{-2}\sim\xcfo^2$, this does not necessarily lead to a smaller final yield than in standard freeze-out. This is because the enhancement, $\langle\sigma v\rangle_{\rm eff}\propto\gamma^{-2}$, rapidly diminishes as the ALP relaxes toward the origin and $\gamma$ approaches unity. For $s$-wave annihilation, the enhanced annihilation removes the parametric $\xcfo$ dependence of the final yield, giving $Y_\chi^{\rm cfo}/Y_\chi^{\rm fo}\sim{\cal O}(1)/x_{\rm fo}$. However, the numerical coefficient $3(c_1+3\Delta x)$ is sizable: for a typical thermal annihilation cross section, $\Delta x\approx7$, yielding $Y_\chi^{\rm cfo}/Y_\chi^{\rm fo}\approx3$. For $p$-wave annihilation, the additional velocity suppression of the bare cross section introduces an extra factor of $\xcfo$, such that $Y_\chi^{\rm cfo}/Y_\chi^{\rm fo}\sim\xcfo/x_{\rm fo}^2$, which can become parametrically larger than unity for sufficiently large $\xcfo$.

Note that the analytical expressions in Eqs.~(\ref{eq:yieldinheavyregion})--(\ref{eq:yieldratiocases}) are accurate only to leading order in $\Delta x/\xcfo$, at which $Y_\chi^{\rm cfo}$ is independent of $\kappa$ for $s$-wave annihilation. Higher-order corrections to the yield are computed in Appendix~\ref{app:higherorderheavywimp}, yielding better agreement with the numerical solution of the Boltzmann equation. Once these corrections are included, $Y_\chi^{\rm cfo}$ for $s$-wave annihilation also acquires a mild dependence on $\kappa$.

In Fig.~\ref{fig:moneyplotFOPT}, we show the $(m_\chi,m_\phi)$ parameter space for coherent freeze-out in the heavier WIMP mass region, including the higher-order corrections discussed above. We assume the mediator mass $m_{\rm M}\sim 100$~GeV, so that the region considered in this subsection sets in at $m_\chi\gsim 10$~TeV. (For reference, the corresponding parameter space for the lighter WIMP mass, $1~{\rm GeV} \lesssim m_\chi \lesssim 10~{\rm TeV}$, has been shown in Fig.~3 of Ref.~\cite{Ferrante:2025avp}.) The solid red (blue) lines denote the enhancement of the $s$-wave ($p$-wave) annihilation cross section relative to standard freeze-out, assuming that both scenarios reproduce the observed DM relic abundance. The red and blue dashed lines show the partial-wave unitarity bounds for $s$-wave and $p$-wave annihilation, respectively (see Appendix~\ref{app:higherorderheavywimp} for details). Before entering the vacuum-dominated regime (green shaded region) or violating the unitarity bounds, the $s$-wave annihilation cross section can be enhanced by at most a factor of roughly two, whereas the $p$-wave cross section can be enhanced by up to $\sim 20$. For both $s$-wave and $p$-wave annihilation, these enhancements are considerably smaller than those in the lighter WIMP mass region, as discussed in Sec.~\ref{subsubsec:lightermass} (see also Fig.~3 of Ref.~\cite{Ferrante:2025avp}).

Finally, we note that in the FOPT regime considered in this section, the coherent ALP abundance would overclose the Universe if the ALP were stable. The ALP must therefore decay into radiation after coherent freeze-out, and its energy must be transferred to the SM radiation before Big Bang Nucleosynthesis (BBN), as discussed in Ref.~\cite{Ferrante:2025avp}.\footnote{The specific decay channel is model dependent. However, as long as coherent freeze-out begins before vacuum domination sets in and the ALP energy is transferred sufficiently rapidly into radiation, the resulting entropy injection leads to at most a mild, ${\cal O}(1)$ dilution of the WIMP yield and does not qualitatively affect the coherent freeze-out predictions; see Appendix~C of Ref.~\cite{Ferrante:2025avp} for further discussion and working examples.} Consequently, unlike in the crossover regime discussed in Sec.~\ref{sec:crossover}, WIMPs effectively constitute the entire present-day DM abundance in the FOPT regime.
 
\section{Vacuum-Dominated Regime}
\label{sec:vacuumdomination}
As discussed in Sec.~\ref{subsec:phasestructure}, for sufficiently small $\kappa$, the ALP remains trapped in the false vacuum long enough for the vacuum energy to dominate the total energy density of the Universe before coherent freeze-out takes place.
This corresponds to $\kappa \lesssim 10^{-9}$, or equivalently $\xcfo \gtrsim 500$ (see the green shaded region in Fig.~\ref{fig:phasediagram}). In this section, we investigate this vacuum-dominated regime in detail.

\subsection{Inflation Dynamics}
We consider the cosmological evolution prior to matter-radiation equality. Including the contribution from the $\phi$ field, the Friedmann equation reads
\begin{align}
H^2 = \frac{8\pi}{3 M_{\rm Pl}^2}\left(\rho_{\rm rad}+\rho_\phi\right),\label{eq:Friedmann}
\end{align}
where $\rho_{\rm rad}$ and $\rho_\phi$ denote the radiation and ALP energy densities, respectively. The temperature dependence of the Hubble expansion rate is therefore given by
\begin{align}
H(T) = 1.66\sqrt{g_*}\frac{T^2}{M_{\rm Pl}}\sqrt{1+\alpha(x)}\;, \label{eq:Hubble-evolution}
\end{align}
where $\alpha(x)\equiv \rho_\phi/\rho_{\rm rad}$. Throughout the regime relevant to this section, $x\gg1$, such that the finite-temperature contribution to the ALP energy density is negligible compared to the vacuum contribution, and $\rho_\phi \approx m_\phi^2m_\chi\Lambda$. Consequently, 
\begin{align}
\alpha(x)=\alpha_0 x^4\;,    
\end{align}
where $\alpha_0 \equiv 15 g_\chi \kappa/(\pi^4 g_*)$ is approximately a temperature-independent constant, as shown in Eq.~(\ref{eq:alphadef}). In the vacuum-dominated limit, $\alpha(x)\gg1$, the Hubble parameter approaches a constant,
\begin{align}
H\to H_{\rm inf}\equiv \sqrt{\frac{8\pi m_\phi^2m_\chi\Lambda}{3M_{\rm Pl}^2}}\;,\label{eq:Hinf}
\end{align}
and the Universe enters an inflationary phase.

Integrating Eq.~(\ref{eq:Friedmann}) over cosmic time $t$, we obtain the evolution of the scale factor,
\begin{align}
a(t) \propto \sqrt{\sinh\left(2 H_{\rm inf}t\right)}\;,    
\end{align}
which in turn gives
\begin{align}
 H(t) = H_{\rm inf}\coth \left(2H_{\rm inf}t\right).\label{eq:Hubble-inf}   
\end{align}
In the limit $H_{\rm inf}t\to0$, this reduces to the radiation-dominated result, $H=1/(2t)$. For $H_{\rm inf}t\gtrsim1$, the Hubble parameter rapidly approaches its asymptotic value, $H\approx H_{\rm inf}$, while the scale factor grows exponentially as $a(t) \propto \exp(H_{\rm inf}t)$. 
Combining Eqs.~(\ref{eq:Friedmann}) and (\ref{eq:Hubble-inf}), we obtain a very simple relation between cosmic time $t$ and the inverse temperature $x\equiv m_\chi/T$:
\begin{align}
\cosh^2\left(2H_{\rm inf}t\right) = 1+\alpha_0 x^4\;.   
\end{align}

As the Universe enters the inflationary phase, its temperature decreases exponentially with time, $T\propto \exp(-H_{\rm inf}t)$. Inflation ends at a temperature $T_e$, when the ALP can tunnel through the potential barrier.\footnote{After tunneling, the ALP rolls toward the origin with an initial field displacement $\phi_i=\phi_*=\sqrt{2m_\chi\Lambda}$. Because $m_\phi/H_{\rm inf} \sim M_{\rm Pl}/\sqrt{m_\chi \Lambda} \gg 1$, the ALP is not damped by Hubble friction after tunneling.
Its subsequent evolution is governed by the quadratic potential. The corresponding slow-roll parameters are given by
$M_{\rm Pl}^2/(4\pi\phi^2) \geq M_{\rm Pl}^2/(4\pi\phi_i^2)=M_{\rm Pl}^2/(8\pi m_\chi\Lambda)\gg1$,
and therefore the subsequent evolution of the ALP does not support a further period of inflation.} The tunneling condition is approximately $T_e^4\,e^{-S_B}\sim H_{\rm inf}^4$, or equivalently,
\begin{align}
S_B\sim4\log\left(\frac{T_e}{H_{\rm inf}}\right).
\end{align}
As discussed in Sec.~\ref{subsec:bounceaction}, tunneling remains highly suppressed until the potential barrier nearly disappears. Using the bounce action derived in Eq.~(\ref{eq:thickwallbouncescaling}) in the thick-wall limit, we find
\begin{align}
\frac{\xcfo - x_e}{\xcfo}  &\approx    \left[\frac{4}{A}\log \left(\frac{T_e}{H_{\rm inf}}\right)\right]^{4/3} \left(\frac{m_\chi}{\Lambda}\right)^{2}\sim \left[\frac{4}{A}\log\left(\frac{M_{\rm Pl}}{\kappa^{1/6}\, m_\chi}\right)\right]^{4/3} \left(\frac{m_\chi}{\Lambda}\right)^{2}  \ll 1\;,
\end{align}
where $x_e \equiv m_\chi/T_e$. In the second step, we have used $T_e\approx T_{\rm cfo}$ together with Eqs.~(\ref{eq:kappadef}) and (\ref{eq:Hinf}). Therefore, owing to the large hierarchy $m_\chi\ll\Lambda$, the end of inflation occurs extremely close to $\xcfo$, and we can safely approximate $x_e\approx\xcfo$.

Let $x_i$ denote the temperature at which the vacuum energy begins to dominate, i.e., $\alpha(x_i)\equiv 1$. Using Eq.~(\ref{eq:alphadef}), we obtain
\begin{align}
x_i = c_i\,\kappa^{-1/4}\;, \qquad \text{with $c_i \equiv \left(\pi^4 g_*/(15 g_\chi)\right)^{1/4}$}\;.
\label{eq:xi}
\end{align}
Here $c_i$ is an ${\cal O}(1)$ number; $c_i \sim 2 - 4$ for $g_* \sim 10 - 90$. On the other hand, inflation ends at
\begin{align}
x_e \approx \xcfo = c_e\, \kappa^{-1/3}\;, \quad \text{with $c_e \approx 0.836$}\;.\label{eq:xe} 
\end{align}
Vacuum domination occurs whenever $x_i < x_e$. During the interval $x_i < x < x_e$, the Universe undergoes a period of inflation. The corresponding number of $e$-folds is
\begin{align}
N = \log\left(\frac{x_e}{x_i}\right)
= \frac{1}{12}\log\left(\frac{1}{\kappa}\right) - \log\left(\frac{c_i}{c_e}\right).
\end{align}
Requiring $N>0$ then determines the maximum value of $\kappa$ for which vacuum domination occurs, reproducing the bound in Eq.~(\ref{eq:kappavac}), which also corresponds to the green shaded region in Fig.~\ref{fig:phasediagram}.\\

During inflation, the temperature, and hence $\mchieff$, evolves rapidly, raising the question of whether the time-dependent WIMP mass could lead to non-adiabatic production of $\chi$ particles. We show below that such production is negligible due to the smallness of $H_{\rm inf}$ relative to the WIMP mass. Using Eq.~(\ref{eq:phistarevolution}), we have $\mchieff=\gamma m_\chi\sim\kappa x^2m_\chi$ for $x<\xcfo$. A direct calculation gives
\begin{align}
\frac{\dot{m}_{\chi,\rm eff}}{\mchieff^2} \sim \frac{H_{\rm inf}}{m_\chi} \frac{1}{\kappa x^2} \sqrt{\frac{1+\alpha_0 x^4}{\alpha_0 x^4}}\;.   
\end{align}
In the vacuum-dominated limit, $\alpha_0x^4\gg1$. Using Eq.~(\ref{eq:Hinf}), we then obtain
\begin{align}
\frac{\dot{m}_{\chi,\rm eff}}{\mchieff^2} \sim \frac{m_\chi}{M_{\rm Pl}}\frac{\kappa^{-1/2}}{x^2} <  \frac{m_\chi}{M_{\rm Pl}}\frac{\kappa^{-1/2}}{x_i^2} \sim \frac{m_\chi}{M_{\rm Pl}} \ll 1\;,  
\end{align}
where in the last step we have used $x_i\sim\kappa^{-1/4}$. Therefore, throughout the inflationary phase, $x_i<x<x_e$, the effective WIMP mass evolves adiabatically, $\dot m_{\chi,\rm eff}\ll\mchieff^2$, and non-adiabatic production of $\chi$ particles can be safely neglected.

\subsection{Reheating}
\label{subsec:reheating}
For $x>x_e$, the ALP rolls toward the origin and subsequently undergoes coherent oscillations, with its energy density redshifting as non-relativistic matter, $\rho_\phi \sim m_\phi^2 \phi_*^2 x_e^3/x^3 \sim m_\chi T^3$. To avoid overclosing the Universe, the ALP condensate must transfer most of its energy to other (relativistic) species, thereby initiating a period of reheating. The WIMP-ALP coupling alone is insufficient to efficiently dissipate the ALP energy: as $\phi$ rolls toward the origin, the effective WIMP mass rapidly approaches its bare value, $m_{\chi,\rm eff}\to m_\chi\gg m_\phi$, kinematically suppressing the perturbative process $\phi\phi\to\chi\chi$. We therefore require additional couplings of $\phi$ to light SM species to achieve efficient reheating.

As an example, we consider the coupling to photons through a Chern-Simons term,
\begin{align}
{\cal L} \supset -\frac{\phi}{4f}F \tilde{F}\;,    
\end{align}
where $f$ denotes the corresponding decay constant. As discussed in Sec.~\ref{subsec:experiments-for-ALP-Miracle}, this coupling does not generate a thermal potential for the ALP and therefore does not affect the ALP dynamics prior to reheating. We further take $f$ to be sufficiently large that $\phi$ neither thermalizes through scattering with photons nor decays perturbatively into photons before reheating.

Once $\phi$ begins to oscillate about the origin, however, its nonzero velocity can induce a tachyonic instability in one of the photon helicity modes, opening a non-perturbative (preheating) channel for transferring the ALP energy into photons (see, e.g., Ref.~\cite{Adshead:2015pva}).
Neglecting the expansion of the Universe over the short timescale of particle production, the photon modes obey
\begin{align}
\ddot{A}_{\pm} + \left(k^2 \mp k\frac{\dot\phi}{f}\right) A_{\pm} = 0\;,    
\end{align}
where $k$ denotes the momentum of a given photon mode. Here we assume that the pre-existing population of charged particles has been sufficiently diluted during inflation such that the photon plasma mass and conductivity can be neglected. This assumption is justified when $T_e$ falls below the electron mass, such that the thermal electron-positron population is Boltzmann suppressed.

For the ALP to transfer most of its energy to the photon bath within one Hubble time, we require the maximum instantaneous growth rate of the tachyonic mode to be much larger than the Hubble rate,
\begin{align}
\frac{|\dot{\phi}|}{2f} \gg H\;.    
\end{align}
Using $|\dot{\phi}|_{\rm max} \approx  m_\phi\phi_*\approx m_\phi\sqrt{2m_\chi\Lambda}$ and $H\approx H_{\rm inf}$, we find
\begin{align}
\frac{|\dot{\phi}|_{\rm max}}{2f H} \approx \sqrt{\frac{3}{16 \pi}}\,\frac{M_{\rm Pl}}{f}\;.    
\end{align}
Interestingly, the dependence on $m_\phi$, $m_\chi$, and $\Lambda$ drops out entirely. Therefore, the condition for effectively instantaneous reheating is parametrically $f\ll M_{\rm Pl}$, which can be readily satisfied for experimentally allowed values of $f$.

Assuming instantaneous reheating, the transfer of energy from the ALP condensate to photons reheats the Universe from $T_e$ to a higher temperature $T_{\rm RH}'$, where we use a prime to distinguish this temperature from the primordial reheating temperature $T_{\rm RH}$. Energy conservation gives
\begin{align}
\frac{\pi^2}{30}g_*(T_{\rm RH}') T_{\rm RH}'^4 =  \rho_\phi (T_e) = m_\phi^2 m_\chi \Lambda\;,  
\end{align}
which yields 
\begin{align}
T_{\rm RH}' &= \left[\frac{30\,m_\phi^2m_\chi \Lambda}{\pi^2 g_*(T_{\rm RH}')}\right]^{1/4} = m_\chi \left(\frac{15 g_\chi}{\pi^4 g_*}\right)^{1/4} \kappa^{1/4}\;,
\end{align}
where in the second equality we have used the definition of $\kappa$ in Eq.~(\ref{eq:kappadef}). Therefore, in the instantaneous-reheating limit, the Universe is reheated back to the temperature at which vacuum domination began, $T_{\rm RH}' = T_i$, where $T_i\equiv m_\chi/x_i$, cf. Eq.~(\ref{eq:xi}). This follows from the fact that the ALP vacuum energy remains approximately constant throughout inflation, $\rho_\phi(x_e)\approx \rho_\phi(x_i)$. The burst of inflation in this scenario is consistent with all observational data as long as the temperature $T_i$ is above a few MeV, allowing for the standard BBN to occur after reheating.

\subsection{WIMP Relic Abundance}
Next, we discuss WIMP freeze-out during inflation. In terms of $x\equiv m_\chi/T$, the Boltzmann equation for the WIMP yield is given by
\begin{align}
\frac{{\rm d}Y_\chi}{{\rm d}x}= -\frac{s}{x H}\langle\sigma v \rangle_{\rm eff}\left(Y_\chi^2-Y_{\chi,{\rm eq}}^2\right).\label{eq:Boltzmann-inflation} 
\end{align}
Note that this equation takes the same form as in a radiation-dominated Universe, with the effect of vacuum energy entering only through the modified Hubble parameter given in Eq.~(\ref{eq:Hubble-inf}). It is useful to recast the Boltzmann equation into the more suggestive form
\begin{align}
\frac{x}{Y_{\chi,{\rm eq}}} \frac{{\rm d}Y_\chi}{{\rm d}x} = -\frac{n_{\chi,{\rm eq}} \langle \sigma v\rangle_{\rm eff}}{H}\left(\frac{Y_\chi^2}{Y_{\chi,{\rm eq}}^2}-1\right),    
\end{align}
where $n_{\chi,{\rm eq}}$ is given by Eq.~(\ref{eq:nchieq}) and $Y_{\chi,{\rm eq}}\equiv n_{\chi,{\rm eq}}/s$. Therefore, freeze-out occurs as the effective annihilation rate $n_{\chi,{\rm eq}} \langle \sigma v\rangle_{\rm eff}$ drops below the Hubble rate.

Before the end of inflation, $x<x_e\approx\xcfo$, WIMPs remain relativistic, $\mchieff\lesssim T$, due to the suppression of their effective mass.   For temperatures below $m_{\rm M}$, where $m_{\rm M}$ is the mass of the mediator responsible for WIMP annihilation to SM particles,
the partial-wave annihilation rate takes the form $\Gamma_{\rm ann}\sim n_{\chi,{\rm eq}} \langle \sigma v\rangle_{\rm eff} \sim T^5/m_{\rm M}^4$. Meanwhile, the Hubble parameter approaches the constant value $H_{\rm inf}$. Therefore, if inflation lasts sufficiently long, WIMPs can freeze out while still relativistic. The relativistic freeze-out temperature $T_{\rm rfo}$ is determined by equating the annihilation rate to the Hubble rate. For a given $T_i$, the temperature at which vacuum domination begins, the relativistic freeze-out temperature is estimated as
\begin{align}
T_{\rm rfo} & \sim \left(\frac{T_i^2 m_{\rm M}^4}{M_{\rm Pl}}\right)^{1/5}\sim  10~{\rm MeV} \left(\frac{T_i}{\rm GeV}\right)^{2/5}\left(\frac{m_{\rm M}}{100~{\rm GeV}}\right)^{4/5}.
\label{eq:Trfo}
\end{align} 
Constraints from BBN require the post-reheating temperature to be above the MeV scale~\cite{Kawasaki:2000en,Giudice:2000ex,Hannestad:2004px}, implying $T_i\gtrsim \text{a few MeV}$. For mediator masses around or below the weak scale, this generally gives $T_{\rm rfo}\lesssim T_i$, such that vacuum domination begins before relativistic freeze-out can occur. Depending on the relative ordering of $T_e$ and $T_{\rm rfo}$, two distinct regimes arise, as discussed below.

\subsubsection{Short-Inflation Branch: $T_{\rm rfo}<T_e<T_i$}
This regime corresponds to a relatively short period of inflation that ends before relativistic freeze-out can occur. Using $T_e=T_i\,e^{-N}$ together with Eq.~(\ref{eq:Trfo}), the condition $T_{\rm rfo}<T_e$ imposes an upper bound on the number of $e$-folds, denoted by $N_*$:
\begin{align}
N_*&\equiv
\log\left(T_i^{3/5} m_{\rm M}^{-4/5}M_{\rm Pl}^{1/5}\right) \approx 5 +\frac{3}{5}\log\left(\frac{T_i}{\rm GeV}\right)-\frac{4}{5}\log\left(\frac{m_{\rm M}}{100~{\rm GeV}}\right).
\end{align}
In terms of $\kappa$, this translates into a lower bound on $\kappa$, denoted by $\kappa_*$:
\begin{align}
\kappa_*&\equiv\left(c_e c_i^{-2/5} m_{\rm M}^{4/5}m_\chi^{-3/5}M_{\rm Pl}^{-1/5}\right)^{30/7}\nonumber\\
&\approx 2 \times 10^{-16}\left(\frac{m_\chi}{100~{\rm GeV}}\right)^{6/7}\left(\frac{m_{\rm M}}{m_\chi}\right)^{24/7}\left(\frac{g_\chi/2}{g_*(x_i)/10}\right)^{3/7}.\label{eq:kapparegion-no-rfo}
\end{align}
Thus, $\kappa$ in this regime is located within 
\begin{align}
 \kappa_{*}  < \kappa < \kappa_{\rm vac}\;,
\end{align}
where the upper bound $\kappa_{\rm vac}$ (cf.~Eq.~(\ref{eq:kappavac})) follows from the requirement of vacuum domination, $T_e<T_i$, which is necessary for an inflationary phase to occur. 

Combining Eqs.~(\ref{eq:kapparegion-no-rfo}) and (\ref{eq:xi}), the corresponding range of $x_i$ is
\begin{align}
x_{\rm vac}<x_i <  \min\left\{ x_{i,{\rm max}}\,,\; \frac{m_\chi}{T_{\rm BBN}} \right\},  
\end{align}
where $x_{\rm vac}$ is given by Eq.~(\ref{eq:xvac}), $T_{\rm BBN}\sim {\rm MeV}$, and 
\begin{align}
 x_{i,{\rm max}} &\approx 2\times 10^4\left(\frac{m_\chi}{100~{\rm GeV}}\right)^{-3/14} \left(\frac{m_\chi}{m_{\rm M}}\right)^{6/7}\left(\frac{g_\chi/2}{g_*(x_i)/10}\right)^{-5/14}. 
\end{align}

In this regime, WIMPs remain in thermal equilibrium with the SM bath until the end of inflation. Subsequent reheating, as discussed in Sec.~\ref{subsec:reheating}, reheats the Universe from $T_e$ back to $T_i$. The characteristic timescale for reheating is
\begin{align}
\tau_{\rm RH} \sim \frac{f}{|\dot{\phi}|} \sim \frac{f}{m_\phi \sqrt{m_\chi \Lambda}}\;.    
\end{align}
On the other hand, WIMPs rapidly become non-relativistic after inflation as $\phi$ begins to roll at $T_e$. Their effective mass subsequently evolves as
\begin{align}
\mchieff = \left(1-\frac{\varphi^2}{2}\right)m_\chi \approx m_\chi\sin^2(m_\phi t)\;, 
\end{align}
where $t$ denotes the time elapsed since the onset of the ALP rolling. 
As a conservative criterion, we take WIMPs to become non-relativistic
once $\mchieff\gtrsim T_i$.\footnote{The actual onset of the non-relativistic regime occurs when
$\mchieff\gtrsim T$, where $T$ is the instantaneous temperature.
Here we instead impose the more conservative sufficient condition
$\mchieff\gtrsim T_i$, since $T\leq T_i$ during this period.
Even with this stronger condition, the timescale for WIMPs to become
non-relativistic is parametrically shorter than the reheating timescale,
as shown in Eq.~(\ref{eq:tauratio}).} 
The corresponding duration of the relativistic phase is therefore
\begin{align}
\tau_{\rm rel} \sim \frac{1}{m_\phi \sqrt{x_i}}\;.  
\end{align}
Comparing the two timescales gives
\begin{align}
\frac{\tau_{\rm rel}}{\tau_{\rm RH}} \sim  \frac{\sqrt{m_\chi \Lambda}}{f}\frac{1}{\sqrt{x_i}}\;.\label{eq:tauratio}   
\end{align}
For $m_\chi$ around the weak scale and sub-Planckian $\Lambda$, the parameter space with $f\lesssim\sqrt{m_\chi\Lambda}$ is already largely excluded (see Fig.~\ref{fig:ALPMiracleWindow}). We therefore have $\tau_{\rm rel}\ll\tau_{\rm RH}$, given that $x_i\sim \kappa^{-1/4} \gg1$. Consequently, the WIMP effective mass approaches its bare value $m_\chi\gg T_i$ well before reheating is completed, making thermal WIMP production during reheating negligible.

WIMPs undergo coherent freeze-out once the ALP starts to roll at $T_e$. 
Since $x_{\rm cfo}\approx x_e\gg {\cal O}(100)$ in the vacuum-dominated regime, the parameter space of interest for a weak-scale mediator typically lies within the lighter-mass region discussed in Sec.~\ref{subsubsec:lightermass}, $m_\chi\lesssim \xcfo m_{\rm M}$. We therefore focus on this region in the following.
The final WIMP yield can then be obtained from the corresponding result in the FOPT regime, with the subsequent dilution from reheating taken into account:
\begin{align}
Y_{\chi}^{\rm vac} = Y_\chi^{\rm cfo}(x_e) \frac{H(x_e)}{H_{\rm RD}(x_e)} \frac{g_{*S}(x_e)}{g_{*S}(x_i)} \left(\frac{x_i}{x_e}\right)^3 = Y_\chi^{\rm cfo}(x_e)\, \Xi\left(x_i,x_e\right) \,e^{-N} \;,
\end{align}
where the superscript ``vac'' denotes the vacuum-dominated regime, $Y_\chi^{\rm cfo}(x_e)$ is given by Eq.~(\ref{eq:yieldinlightregion}) with the replacement $\xcfohat\to x_e$, and $\Xi$ is defined in Eq.~(\ref{eq:Xifunction}).
Unlike in the radiation-dominated era, the Hubble parameter remains approximately constant during vacuum domination rather than decreasing with temperature. As a result, the WIMP yield at decoupling is enhanced by a factor $H(x_e)/H_{\rm RD}(x_e) \sim e^{2N}$ relative to the corresponding coherent freeze-out result in a radiation-dominated Universe. This enhancement partially compensates the entropy dilution from reheating, $x_i^3/x_e^3 = e^{-3N}$, so that the net dilution scales only as $e^{-N}$. 

Altogether, this leads to a parametrically different WIMP yield compared with standard freeze-out. Assuming the same annihilation cross section in the two scenarios, $\sigma_i^{\rm cfo}=\sigma_i^{\rm fo}$, we obtain
    \begin{align}
        \frac{Y_\chi^{\rm vac}}{Y^{\rm fo}_\chi}\Bigg|_\text{$s$-wave}&= \frac{14}{9}\,\Xi\left(x_i,x_{\rm fo}\right)  \frac{x_e}{x_{\rm fo}}\,e^{-N} = \frac{14}{9}\,\Xi\left(x_i,x_{\rm fo}\right) \frac{x_i} {x_{\rm fo}}\;,\\
        \frac{Y_\chi^{\rm vac}}{Y^{\rm fo}_\chi}\Bigg|_\text{$p$-wave} &= \frac{20}{9}\,\Xi\left(x_i,x_{\rm fo}\right)  \frac{x_e^2}{x_{\rm fo}^2}\,e^{-N} = \frac{20}{9}\,\Xi\left(x_i,x_{\rm fo}\right) \frac{x_i x_e} {x_{\rm fo}^2}\;,\label{eq:ratio-short-inflation} 
    \end{align}
where $x_i$ and $x_e$ are given by Eqs.~(\ref{eq:xi})--(\ref{eq:xe}). Although the WIMP abundance is exponentially diluted during inflation, these ratios are parametrically larger than unity. This is because coherent freeze-out occurs at a much lower temperature, corresponding to a substantially larger value of $\xcfo$.
To make this enhancement more transparent, we rewrite the result in terms of $x_{\rm vac}$ defined in Eq.~(\ref{eq:xvac}), which denotes the largest value of $\xcfo$ for which vacuum domination is avoided:
\begin{align}
\frac{Y_\chi^{\rm vac}}{Y^{\rm fo}_\chi}\Bigg|_\text{$s$-wave} &=  \frac{14}{9}\,\Xi\left(x_i,x_{\rm fo}\right)  \frac{x_{\rm vac}}{x_{\rm fo}}\, e^{3N}\;,\\
\frac{Y_\chi^{\rm vac}}{Y^{\rm fo}_\chi}\Bigg|_\text{$p$-wave} &=  \frac{20}{9}\,\Xi\left(x_i,x_{\rm fo}\right)  \frac{x_{\rm vac}^2}{x_{\rm fo}^2}\, e^{7N}\;.
\end{align}
The factors preceding $e^{3N}$ and $e^{7N}$ correspond to the maximal yield attainable in the FOPT regime without vacuum domination, obtained near $\xcfo\approx x_{\rm vac}$. Thus, relative to this maximal non-inflating CFO yield, vacuum domination provides an additional enhancement by a factor of $e^{3N}$ for $s$-wave annihilation and $e^{7N}$ for $p$-wave annihilation, up to the variation in the effective relativistic degrees of freedom. Equivalently, the annihilation cross section required to reproduce the observed relic abundance can be enhanced by the same factors, as shown in Fig.~\ref{fig:moneyplotvac}.

\subsubsection{Long-Inflation Branch: $T_e < T_{\rm rfo} < T_i$}
For a longer period of inflation, $N>N_*$ (or equivalently, $\kappa<\kappa_*$), WIMPs can undergo relativistic freeze-out before the end of inflation. Their yield at freeze-out is given by the equilibrium value in the relativistic limit,
\begin{align}
Y_\chi (T_{\rm rfo}) = Y_{\chi,{\rm eq}}(T_{\rm rfo}) \approx  0.2\,\frac{g_\chi}{g_{*S}(T_{\rm rfo})}\;.   
\end{align}
After freeze-out, the WIMP yield remains constant throughout the remainder of the inflationary period.\footnote{Strictly speaking, after relativistic freeze-out,
the WIMP distribution becomes non-thermal, taking the form
$f(p/T)$ rather than the equilibrium form $f(E/T)$. This can
modify the ALP potential and hence $\xcfo$, whose derivation
assumes thermal equilibrium. Nevertheless, since WIMPs remain
relativistic until near the end of inflation, we expect only
an ${\cal O}(1)$ correction to $\xcfo$. We therefore retain
the equilibrium result and neglect this additional non-thermal correction
in the following analysis.}

At the end of inflation, the Universe is reheated from $T_e$ back to $T_i$, while the WIMP mass is restored from $\mchieff$ to its bare value $m_\chi$. The entropy injection associated with reheating dilutes the WIMP yield by a factor
$[s(T_e)/s(T_i)] = e^{-3N} g_{*S}(T_e)/g_{*S}(T_i)$. The yield immediately after reheating is therefore
\begin{align}
Y_d &= Y_{\chi}(T_{\rm rfo})\,\frac{g_{*S}(T_e)}{g_{*S}(T_i)}\, e^{-3N}\nonumber\\
& \approx 2.2\,\frac{g_*(T_i)}{g_{*S}(T_{\rm rfo})}\frac{g_{*S}(T_e)}{g_{*S}(T_i)}\, \frac{T_i}{m_\chi}\;,\label{eq:Yd}
\end{align}
which is almost entirely determined by $T_i/m_\chi = 1/x_i$, independent of the annihilation cross section.

If WIMPs do not recouple with the SM bath after reheating, their present-day energy density is simply given by $\rho_0=m_\chi Y_d s_0\sim T_i s_0$, where $s_0 \approx 2891~{\rm cm}^{-3}$ is the present-day entropy density. The resulting WIMP abundance exceeds the observed DM abundance for $T_i\gtrsim0.2~{\rm eV}$, implying DM overproduction by roughly seven orders of magnitude or more under the requirement $T_i\gtrsim T_{\rm BBN}\sim{\rm MeV}$. Below, we show that the increase in the WIMP annihilation cross section after inflation can indeed lead to recoupling with the SM bath and substantially reduce the relic abundance, which prevents overclosure.

The Hubble expansion rate immediately after reheating is of the same order as that at relativistic freeze-out during inflation, $H \sim \Gamma_{\rm ann}(T_{\rm rfo}) \sim T_{\rm rfo}^5/m_{\rm M}^4$ for $T_{\rm rfo} \ll m_{\rm M}$. 
After inflation, the effective WIMP mass approaches its bare value, $\mchieff\to m_\chi$. As a result, the cross section changes from 
$\langle \sigma v\rangle\big|_{T_{\rm rfo}} \sim T_{\rm rfo}^2/m_{\rm M}^4$ to $\langle \sigma v\rangle\big|_{T_i} \sim \min \left\{ m_\chi^2/(m_{\rm M}^4 x_i^\ell), 1/(m_\chi^2 x_i^\ell) \right\}$, where $\ell=0$ and $1$ correspond to $s$-wave and $p$-wave annihilation, respectively. 
Neglecting annihilation during the short reheating interval, the surviving WIMP number density immediately after reheating is $n_\chi \sim (T_e/T_{\rm rfo})^3 n_\chi(T_{\rm rfo}) \sim T_e^3$. Comparing the annihilation rate with the Hubble rate at $T_i$, we obtain
\begin{align}
{\cal R}_s\equiv\frac{n_\chi \langle \sigma v\rangle_s}{H}\Bigg|_{T_i} &\sim \frac{c_i^4}{c_e^3}\left(\frac{T_i}{T_{\rm rfo}}\right)^5 \frac{m_\chi}{T_i} \min\left\{1,\frac{m_{\rm M}^4}{m_\chi^4}
\right\},\\
{\cal R}_p \equiv\frac{n_\chi \langle \sigma v\rangle_p}{H}\Bigg|_{T_i} &\sim \frac{c_i^4}{c_e^3}\left(\frac{T_i}{T_{\rm rfo}}\right)^5  \min\left\{1,\frac{m_{\rm M}^4}{m_\chi^4}
\right\}.
\end{align}
Using Eq.~(\ref{eq:Trfo}) and taking $m_{\rm M}\sim100~{\rm GeV}$, we find
\begin{align}
{\cal R}_s &\sim 10^{9} \left(\frac{T_i}{{\rm MeV}}\right)^2
\begin{cases}
\dfrac{m_\chi}{100~{\rm GeV}} & \text{for $m_\chi \lesssim 100~{\rm GeV}$}\;, \\[5pt]
\left(\dfrac{100~{\rm GeV}}{m_\chi}\right)^3 &  \text{for $m_\chi \gtrsim 100~{\rm GeV}$}\;,
\end{cases}\\
{\cal R}_p &\sim 10^{4} \left(\frac{T_i}{{\rm MeV}}\right)^3 
\begin{cases}
1 & \text{for $m_\chi \lesssim 100~{\rm GeV}$}\;, \\[5pt]
\left(\dfrac{100~{\rm GeV}}{m_\chi}\right)^4 &  \text{for $m_\chi \gtrsim 100~{\rm GeV}$}\;.
\end{cases}
\end{align}

These estimates imply efficient $s$-wave recoupling throughout the WIMP mass range considered here, from 1~GeV to 10~TeV. For $p$-wave annihilation, the benchmark estimates allow recoupling for $m_\chi\lesssim{\rm TeV}$ at $T_i \sim T_{\rm BBN}$, with heavier WIMPs requiring a higher $T_i$. However, these estimates assume a benchmark weak-scale annihilation cross section, whereas the CFO scenario can accommodate substantially larger cross sections to reproduce the observed relic abundance. An enhancement of the annihilation cross sections lowers $T_{\rm rfo}$ and increases ${\cal R}_p$. Consequently, efficient recoupling with ${\cal R}_p\gg1$ can also occur for $m_\chi>{\rm TeV}$, provided the ordering $T_e<T_{\rm rfo}<T_i$ is maintained.

Following recoupling at $T_i$, the equilibrium WIMP abundance is exponentially suppressed, $Y_{\chi,{\rm eq}}\propto e^{-x_i}$, since $x_i\gg1$. Inverse annihilations of SM particles into WIMPs can therefore be neglected, while the existing WIMP population can still annihilate efficiently. The subsequent evolution in the radiation-dominated era is governed by
\begin{align}
\frac{{\rm d}Y_\chi}{{\rm d}x}= -\frac{s}{x H}\langle\sigma v \rangle\,Y_\chi^2\;.
\end{align}
Neglecting annihilation during the short reheating interval and keeping terms up to $p$-wave order, we integrate the Boltzmann equation from $x_i$ to obtain
\begin{align}
\frac{1}{Y_\chi^{\rm vac}}  &\approx  \frac{1}{Y_d}+\sqrt{\frac{\pi}{45}}\frac{g_{*S}(T_i)}{\sqrt{g_*(T_i)}} m_\chi M_{\rm Pl} \left(\frac{\sigma_0}{x_i} + \frac{3\sigma_1}{x_i^2}\right).\label{eq:yieldrecoupling}  
\end{align}
For the $p$-wave contribution, we have assumed that WIMPs remain in kinetic equilibrium with the thermal bath during freeze-out, so that their mean squared relative velocity satisfies $\langle v^2\rangle\approx6/x$.

For efficient recoupling, ${\cal R}_{s,p}\gg1$, the first term on the right-hand side of Eq.~(\ref{eq:yieldrecoupling}) is negligible. The final WIMP yield then becomes insensitive to the initial relativistic freeze-out yield and approaches the attractor solution
\begin{align}
Y_\chi^{\rm vac} \approx  \sqrt{\frac{
45}{\pi}} \frac{\sqrt{g_*(T_i)}}{g_{*S}(T_i)}\frac{1}{m_\chi M_{\rm Pl}} \frac{x_i}{\sigma_0+3 \sigma_1/x_i}\;.
\end{align}
Comparing with standard freeze-out for the same annihilation cross section, we obtain
\begin{align}
\frac{Y_\chi^{\rm vac}}{Y_\chi^{\rm fo}} = \Xi\left(x_i,x_{\rm fo}\right)\times\begin{cases}
\dfrac{x_i}{x_{\rm fo}}&\text{for $s$-wave}\;, \\[10pt]
\dfrac{x_i^2}{x_{\rm fo}^2}&\text{for $p$-wave}\;.\label{eq:ratio-long-inflation} 
\end{cases}
\end{align}

In contrast to the short-inflation branch result in Eq.~(\ref{eq:ratio-short-inflation}), the final yield in this limit is insensitive to the temperature $T_e$ at which inflation ends. Instead, it depends on $T_i$, the temperature at the onset of vacuum domination and immediately after reheating. For fixed $m_\chi$, lowering $T_i$ therefore increases the annihilation cross section required to reproduce the observed relic abundance. The maximal enhancement is consequently controlled by the lowest $T_i$ consistent with the long-inflation branch condition and the BBN requirement $T_i\gtrsim T_{\rm BBN}$.

\begin{figure}[t!]
	\centering
	\includegraphics[width=6.0in]
	{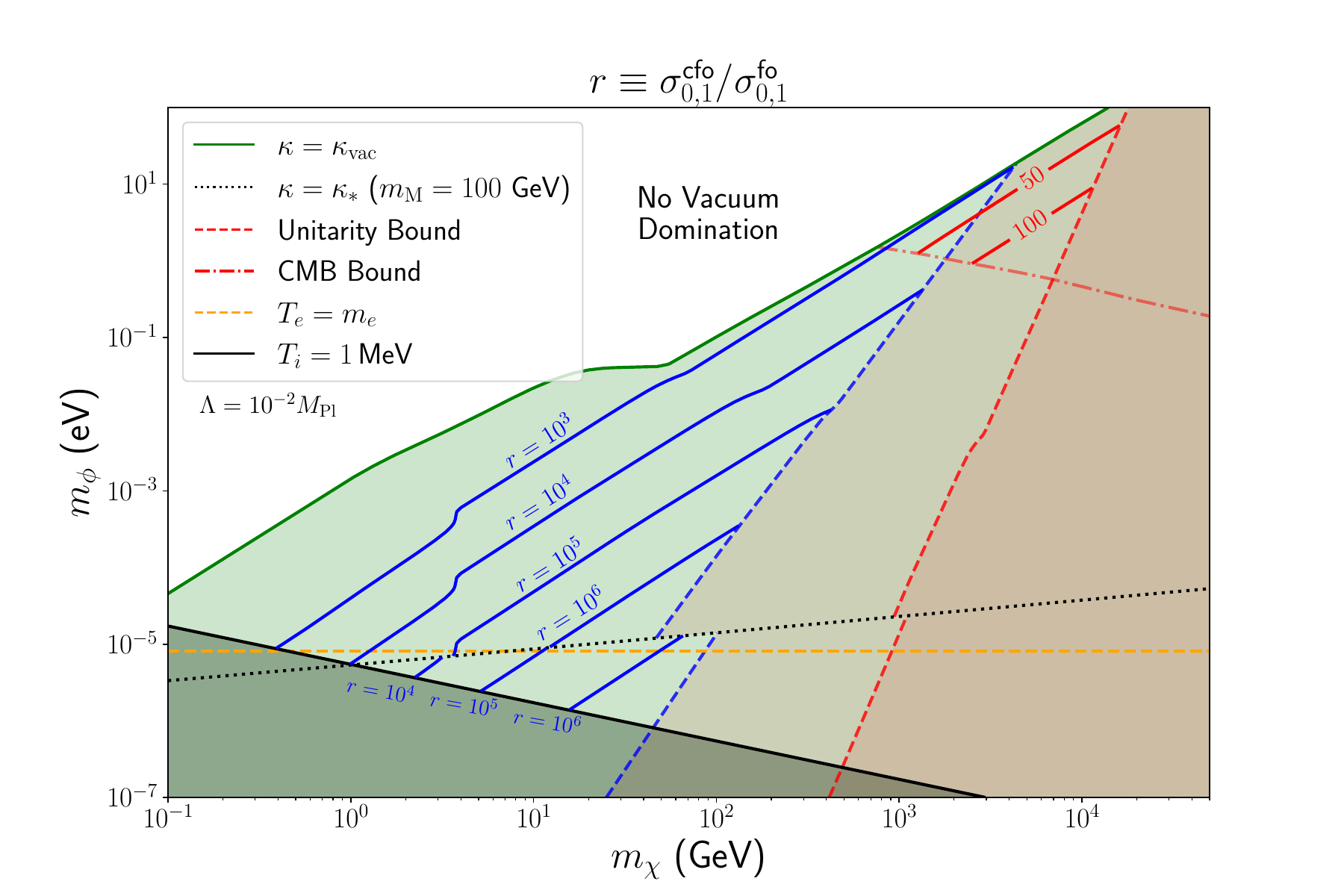}
   \caption{Parameter space of the coherent freeze-out mechanism in the vacuum-dominated regime. The contour lines and unitarity bounds follow the same conventions as in Fig.~\ref{fig:moneyplotFOPT}. The green solid line marks $\kappa=\kappa_{\rm vac}$, above which vacuum domination does not occur. The red dash-dotted line shows the CMB constraint on the $s$-wave annihilation cross section, excluding the region below it. The black dotted line marks $\kappa=\kappa_*$, separating the short- and long-inflation branches. The yellow dashed line marks $T_e=m_e$; below this line, the reheating scenario assumed in Sec.~\ref{subsec:reheating} applies. The black shaded region is excluded by the BBN lower bound on the reheating temperature.}
    \label{fig:moneyplotvac}
\end{figure}

In Fig.~\ref{fig:moneyplotvac}, we show the enhancement of the annihilation cross section relative to standard freeze-out required to reproduce the observed DM relic abundance in the vacuum-dominated regime, for both $s$-wave (red contours) and $p$-wave (blue contours) annihilation. The regions below the red and blue dashed lines violate partial-wave unitarity for $s$-wave and $p$-wave annihilation, respectively. The red dash-dotted line shows the CMB constraint~\cite{Slatyer:2015jla,Slatyer:2015kla,Planck:2018vyg}, which applies only for $s$-wave annihilation. The black dotted line marks $\kappa=\kappa_*$ for $g_\chi=2$ and $m_{\rm M}=100~{\rm GeV}$ [see Eq.~(\ref{eq:kapparegion-no-rfo})], separating the short-inflation branch above the line from the long-inflation branch below it.
The yellow dashed line shows a model-dependent upper bound on $T_e$, below which the instantaneous reheating scenario assumed in Sec.~\ref{subsec:reheating} may apply; this bound could be relaxed in alternative reheating models. The black solid curve marks the BBN lower bound on $T_i$, excluding the region below it. 
To summarize, in the vacuum-dominated regime, the WIMP annihilation cross section can be enhanced by factors of approximately $10^2$ for $s$-wave and $10^6$ for $p$-wave annihilation before reaching the unitarity bounds, while still yielding a viable cosmology and a relic density consistent with observations. These enhancements are substantially larger than those attainable in the non-inflating regime discussed in Sec.~\ref{sec:FOPT}.

\section{Conclusions}
\label{sec:conclusion}

WIMP and ALP are the two most popular particle dark matter candidates, and each is also independently motivated by considerations outside of DM. What if {\it both} particles are present in the theory? 

Our investigation, which began in Ref.~\cite{Ferrante:2025avp} and continued in this paper, shows that even a tiny (e.g. Planck-suppressed) coupling between the WIMP and the ALP can have dramatic effects on their cosmological evolution. Specifically, we consider the coupling in Eq.~(\ref{eq:Leff}), which is well motivated as the dark sector counterpart of the familiar QCD axion-nucleon coupling. We find that at high temperatures in the early Universe, this coupling leads to a spontaneous symmetry breaking in which the ALP acquires a vev due to coherent scattering on the thermal bath of WIMPs. The ALP vev in turn leads to a large, temperature-dependent renormalization of the WIMP mass. As the Universe expands and cools, it goes through an inverse phase transition, in which the ALP vev disappears. The WIMP remains relativistic until the phase transition occurs, and rapidly freezes out at that time -- the phenomenon we called {\it coherent freeze-out}. This resulting WIMP yield is parametrically different from the standard freeze-out case, and as a result the annihilation cross section required to obtain the observed DM may be enhanced by up to several orders of magnitude from the canonical thermal relic value. 
The enhancement can be especially large in scenarios where the phase transition is accompanied by a short burst of inflation (see Sec.~\ref{sec:vacuumdomination}).

The ALP dynamics is also quite different from the usual misalignment scenario: the ALP is stuck at its vev at high temperature, and begins coherent oscillations only after the phase transition. The amplitude of these oscillations is fully determined by the coupled dynamics of the WIMP-ALP system, and any dependence on the incalculable initial value of the ALP field produced during inflation is erased. Intriguingly, the ALP relic density predicted by our scenario matches the observed DM density for very reasonable parameter values motivated by other known scales in physics, namely the weak and the Planck scales. This {\it ALP miracle} provides an attractive alternative mechanism for the production of ALP dark matter in the early Universe. This scenario predicts ALP masses within a specific range, 
well above the canonical axion DM masses; see Eq.~(\ref{eq:minimalALPmass}). We also find that there is a distinct region of parameter space where the ALP miracle scenario is consistent with the canonical QCD axion solution to the strong CP problem (see Sec.~\ref{subsub:QCDaxion}). Moreover, a large fraction of this region lies within the projected sensitivity reach of the IAXO experiment~\cite{IAXO:2019mpb,IAXO:2025ltd} (see Fig.~\ref{fig:ALPMiracleWindow}), making this scenario experimentally testable in the near future. 

Our studies of the WIMP-ALP coupled dynamics in the early Universe, both in Ref.~\cite{Ferrante:2025avp} and in this paper, were based on the effective coupling in Eq.~(\ref{eq:Leff}). In the future, it would be interesting to investigate how this coupling can be incorporated in a more comprehensive theoretical framework for the dark sector. The most appealing avenue is to model the dark sector as a gauge theory with QCD-like confining dynamics. The WIMP field $\chi$ would appear as a fermionic bound state, analogous to the proton in QCD, with a conserved dark baryon number responsible for its stability. If an axion is then added to this theory, the coupling in Eq.~(\ref{eq:Leff}) is generated automatically by the same mechanism as the corresponding axion-nucleon coupling in QCD. An additional direction for model-building is to reduce or eliminate the 
fine-tuning in the ALP mass which occurs in the minimal effective theory for $\kappa\ll 0.1$ (see Appendix~\ref{app:naturalness}). This may be achieved, for example, by introducing additional discrete symmetries, along the lines of recent models of ultralight dark matter and dynamical dark energy~\cite{Hook:2018jle,Brzeminski:2020uhm,Delaunay:2025pho,Delaunay:2026jto,Delaunay:2026fse,Delaunay:2026pnu}. We plan to pursue these questions in future work.

\section*{Acknowledgement}
We are grateful to Daniel Carney for a useful conversation. This work was supported by the NSF grants PHY-2309456 and PHY-2609553. S.F. acknowledges support from the Eric and Wendy Schmidt AI in Science Postdoctoral Fellowship, a program of Schmidt Sciences. 

\begin{appendix}
\section{Naturalness}
\label{app:naturalness}
In the main text, we have assumed that the zero-temperature potential of $\phi$ is well described by a simple quadratic term $m_\phi^2 \phi^2/2$, where $m_\phi$ is understood as the \emph{physical} zero-temperature mass.
However, the quadratic coupling introduced in Eq.~(\ref{eq:Leff}) also induces radiative corrections to the zero-temperature mass of $\phi$. These are captured by the one-loop Coleman-Weinberg potential:
\begin{align}
V_{\rm CW}(\phi) = - \frac{g_\chi}{64\pi^2} \mchieff^4(\phi) \left[\log\left(\frac{\mchieff^2(\phi)}{\mu^2}\right) - \frac{3}{2}
\right],      
\end{align}
where $\mu$ is the renormalization scale and $\mchieff(\phi) = |m_\chi - \phi^2/(2\Lambda)|$. Expanding around $\phi=0$ up to quadratic order yields the radiative correction to the $\phi$ mass:
\begin{align}
\delta m_\phi^2 \sim \frac{g_\chi}{16\pi^2}\frac{m_\chi^3}{\Lambda}\;.\label{eq:radiativecorrection}    
\end{align}
Requiring naturalness, $\delta m_\phi^2 \lesssim m_\phi^2$, implies a lower bound on $m_\phi$:
\begin{align}
m_\phi \gtrsim  \frac{\sqrt{g_\chi}}{4\pi} \frac{m_\chi^{3/2}}{\Lambda^{1/2}} \sim 1\,{\rm eV} \left(\frac{m_\chi}{10~{\rm GeV}}\right)^{3/2}\left(\frac{\Lambda}{M_{\rm Pl}}\right)^{-1/2},    
\end{align}
where we have taken $g_\chi =2$.

The naturalness constraint can be directly translated into a lower bound on $\kappa$. Combining Eqs.~(\ref{eq:kappadef}) and (\ref{eq:radiativecorrection}), we find
\begin{align}
\frac{\delta m_\phi^2}{m_\phi^2} \sim \frac{1}{8\kappa}\;.    
\end{align}
Requiring $\delta m_\phi^2 \lesssim m_\phi^2$ therefore implies
\begin{align}
\kappa \gtrsim 0.1\;.    
\end{align}

Since the critical value separating the crossover and FOPT regimes is $\kappa_c \approx 0.2$, the entire crossover regime is naturally consistent with this naturalness criterion.
In contrast, the strong FOPT regime with $\kappa \ll 0.1$ requires some degree of tuning between the bare ALP mass and the radiative correction. While this conclusion holds in the simplest model considered here, it is possible to introduce additional fields and symmetries that can be used to suppress the Coleman-Weinberg potential. Such model-building can use some of the same ingredients as the recently proposed natural models of ultralight dark matter and dynamical dark energy, e.g.~\cite{Hook:2018jle,Brzeminski:2020uhm,Delaunay:2025pho,Delaunay:2026jto,Delaunay:2026fse,Delaunay:2026pnu}.

\section{False Vacuum Evolution: Analytical Details}
\label{app:falsevacuum}

Starting from the potential in Eq.~(\ref{eq:Vfullseries}), we compute the first and second derivatives:
\begin{align}
\frac{\partial{\cal V}(\varphi,x)}{\partial \varphi} &=\varphi\left[\kappa - \sum_{n=1}^\infty\frac{(-1)^{n-1}}{n} \frac{\gamma^2 K_1(n \gamma x)}{x}
\right],\label{eq:DV}\\
\frac{\partial^2{\cal V}(\varphi,x)}{\partial \varphi^2} &=\kappa-\sum_{n=1}^\infty \frac{(-1)^{n-1}}{n}\frac{\gamma}{x} \left[2 n \gamma x \left(1 -\gamma\right)K_0(n\gamma x) - \left(2 -3 \gamma\right)K_1(n \gamma x) 
\right].\label{eq:DV2}
\end{align}
The symmetry-breaking vacuum $\varphi_*$ is determined by the nonzero solution of Eq.~(\ref{eq:DV}):
\begin{align}
\frac{\partial{\cal V}(\varphi,x)}{\partial \varphi}\Bigg|_{\varphi=\varphi_*}=0 \quad \Rightarrow \quad
\kappa x^3 = (\gamma_* x)^2 \sum_{n=1}^\infty \frac{(-1)^{n-1}}{n} K_1  (n \gamma_* x)\;,\label{eq:phistarEOM}
\end{align}
where $\gamma_* \equiv |1-\varphi_*^2/2|$. For any fixed $\kappa$, one can obtain the corresponding $\varphi_*(x)$ or $\gamma_*(x)$ by solving Eq.~(\ref{eq:phistarEOM}). First, let us consider the relativistic limit $\gamma_*x\ll 1$, then Eq.~(\ref{eq:phistarEOM}) is reduced to
\begin{align}
\gamma_* x = \frac{12}{\pi^2} \kappa x^3\quad \Rightarrow \quad \varphi_*(x) = \sqrt{2\left(1-\frac{12}{\pi^2} \kappa x^2\right)}\;, \quad \text{for $x \ll \kappa^{-1/3}$}\;.\label{eq:phistarrelativistic}
\end{align}
The key observation is that, for a fixed $\kappa$, one has $\gamma_* x \gtrsim {\cal O}(1)$ once $x\gtrsim \kappa^{-1/3}$. At this point, the right-hand side of Eq.~(\ref{eq:phistarEOM}) becomes Boltzmann suppressed, and the equation can no longer be satisfied. This corresponds to the temperature at which the local minimum at $\varphi_*$ disappears; we denote this special temperature by $\xcfo$, which also marks the onset of coherent freeze-out.

From Eq.~(\ref{eq:phistarrelativistic}), we can already see that $\xcfo \sim \kappa^{-1/3}$. In order to determine the coefficient more precisely, we use the definition of $\xcfo$:
\begin{align}
\frac{\partial^2{\cal V}(\varphi,\xcfo)}{\partial  \varphi^2}\Bigg|_{\varphi=\varphi_*}=0  \quad \Rightarrow \quad \kappa =\frac{\tilde{\gamma}_*}{\xcfo}\sum_{n=1}^\infty \frac{(-1)^{n-1}}{n} \left[2 n \tilde{\gamma}_* \xcfo \left(1 -\tilde{\gamma}_*\right)K_0(n\tilde{\gamma}_* \xcfo) - \left(2 -3 \tilde{\gamma}_*\right)K_1(n \tilde{\gamma}_* \xcfo) 
\right],\label{eq:DV2vanish}
\end{align}
where $\tilde{\gamma}_* \equiv \gamma_*(\xcfo)$.
Substituting Eq.~(\ref{eq:DV2vanish}) into Eq.~(\ref{eq:phistarEOM}) and evaluating at $x=\xcfo$, we arrive at
\begin{align}
\sum_{n=1}^\infty (-1)^{n-1} \left[K_0(n\tilde{\gamma}_* \xcfo) - \frac{K_1(n \tilde{\gamma}_* \xcfo)}{n \tilde{\gamma}_*\xcfo }
\right]=0\;.\label{eq:transcendentalgammastarxcfotilde}   
\end{align}
As a transcendental equation of $\tilde{\gamma}_* \xcfo$, Eq.~(\ref{eq:transcendentalgammastarxcfotilde}) can be solved numerically. The series appearing in Eq.~(\ref{eq:transcendentalgammastarxcfotilde}) converges rapidly and is effectively stable for $n \gtrsim 5$. Numerically, we find the solution:
\begin{align}
\tilde{\gamma}_*\xcfo \equiv c_1 \approx  1.429\;.\label{eq:c1def}   
\end{align}
Substituting it back into Eq.~(\ref{eq:phistarEOM}) and evaluating at $x=\xcfo$, we obtain
\begin{align}
\kappa \xcfo^3 = c_1^2 \sum_{n=1}^\infty \frac{(-1)^{n-1}}{n} K_1  (n c_1) \equiv c_2 \approx 0.585\;,\label{eq:c2def}
\end{align}
which leads to 
\begin{align}
\xcfo = (c_2/\kappa)^{1/3} \approx 0.836\, \kappa^{-1/3}\;.\label{eq:xcfokappa}
\end{align}

\begin{figure}[t!]
\centering
\includegraphics[width=0.6\textwidth]
{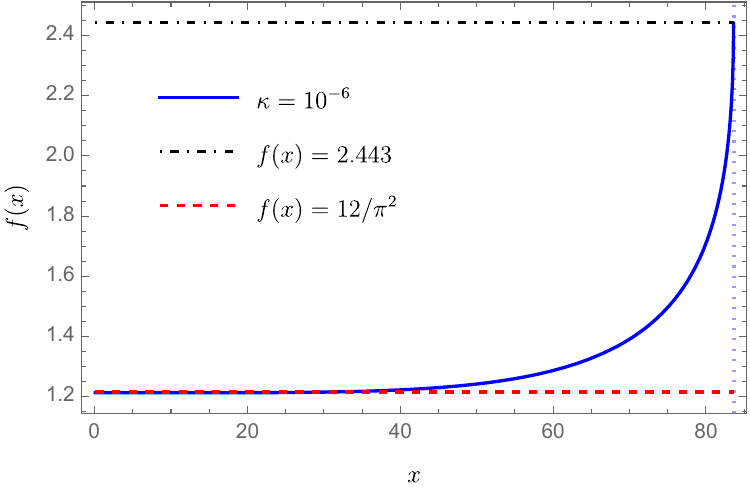}
\caption{\label{fig:fx} The function $f(x)$ in Eq.~(\ref{eq:phistarapp}), obtained by numerically solving Eq.~(\ref{eq:phistarEOM}). For illustration, we fix $\kappa = 10^{-6}$. The vertical dotted line indicates $x = \xcfo$, beyond which the local minimum $\varphi_*$ disappears and $f(x)$ is no longer well defined. The function approaches $12/\pi^2 \approx 1.216$ (red dashed line) for $x \ll \xcfo$  and saturates at $c_1/c_2 \approx 2.443$ (black dash-dotted line) at $x = \xcfo$.}
\end{figure}

Therefore, the evolution of the false vacuum can be described by
\begin{align}
\varphi_*(x) = \sqrt{2\left(1-f(x)\,\kappa x^2\right)}\;,\label{eq:phistarapp}    
\end{align}
where $f(x)$ is a monotonically increasing function with the asymptotic behavior (see Fig.~\ref{fig:fx})
\begin{align}
f(x \ll \xcfo) = \frac{12}{\pi^2} \approx 1.216 \;, \qquad
f(\xcfo) = \frac{c_1}{c_2} \approx 2.443\;.
\end{align}

Note that the derivative of $\varphi_*(x)$ becomes divergent as $x \to \xcfo$ from below, even though the value of $\varphi_*(x)$ itself remains finite. More specifically, expanding  Eq.~(\ref{eq:phistarEOM}) around $\xcfo$, we find the following scaling behavior:
\begin{align}
\gamma_*(x \to \xcfo^{-}) \approx \gamma_*(\xcfo) - c_4\,\kappa^{1/2}\, \left(\xcfo-x\right)^{1/2}\;, \label{eq:gammaxcfoscaling}
\end{align}
where the prefactor is given by
\begin{align}
c_4 \equiv \sqrt{\frac{6}{3\,c_2/c_1^2-c_1^2 c_3}} \approx 3.955\;,\qquad
 c_3 \equiv \sum_{n=1}^{\infty}(-1)^{n-1} n K_1(nc_1) \approx 0.233\;.
\end{align}
Using $\gamma_*'(x) = -\varphi_*(x)\, \varphi_*'(x)$, this implies
\begin{align}
\varphi_*'(x \to \xcfo^{-}) \approx  - \frac{c_4}{2\sqrt{2\left(1-c_1/\xcfo\right)}}\,\frac{\kappa^{1/2}}{\sqrt{\xcfo -x}}\;.   
\end{align}
Therefore, although both $\gamma_*(x)$ and $\varphi_*(x)$ remain finite at $x=\xcfo$,
their derivatives diverge as $(\xcfo-x)^{-1/2}$ in the limit $x \to \xcfo$ from below. 

For $x>\xcfo$, the local minimum at $\varphi_*$ disappears and the field relaxes to the origin, exhibiting matter-like redshift behavior. Since the physical field value $\varphi(x)$ must be continuous as a function of temperature, it takes the following piecewise form:
\begin{align}
\varphi(x) = 
\begin{cases}
\sqrt{2\left(1-f(x)\,\kappa x^2\right)}  & \text{for}\;x < \xcfo\;,\\[10pt]
\sqrt{2\left(1-\dfrac{c_1}{\xcfo}\right)}\left(\dfrac{\xcfo}{x}\right)^{3/2}  & \text{for}\;x > \xcfo\;.
\end{cases}
\end{align}

\section{Analytical Derivation of the Bounce Action in the Thick-Wall Limit}
\label{app:thickwall}

The bounce action in the thick-wall limit rapidly decreases as $x$ approaches $\xcfo$. We analytically compute the bounce action near $\xcfo$ by expanding $M^2$ and $\delta$ in terms of the small parameter $\Delta\equiv \xcfo -x$ to leading order,
\begin{align}
    M^2(\Delta) &= a\ \Delta^{p}\,,\label{eq:DV2power}\\
    \delta(\Delta) &= \frac{b}{2}\,.\label{eq:DV3power}
\end{align}
The bounce action in Eq.~(\ref{eq:thickbounce}) then has power-law scaling,
\begin{equation}
    S_B = \tilde{S}_3 \frac{9\xcfo a^{\frac{3}{2}}\Delta^{\frac{3}{2}p}}{m_\chi b^2}\,.
\end{equation}
To find the scaling power $p$, we substitute Eq.~(\ref{eq:gammaxcfoscaling}) into Eq.~(\ref{eq:DV2}),
\begin{align*}
    \frac{\partial^2{V}}{\partial \phi^2} \Bigg |_{\phi=\phi_*} &= \frac{g_\chi m_\chi^3}{2\pi^2 \Lambda} \frac{\partial^2{\cal V}(\varphi,x)}{\partial \varphi^2}\Bigg |_{\varphi=\varphi_*}\\
    &= \frac{g_\chi m_\chi^3}{2\pi^2 \Lambda} \Bigg(\kappa - \sum_{n=1}^{\infty} \frac{(-1)^{n-1}}{n\xcfo}\bigg(2n\xcfo(1-\tilde{\gamma}_*+c_4\kappa^{\frac{1}{2}}\Delta^{\frac{1}{2}})(\tilde{\gamma}_*-c_4\kappa^{\frac{1}{2}}\Delta^{\frac{1}{2}})^2 K_0(n\xcfo(\tilde{\gamma}_*-c_4\kappa^{\frac{1}{2}}\Delta^{\frac{1}{2}})) -\\
    &\qquad\qquad (\tilde{\gamma}_*-c_4\kappa^{\frac{1}{2}}\Delta^{\frac{1}{2}})(2-3\tilde{\gamma}_*+3c_4\kappa^{\frac{1}{2}}\Delta^{\frac{1}{2}})K_1(n\xcfo(\tilde{\gamma}_*-c_4\kappa^{\frac{1}{2}}\Delta^{\frac{1}{2}}))\bigg) \Bigg)\,.
\end{align*}
The zeroth-order (constant) term vanishes exactly as Eq.~(\ref{eq:DV2vanish}), so the leading-order term scales like $\Delta^{\frac{1}{2}}$ ($p=\frac{1}{2})$, and the constant $a$ of Eq.~(\ref{eq:DV2power}) is given by
\begin{equation}
    a = \frac{g_\chi m_\chi^3}{2\pi^2 \Lambda}\sum_{n=1}^{\infty} \frac{(-1)^{n-1}}{n\xcfo}\bigg(3 n \xcfo (2-3\tilde{\gamma}_*)K_0(n \xcfo \tilde{\gamma}_*)+(3-2n^2\xcfo^2(1-\tilde{\gamma}_*)\tilde{\gamma}_*)K_1(n \xcfo \tilde{\gamma}_*)\bigg)\tilde{\gamma}_* c_4 \kappa^{\frac{1}{2}}\,.
\end{equation}
The bounce action then rapidly decreases as $x$ approaches $\xcfo$ with scaling behavior $S_B \propto \Delta^{\frac{3}{4}}$.\\

To find $b$, we take the third derivative of the potential,
\begin{align*}
    \frac{\partial^3{V}}{\partial \phi^3}\Bigg |_{\phi=\phi_*} &= \frac{g_\chi m_\chi^4}{2\pi^2 (m_\chi \Lambda)^{3/2}} \frac{\partial^3{\cal V}(\varphi,x)}{\partial \varphi^3}\Bigg |_{\varphi=\varphi_*}\\
    &= \frac{g_\chi m_\chi^4}{2\pi^2 (m_\chi \Lambda)^{3/2}} \left[\varphi_* \sum_{n=1}^{\infty}\frac{(-1)^{n-1}}{nx}\gamma_*\bigg(3nx(2-3\gamma_*)K_0(nx\gamma_*)+(3-2n^2x^2\gamma_*(1-\gamma_*))K_1(nx\gamma_*)\bigg)\right]\,.
\end{align*}
Again substituting in Eq.~(\ref{eq:gammaxcfoscaling}) near $\xcfo$, we find the constant $b$ of Eq.~(\ref{eq:DV3power}) is 
\begin{equation}
    b = \frac{g_\chi m_\chi^4}{2\pi^2 (m_\chi\Lambda)^{3/2}}\sum_{n=1}^{\infty} \frac{(-1)^{n-1}}{n\xcfo} \tilde{\varphi}_*\tilde{\gamma}_*\bigg(3n\xcfo(2-3\tilde{\gamma}_*)K_0(n\xcfo\tilde{\gamma}_*) + (3-2n^2\xcfo^2(1-\tilde{\gamma}_*)\tilde{\gamma}_*)K_1(n\xcfo\tilde{\gamma}_*)\bigg)\,.
\end{equation}

Additionally, we find the leading-order dependence of $S_B$ on the model parameters $\kappa$, $m_\chi$ and $\Lambda$. The  coefficients $a$ and $b$ depend implicitly on $\kappa$ through $\tilde{\gamma}_*$ and $\xcfo$. We use the scaling relations Eqs.~(\ref{eq:c1def}), (\ref{eq:xcfokappa}) and find to leading order,
\begin{align}
\left.
\begin{array}{ll}
a &\propto \dfrac{m_\chi^3}{\Lambda}\kappa^{5/6}\\
\\
b &\propto \dfrac{m_\chi^{5/2}}{\Lambda^{3/2}}\kappa^{1/3}
\end{array}
\right\} \implies S_B \propto \left(\dfrac{\Lambda}{m_\chi}\right)^{3/2}\kappa^{1/4}\,.
\end{align}

We also collect the leading model-independent constants in $a$ and $b$, and compute the overall constant $A$ of Eq.~(\ref{eq:thickwallbouncescaling}) numerically, with the same precision up to the $n=5$ term in the infinite sum for consistency:
\begin{align}
    A \approx \frac{9\tilde{S}_3 c_2^{1/3}a_0^{3/2}}{b_0^2} \approx 
\begin{cases}
1300 & \text{for $g_\chi=4$}\;,\\ 
1900 & \text{for $g_\chi=2$}\;,
\end{cases}    
\end{align}
with $a_0, b_0$ denoting the model-independent constants of $a, b$ respectively,
\begin{equation}
    a \equiv a_0 \dfrac{m_\chi^3}{\Lambda}\kappa^{5/6}\,, \quad b \equiv b_0 \dfrac{m_\chi^{5/2}}{\Lambda^{3/2}}\kappa^{1/3}\,.
\end{equation}

\section{Higher-order Corrections to Coherent Freeze-Out in the Heavier WIMP Mass Region}
\label{app:higherorderheavywimp}
The derivation of $\Delta x$ in the heavier-WIMP mass region, Eq.~(\ref{eq:xcfohatheavy}), relies on an approximate expression for $Y'_{\chi,{\rm eq}}/Y_{\chi,{\rm eq}}$ [see Eq.~(C16) of Ref.~\cite{Ferrante:2025avp}]. In this appendix, we refine the derivation by including higher-order corrections.

For $x>\xcfo$, using the non-relativistic expansion $Y_{\chi,{\rm eq}} = a (\gamma x)^{3/2} e^{-\gamma x}$ at $\gamma x \gg 1$, where $a\equiv 45 g_\chi/(4\sqrt{2} \pi^{7/2}  g_{*S})$, we have
\begin{align}
    \frac{Y'_{\chi,\text{eq}}}{Y_{\chi,\text{eq}}} 
    = 
    \frac{3}{2}\frac{(\gamma x)'}{\gamma x} 
    - 
    (\gamma x)'\;,
\end{align}
where the prime denotes the derivative with respect to $x$. Thus, for $ \xcfo < x < \xcfohat$, the Boltzmann equation (\ref{eq:BoltzmannEquationheavyregion}) is reduced to
\begin{align}
    \Delta \equiv Y_\chi-Y_{\chi,{\rm eq}} \approx -\frac{\gamma^2 x^2}{2\lambda} \frac{Y'_{\chi,{\rm eq}}}{Y_{\chi,{\rm eq}}} 
    = \frac{ \gamma^{2} x^{2}}{2\lambda} 
    \bigg(1-\frac{3}{2\gamma x}\bigg) 
    (\gamma x)'\;.
\end{align}
The coherent freeze-out temperature $\xcfohat \equiv \xcfo + \Delta x$ is then found by setting $\Delta(\xcfohat)=Y_{\chi,\text{eq}}(\xcfohat)$, yielding
\begin{align}
    e^{\gamma x}\Bigg|_{\xcfohat} = 
    2 a \lambda 
    (\gamma x)^{-1/2} (\gamma x)'^{-1}
    \left(1 - \frac{3}{2\gamma x}\right)^{-1}
    \Bigg|_{\xcfohat}.
\end{align}

On the other hand, we have $\gamma x = x- \xcfo^3/x^2 + c_1 \xcfo^2/x^2$ for $x>\xcfo$ [cf. Eq.~(\ref{eq:mchieffoverT})]. Expanding $\gamma x$ and its derivative around $x = \xcfohat$, 
\begin{align}
    \gamma x\bigg|_{\xcfohat}
    &= 3\Delta x + c_1 
    - \frac{3\Delta x^2 + 2c_1 \Delta x}{\xcfohat}
    + \mathcal{O}(\xcfohat^{-2}) \equiv  A + B\epsilon + \mathcal{O}\left(\epsilon^2\right),
    \\
    (\gamma x)'\Bigg|_{\xcfohat}
    &= 3 
    - \frac{6\Delta x + 2c_1}{\xcfohat}
    + \mathcal{O}(\xcfohat^{-2})\equiv A' + B'\epsilon + \mathcal{O}\left(\epsilon^2\right),
\end{align}
where we have defined the shorthand $\epsilon \equiv \Delta x/\xcfohat$, $A\equiv 3\Delta x + c_1$, $A' = 3$, $B = -(3\Delta x + 2c_1)$, and $B'=-(6+2c_1/\Delta x)$. The condition to find $\xcfohat$ then becomes 
\begin{align}
    A + B\epsilon 
    = 
    \underbrace{
    \log \left(2a\lambda\right)
    - \frac{1}{2}\log\left(A+B\epsilon\right) 
    - \log\left(A' + B'\epsilon\right) 
    - \log\left(1 - \frac{3}{2\left(A+B\epsilon\right)}\right)
    }_{R},
    \label{eq:xcfocondition}
\end{align}
where both $a$ and $\lambda$ are evaluated at $x=\xcfohat$. Defining $R$ as the right-hand side of Eq.~(\ref{eq:xcfocondition}) and applying the quadratic formula, we obtain
\begin{align}
    \Delta x = \frac{\xcfohat}{6} 
    \left( 
    3 - \frac{2 c_1}{\xcfohat}
    \mp 
    \sqrt{\left(3 - \frac{2 c_1}{\xcfohat}\right)^2 - \frac{12}{\xcfohat}\left(R-c_1\right)}
    \right). 
\end{align}
Only the branch with the minus sign reproduces the expected limit $\Delta x\to(R-c_1)/3$ for $\xcfohat\gg1$ and therefore corresponds to the physical solution. Thus, we end up with
\begin{align}
\Delta x = \frac{2\left(R-c_1\right)}{3 - \dfrac{2 c_1}{\xcfohat}+\sqrt{\left(3 - \dfrac{2 c_1}{\xcfohat}\right)^2 - \dfrac{12\left(R-c_1\right)}{\xcfohat}}}\;. \label{eq:deltaxhigherorder}    
\end{align}
Since both $R$ and $\xcfohat=x_{\rm cfo}+\Delta x$ depend on $\Delta x$, Eq.~(\ref{eq:deltaxhigherorder}) is implicit. It can be solved iteratively using Eq.~(\ref{eq:xcfohatheavy}) to initialize $\Delta x$ and updating both $R$ and $\xcfohat$ at each iteration.

For $x>\xcfohat$, one can ignore the equilibrium yield in Eq.~(\ref{eq:BoltzmannEquationheavyregion}), and integrate it from $\xcfohat$ to find the $\chi$ yield today ($Y_\chi^\text{cfo}\equiv Y_\chi(\infty))$: 
\begin{align}
\frac{1}{Y_\chi^\text{cfo}} - \frac{1}{Y_\chi(\xcfohat)}
    &=
    \int_{\xcfohat}^\infty{\rm d}x\,
    \frac{ \lambda}{\gamma^2 x^2}\;,   
\end{align}
where $\lambda$ is defined in Eq.~(\ref{eq:lambdadef}). Using the expression of $\gamma x$ in Eq.~(\ref{eq:mchieffoverT}) to perform the integral and expanding up to $\mathcal{O}(\Delta x / \xcfohat)$ gives
\begin{align}
    \frac{1}{Y_\chi^\text{cfo}}\Bigg|_\text{$s$-wave} 
    &\approx \frac{1}{Y_\chi(\xcfohat)}+
    \lambda_s 
    \left[ 
    \frac{1}{3A} + \frac{1}{\xcfohat}
    \left( 
    \frac{2}{9}\log\frac{3\sqrt{3}\,\xcfohat}{A} + \frac{\pi}{9\sqrt{3}}
    + 
    \frac{2\Delta x}{3A} - \frac{\Delta x^2}{A^2}
    \right)
    \right],\\
    \frac{1}{Y_\chi^\text{cfo}}\Bigg|_\text{$p$-wave} 
    &\approx \frac{1}{Y_\chi(\xcfohat)}+
    \frac{6\lambda_p}{\xcfohat} 
    \left[
    \frac{1}{3A} + \frac{1}{\xcfohat}
    \left( 
    \frac{1}{9}\log\frac{3\sqrt{3}\,\xcfohat}{A} - \frac{\pi}{18\sqrt{3}}
    + \frac{\Delta x}{3A} +
    \frac{c_1 \Delta x}{3A^2}
    \right)
    \right],
\end{align}
where 
\begin{align}
\lambda_{s,p} \equiv \sqrt{\frac{\pi}{45}}\frac{g_{*S}(\xcfohat)}{\sqrt{g_*(\xcfohat)}} m_\chi M_{\rm Pl} \sigma_{0,1}\;.
\end{align}
Using $Y_\chi(\xcfohat)=2Y_{\chi,{\rm eq}}(\xcfohat)$, we find that the matching contribution $1/Y_\chi(\xcfohat)$ is smaller than $1/Y_\chi^{\rm cfo}$ by a factor of approximately $3\Delta x+c_1\sim{\cal O}(20)$ and can therefore be neglected. Taking $\sigma_i^{\rm fo}=\sigma_i^{\rm cfo}$, the yield ratio between coherent freeze-out and ordinary freeze-out is then: 
\begin{align}
    \frac{Y_\chi^\text{cfo}}{Y_\chi^\text{fo}}\Bigg|_\text{$s$-wave} 
    &= 
    \frac{\Xi\left(\xcfohat,x_{\rm fo}\right)}{x_\text{fo}} 
    \left[ 
    \frac{1}{3A} 
    + \frac{1}{\xcfohat}
    \left(
    \frac{2}{9}\log\frac{3\sqrt{3}\,\xcfohat}{A} + \frac{\pi}{9\sqrt{3}}
    + 
    \frac{2\Delta x}{3A} - \frac{\Delta x^2}{A^2}
    \right)
    \right]^{-1}, \label{eq:rsapp}\\
    \frac{Y_\chi^\text{cfo}}{Y_\chi^\text{fo}}\Bigg|_\text{$p$-wave} 
    &= \frac{\xcfohat\, \Xi\left(\xcfohat,x_{\rm fo}\right)}{2x_{\rm fo}^2}    
    \left[
    \frac{1}{3A} + \frac{1}{\xcfohat}
    \left( 
    \frac{1}{9}\log\frac{3\sqrt{3}\,\xcfohat}{A} - \frac{\pi}{18\sqrt{3}}
    + \frac{\Delta x}{3A} +
    \frac{c_1 \Delta x}{3A^2}
    \right)
    \right]^{-1},\label{eq:rpapp}
\end{align}
with $\Delta x$ determined by Eq. (\ref{eq:deltaxhigherorder}). It is easy to check that their leading-order results in the large-$\xcfohat$ limit reduce to Eq.~(\ref{eq:yieldratiocases}).

To reproduce the observed relic abundance, the required enhancements of the $s$-wave and $p$-wave annihilation cross sections relative to ordinary freeze-out are 
\begin{align}
    r_s  =
        \frac{Y_\chi^\text{cfo}}{Y_\chi^\text{fo}} 
    \Bigg|_\text{$s$-wave}\;, \qquad\qquad
    r_p =
        \frac{Y_\chi^\text{cfo}}{Y_\chi^\text{fo}} 
    \Bigg|_\text{$p$-wave}\;,
\end{align}
where the yield ratios are evaluated at equal annihilation cross sections, given by Eqs.~(\ref{eq:rsapp})-(\ref{eq:rpapp}). These expressions provide more accurate predictions for the solid constant-$r$ contours in Fig.~\ref{fig:moneyplotFOPT}. The enhancement factors are also constrained by partial-wave unitarity:
\begin{align}
    r_s < \frac{4\pi}{m_\chi^2v \sigma_0}\;, \qquad
    r_p < \frac{12\pi}{m_\chi^2 v^3 \sigma_1}\;,
\end{align}
where $v$ is the relative velocity of the annihilating WIMPs at freeze-out, and $\sigma_0$ and $\sigma_1$ are the reference cross-section coefficients for ordinary freeze-out. Although $\xcfo$ depends on $m_\chi$ and $m_\phi$ through a power law, the enhancement factors contain additional logarithmic dependence on $\xcfo$. Consequently, the unitarity boundaries in Fig.~\ref{fig:moneyplotFOPT} are curved rather than straight in logarithmic coordinates. To obtain the unitarity bounds shown in Fig.~\ref{fig:moneyplotFOPT}, we take $v=0.5$ and adopt the thermal relic values $\sigma_0 = 2.2\times 10^{-26}~{\rm cm^3/s}= 1/(23 \,\text{TeV})^2$ and $\sigma_1 = 1.9 \times 10^{-25}~{\rm cm^3/s} = 1/(7.8\,\text{TeV})^2$.

\end{appendix}

\bibliographystyle{JHEP}
\bibliography{ref}
\end{document}